\documentclass[preprint,journal]{vgtc} 

\onlineid{1823}

\vgtccategory{Research}

\title{A Query-Efficient Stochastic Volume Rendering Framework for Time-Varying Implicit Neural Volumes}

\author{%
  \authororcid{Alper Sahistan}{0000-0002-3480-7713},
  \authororcid{Haichao Miao}{0000-0001-6580-2918},
  \authororcid{Zhimin Li}{0000-0003-4324-741X},
  \authororcid{Peer-Timo Bremer}{0000-0003-4107-3831},
  \authororcid{Joshua A. Levine}{0000-0002-4302-1704},
  \authororcid{Valerio Pascucci}{0000-0002-8877-2042}
}

\authorfooter{
  \item
  	Alper Sahistan and Valerio Pascucci are with the University of Utah.
  \item
  	Haichao Miao and Peer-Timo Bremer are with the Lawrence Livermore National Laboratory.
  \item Zhimin Li is with Vanderbilt University.
  \item Joshua A. Levine is with the University of Arizona.
}

\abstract{%
Time-varying implicit neural representations (INRs) provide a compact representation of scientific volumes and, for modalities such as dynamic X-ray computed tomography (CT), are often the only practical way to represent the data.
However, interactive volume rendering of INRs is challenging, as cheap memory lookups are replaced by expensive neural inferences, hindering the performance. Therefore, conventional volume rendering methods such as ray marching with dense sampling are often impractical. While resampling, caching, and retraining can mitigate this cost, they compromise convenience and accuracy and become impractical for time-varying data. We tackle these challenges using a query-efficient stochastic volume rendering framework based on delta tracking. Our system employs a four-stage pipeline that exploits heterogeneous parallelism, using ray tracing cores for traversal and tensor cores for batched neural evaluation. Furthermore, we present strategies to reduce INR queries via ray budgeting and query pruning, thereby increasing per-frame performance. Using our renderer, many time-varying INRs can be rendered directly from their original representation. The system achieves $\sim30$--$40$ FPS at $1024^2$ resolution on an RTX 4090 GPU and converges to high-fidelity images. Moreover, the system enables interactive temporal exploration of the continuous domain, with timestep updates taking approximately $1$--$2$ ms.
}

\keywords{Volume rendering, ray tracing, implicit neural representations, time series data, scientific visualization.}

\newlength{\imgshift}
\teaser{
\centering
\begin{minipage}{1.15\linewidth}
\hspace{-1.2em}
\fbox{%
  \begin{minipage}{0.18\linewidth}
    \centering
    {$t=-0.9$}\\
    \noindent\hbox to \linewidth{\hfil\hspace{\imgshift}%
        \hfil
      \includegraphics[width=0.99\linewidth]{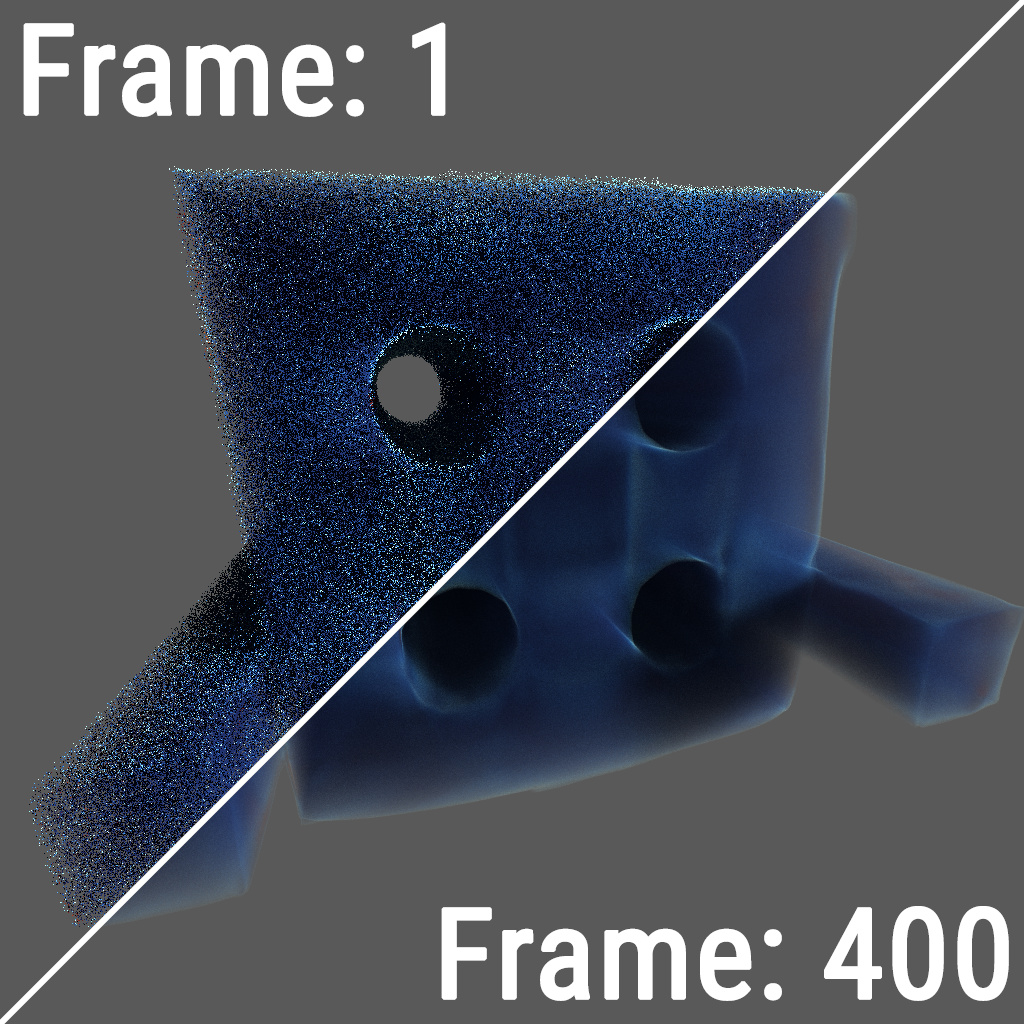}%
    \hfil}\\[-2pt]
    {\small 30.77 FPS}
  \end{minipage}
}
\begin{minipage}[c]{0.1\linewidth}
  \centering
  \includegraphics[width=\linewidth, height=0.3\linewidth]{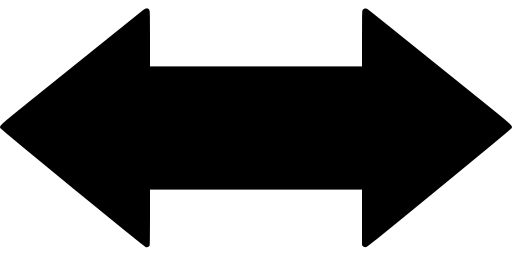}
  {\small 
  \textbf{update:} 1.10 ms}
\end{minipage}
\fbox{%
  \begin{minipage}{0.18\linewidth}
    \centering
    {$t=0.33$}\\
    \noindent\hbox to \linewidth{\hfil\hspace{\imgshift}%
    \hfil
      \includegraphics[width=0.99\linewidth]{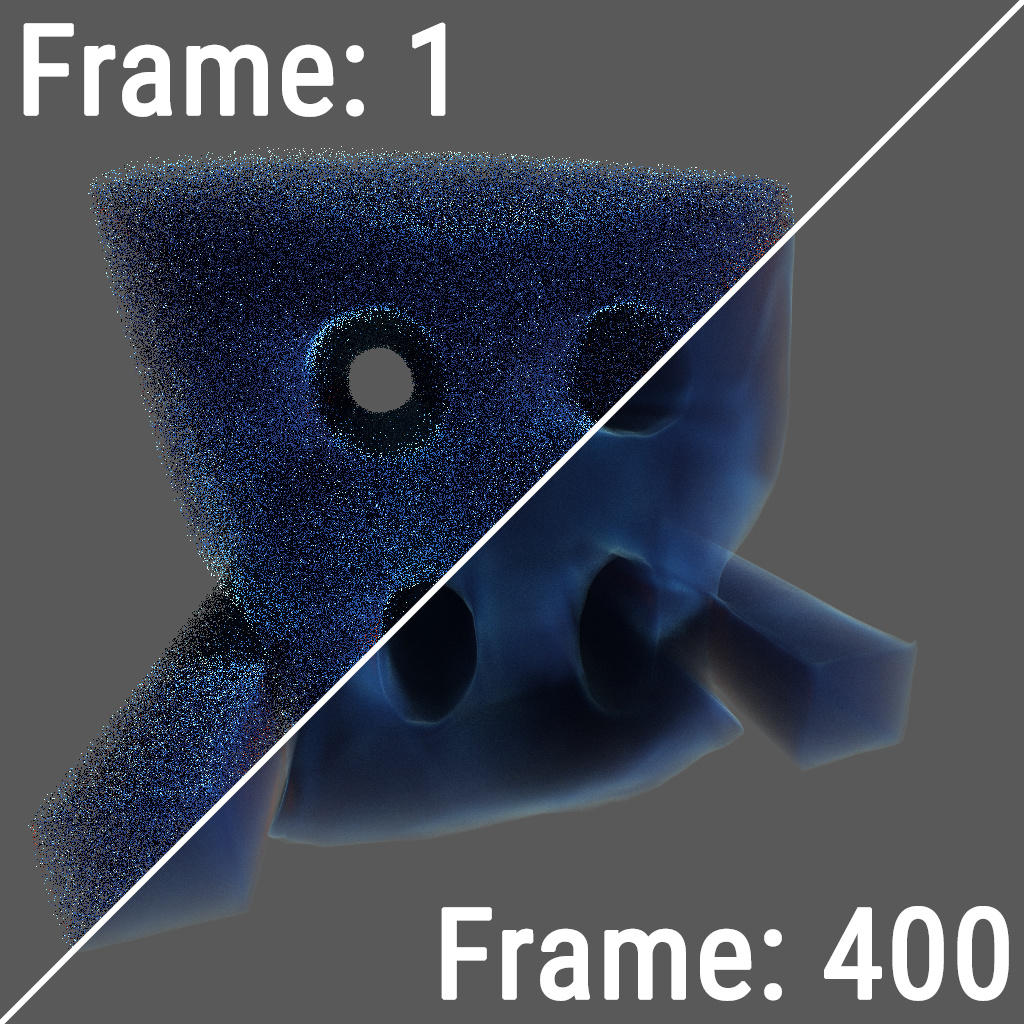}%
    \hfil}\\[-2pt]
    {\small 29.85 FPS}
  \end{minipage}
}
\begin{minipage}[c]{0.1\linewidth}
  \centering
  \includegraphics[width=\linewidth, height=0.3\linewidth]{figs/arrow.png}
  {\small 
  \textbf{update:} 1.06 ms}
\end{minipage}
\fbox{%
  \begin{minipage}{0.18\linewidth}5
    \centering
    {$t=0.9$}\\
    \noindent\hbox to \linewidth{\hfil\hspace{\imgshift}%
    \hfil
      \includegraphics[width=0.99\linewidth]{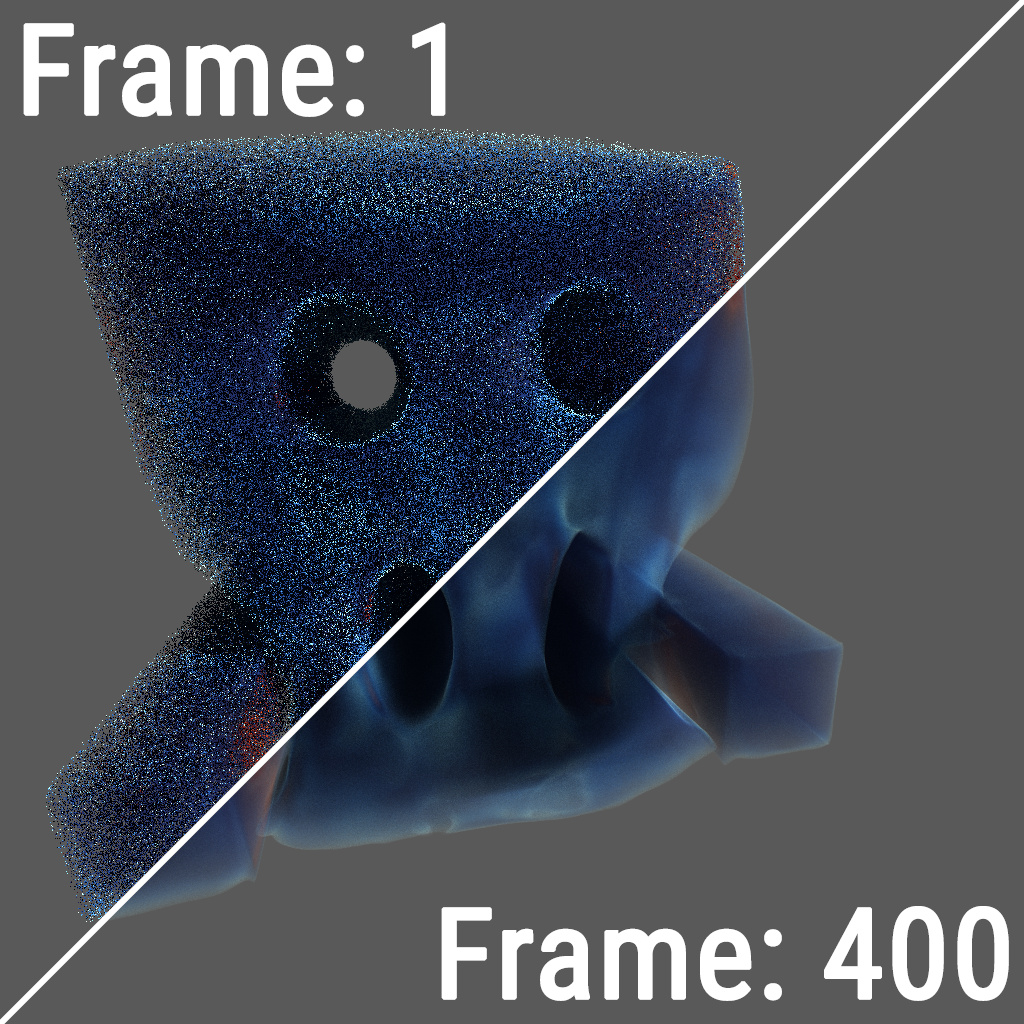}%
    \hfil}\\[-2pt]
    {\small 31.45 FPS}
  \end{minipage}
}
\begin{minipage}[c]{0.2\linewidth}
  \includegraphics[width=0.7\linewidth]{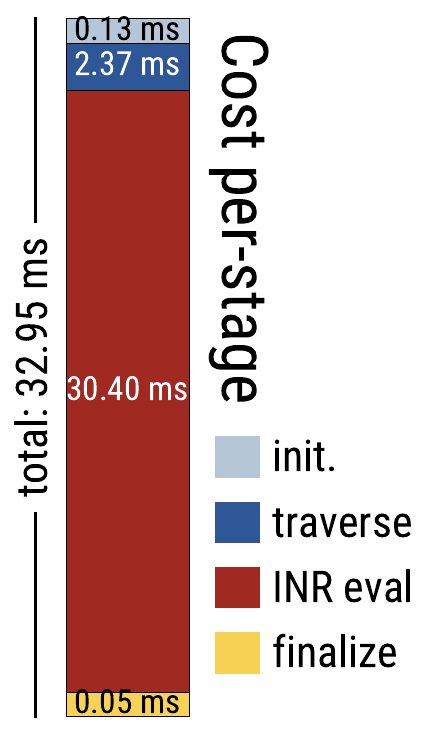}
\end{minipage}
\end{minipage}

\caption{
Volume-rendered frames of a time-varying implicit neural representation (INR) encoding X-ray tomography of a physically deforming object, in which Fourier-feature-encoded coordinates $(x,y,z,t)$ are evaluated by a six-layer fully connected network to produce the scalar field. Despite neural inference dominating runtime (as shown in the bar chart on the right), our query-efficient ray tracing framework renders the original network directly, without retraining or modification, at $\sim30.7$ FPS and $1024^2$ resolution with ray-traced shadows, and enables interactive animation between timesteps ($\sim$1 ms) by leveraging the continuous nature of the INR.
}
  \label{fig:teaser}

}

\graphicspath{{figs/}{figures/}{pictures/}{images/}{./}} 

\usepackage{tabu}                      
\usepackage{booktabs}                  
\usepackage{lipsum}                    
\usepackage{mwe}                       
\usepackage{ccicons}                   

\usepackage{algorithm}
\usepackage{algpseudocode}
\usepackage{stackengine}
\usepackage{framed}
\usepackage{afterpage}
\usepackage{amsfonts}
\usepackage{calc}
\usepackage{enumitem}
\usepackage{diagbox}
\usepackage{graphicx} 

\usepackage{mathptmx}                  
\usepackage{multirow}
\usepackage{amsmath}
\usepackage[svgnames,dvipsnames]{xcolor}   
\usepackage{listings}
\usepackage{adjustbox}
\usepackage{makecell}
\usepackage{wrapfig}

\usepackage{tikz}
\usetikzlibrary{arrows.meta}

\usepackage[moderate,lists=normal]{savetrees}
\lstdefinestyle{rtKernel}{
  language=C++,
  basicstyle={\footnotesize\ttfamily\color{black}},
  commentstyle={\color{ForestGreen}},
  frame=single,
  framerule=0.4pt,
  framesep=3pt,
  xleftmargin=0pt,
  xrightmargin=0pt,
  numbers=none,
  numbersep=4pt,
  keywordstyle={\color{blue}},
  keywordstyle=[2]{\color{magenta}},
  numberstyle={\tiny\color{teal}},
  showspaces=false,
  showstringspaces=false,
  stringstyle={\color{BrickRed}},
  tabsize=2,
  breaklines=true,
  breakatwhitespace=true,
  postbreak={\mbox{\textcolor{gray}{$\hookrightarrow$}\space}},
  columns=fullflexible,
  captionpos=t,
  morekeywords=[1]{float2, float3, float4, uint8_t, log},
  morekeywords=[2]{Ray, HitRecord},
  escapeinside={<@}{@>},
  linewidth=0.9\columnwidth
}

\lstdefinestyle{paperPseudo}{
  language={},                      
  basicstyle=\scriptsize\ttfamily,  
  frame=single,
  framerule=0.4pt,
  framesep=3pt,
  linewidth=\linewidth,             
  breaklines=true,
  breakatwhitespace=true,
  columns=fullflexible,
  showstringspaces=false,
  captionpos=t,
  keywordstyle=\color{blue},
  commentstyle=\color{ForestGreen},
  stringstyle=\color{BrickRed},     
  morekeywords={function, return, if, then, else, end, while, do, break},
  postbreak=\mbox{\textcolor{gray}{$\hookrightarrow$}\ } 
}

\newcommand{\code}[1]{\texttt{#1}}
\def\optix{\emph{OptiX}\xspace}

\def\paratitle#1{\par\medskip\noindent\textbf{#1}\xspace}
\begin{document}


\firstsection{Introduction}
\maketitle

Beyond spatial scale, many scientific datasets are time-varying. Rather than analyzing a single static volume, scientists increasingly seek to understand how complex phenomena evolve, such as fluid dynamics, material deformation, or energy transport. Furthermore, INRs can encode dynamic phenomena when explicit reconstruction is not feasible~\cite{Mohan2025DINR}. Interactive exploration of these datasets, therefore, requires not only rendering large static volumes but also supporting temporal navigation to analyze continuous changes. However, explicit data representations such as meshes restrict navigation to discrete timesteps and require additional processing to interpolate or transition between them.

These challenges have motivated the adoption of implicit neural representations (INRs), which encode volumetric data as continuous neural functions, thereby significantly reducing storage requirements from terabytes to megabytes. Beyond compression, INRs are also well suited for data representation in computed tomography of dynamic or deforming objects, where measurements acquired at different projection angles do not correspond to a single static volume due to deformations during acquisition~\cite{Mohan2025DINR}. Moreover, by selecting a time parameter, a single model can generate arbitrary timesteps without loading, caching, or managing separate timestep volumes, enabling direct access to any point in continuous time. Furthermore, the continuous domain also lends itself to analysis, making the data format valuable beyond purely visualization. 

Despite these advantages, interactive visualization of INRs remains challenging, as sampling a point's value requires evaluating multiple layers of neural network inference, which is substantially more expensive than the memory accesses used in conventional volume rendering. Resampling an INR onto a grid is problematic: coarse grids may miss features, fine grids could exceed memory limits, and selecting an optimal resolution is nontrivial. More broadly, such approaches introduce additional complexity and approximation. Rendering through intermediate grids or retrained models adds extra preprocessing steps, reduces fidelity due to resampling, and becomes increasingly impractical in the temporal domain. While caching is effective for static volumes, extending it to time-varying data requires storing separate caches for each timestep, leading to frequent invalidations during temporal navigation.

In recent volume visualization literature, Monte Carlo–based rendering techniques have gained popularity due to their ability to reduce computational cost by taking fewer but more important samples. While this approach introduces variance into individual frames, the resulting images converge over time as additional samples are accumulated. Among these techniques, delta tracking~\cite{woodcock:1965, Gunther:2016, Martschinke:2019, hofmann:2020}---also known as Woodcock tracking---has emerged as one of the most widely used methods for stochastic volume rendering. In this work, we reinterpret delta tracking as a query-reduction mechanism for implicit neural fields, where the dominant cost is network inference rather than texture fetch.

Building on this insight, we present a query-efficient stochastic volume rendering framework that operates directly on continuous neural representations, minimizing and amortizing neural field evaluations to achieve interactive frame rates and low-latency temporal interaction while preserving visual quality (\autoref{fig:teaser}). Our contributions are:
\begin{itemize}
\item A four-stage rendering pipeline that decouples traversal from neural inference and coordinates ray tracing cores, tensor cores, and general compute units to maximize GPU utilization while operating directly on the original INRs.
\item Three adaptive ray budgeting schemes that prioritize sampling in temporally and perceptually changing regions, using inter-frame color variation and high-frequency cues to allocate computation efficiently while preserving full coverage over time.
\item A homogeneity-based query pruning strategy that detects near-uniform regions and probabilistically avoids unnecessary neural evaluations while minimizing visual artifacts.
\end{itemize}


\section{Related Work}
This work builds on two important bodies of literature: volume rendering and implicit neural representations.
\paratitle{Volume Rendering.}
Ray marching has been the de facto standard approach for volume rendering, as it is capable of producing images with virtually no variance by deterministically accumulating many (typically equally spaced) partial samples along each ray~\cite{Perlin1989Hypertexture,kruger2003accelerationgpu,Rottger2003SmartHAVR,Hadwinger2005RealtimeIsosurface}. Because these deterministic methods require dense sampling patterns, substantial effort has focused on reducing traversal cost through space skipping, bricking, and multi-resolution acceleration structures~\cite{ljung2006multiresolution,morrical2019spaceskip,wald2021exabrick}. However, even with adaptive sampling strategies, achieving real-time rates proves challenging with this family of methods.

Another popular school of methods for volume rendering is tracking-based Monte Carlo estimators, which provide an alternative to dense integration by importance sampling the medium. A well-known tracking method, delta tracking~\cite{woodcock:1965}, performs stochastic sampling of the medium based on an upper bound of its  density~\cite{fong:2017,Yue:2010,Kalos2010EfficientFreePath}. There are several works that improve efficiency through adaptive traversal schemes~\cite{Gunther:2016, Martschinke:2019, hofmann:2020, Zellmann2024Beyond} over multiple data formats, but these works are designed for non-neural volumes. Similar to our work, Sahistan et al.~\cite{sahistan2025MDWT,sahistan2026MDWT} use multiple coarse grids for multi-channel rendering rather than for time-varying data. Other tracking variants, such as Novak et al.'s residual-ratio tracking~\cite{Novak2014Residual} or Kutz et al.'s spectral/decomposition tracking~\cite{Kutz2017SDT}, split \emph{extinction} across independently parameterized media, assumptions impractical for scientific visualization, where color is derived from a transfer function over a single scalar field.

Beyond algorithmic improvements, recent work has explored leveraging the dedicated, specialized hardware in modern GPUs for scientific volume rendering. Ray tracing (RT) cores have been used to accelerate point-location queries in unstructured meshes~\cite{wald2019rtxpointloc, morrical2020rtxpointlocext,zellmann:2022, Sahistan2024VisNonTrivialPart}, enabling efficient traversal and containment tests directly on hardware designed for ray tracing. Building on this capability, Quick Clusters demonstrate GPU-parallel partitioning to support efficient ray tracing of unstructured volumetric grids~\cite{Morrical2022QuickClusters}.



Human perception is not uniform over a rendered frame. Exploiting this fact to improve performance, researchers used various perception-optimized sampling strategies~\cite{Viola2004ImportanceVolume}, such as foveated rendering~\cite{Guenter2012Foveated3D, Koskela2017FoveatedInstant} and blue-noise masks~\cite{Georgiev2016BlueNoiseDithering}, to distribute rendering effort where it matters most visually. Spatio-temporal blue-noise sampling has been used to spread work progressively over space and time while preserving high-frequency noise characteristics~\cite{Wolfe2022STBN}. Although several applications of these ideas live in the scientific visualization domain~\cite{Morrical2023AARBF, Bauer2023FoVoNet}, most research effort has primarily targeted virtual reality applications~\cite{Patney2016FoveatedVR, Ye2022RectFoveated}.


\paratitle{Neural Representations.}
Implicit neural representations (INRs) have gained significant traction across both graphics and scientific visualization. In graphics, neural radiance fields (NeRFs)~\cite{Mildenhall2021NeRF} model view-dependent image synthesis with pre-integrated color, whereas in scientific visualization, INRs represent scalar fields with color defined by a transfer function during integration. Although they are similar in principle, INRs in SciVis are also considered a form of compact neural encoding for lossy compression~\cite{lu2021compressive, han2022coordnet}. 
By enabling random access to arbitrary spatial regions of a volume, INRs eliminate the need to decompress entire volumes compared with previous state-of-the-art lossy compression techniques~\cite{di2016fast, lindstrom2014fixed}.

The NeRF literature in computer graphics has explored model- and system-level optimizations. KiloNeRF~\cite{reiser2021kilonerf} replaces a large and deep MLP with thousands of tiny MLPs, each modeling a small local region of volume, to reduce per-query inference cost. MINER~\cite{saragadam2022miner} leverages a Laplacian pyramid for high-resolution and large-scale signals modeling, enabling a coarse-to-fine scale query. Alternative approaches improve inference performance by optimizing computational memory usage. For example, MERF~\cite{reiser2023merf} replaces dense 3D feature grid with high-resolution 2D planes and sparse feature grid. However, these methods achieve efficiency by modifying the underlying representation, which requires retraining or restructuring the base model.

Instead of learning a function mapping from a coordinate index to density and color, as in NeRF, Gaussian splatting~\cite{kerbl20233d} learns \emph{explicit} Gaussian blobs to represent the data. 
Sewell et al.~\cite{dyken2025volume} highlight the importance of robust point cloud initialization using Gaussian splatting for scientific data visualization.
Bauer et al.~\cite{bauer2025gscache} apply Gaussian splatting as a cache mechanism to improve path tracing performance in scientific volume rendering.
Han et al.~\cite{han2025toward} extend 3D Gaussian splatting over multiple GPU training for scientific data and demonstrate significant performance improvement over a single GPU run.
While Gaussian splatting-based methods achieve high rendering performance and compression, their representations are primarily optimized for image synthesis and do not readily support non-visualization analysis tasks that require direct access to the underlying scalar field.

When applied to interactive scientific visualization, INRs fundamentally shift the performance bottleneck from memory access to neural inference.
While previous works~\cite{han2022coordnet,11264349,tang2023ecnr,han2025dcinr} primarily focus on improving the compression ratio and reconstruction quality, inference cost remains the dominant factor affecting rendering performance.
Among the few works addressing interactive INR rendering, Zavorotny et al.~\cite{zavorotny2025CacheINR} reduce the query cost through advanced caching mechanisms.

\section{Background on Delta Tracking}
\label{sec:delta_tracking_background}
Our framework builds on \emph{delta tracking}~\cite{woodcock:1965}. We briefly summarize its formulation and common design practices~\cite{Morrical2022QuickClusters, sahistan2026MDWT}, using notation similar to that of~\cite{fong:2017}. As this is not a comprehensive guide, we point readers to~\cite{Pharr2023Volume} for further details.

Delta tracking simulates interactions between rays and a participating medium via Monte Carlo sampling. For a homogeneous medium with constant extinction $\sigma_t$, the \emph{free-flight} distance (the distance a ray travels before hitting a particle) follows an exponential distribution:
\begin{equation}
\label{eq:freeflight}
p(t) = \sigma_t \exp(-\sigma_t t),
\end{equation}
where $\sigma_t$ denotes the extinction coefficient. 

The caveat with scientific volumes is that the extinction coefficient $\sigma_t(x)$ varies spatially. Typically, a user-adjusted transfer function maps scalar values to optical properties, including emission $L_e(x,\omega)$ and extinction $\sigma_t(x)$. Because $\sigma_t(x)$ is not constant, the exponential form in~\autoref{eq:freeflight} cannot be sampled directly. Delta tracking instead introduces a constant \emph{majorant} $\bar{\sigma} \geq \sigma_t(x)$ and samples free-flight distances from the homogeneous distribution obtained by replacing $\sigma_t$ in~\autoref{eq:freeflight} with $\bar{\sigma}$. Importance sampling this distribution with a uniform random variable $\xi \sim \mathcal{U}(0,1)$ yields
\begin{equation}
\label{eq:sample_majorant}
t = \frac{-\ln(1-\xi)}{\bar{\sigma}}.
\end{equation}

In practice,~\autoref{eq:sample_majorant} reduces to a single exponential inversion (\autoref{lis:free_flight}), producing the next free-flight step.


\begin{lstlisting}[style=paperPseudo, label=lis:free_flight, caption={Free-flight sampling under a majorant extinction coefficient.}]
function FreeFlightStep(majorant)
    return -log(1 - UniformRandom(0, 1)) / majorant;
end
\end{lstlisting}

At each sampled location $x_t = x + t\omega$, the interaction is classified as either a \emph{real} or \emph{null} collision, occurring with the following probabilities respectively:
\begin{equation}
\label{eq:accept}
P_{\text{real}}(x_t) = \frac{\sigma_t(x_t)}{\bar{\sigma}},
\qquad
P_{\text{null}}(x_t) = 1 - \frac{\sigma_t(x_t)}{\bar{\sigma}}.
\end{equation}
This could be represented as a Bernoulli test with acceptance probability $\sigma_t(x_t)/\bar{\sigma}$ (~\autoref{lis:real_col}).

\begin{lstlisting}[style=paperPseudo, label=lis:real_col, caption={Bernoulli acceptance test for real collisions.}]
function AcceptCollision(majorant, localExtinction)
    if UniformRandom(0, 1) < localExtinction / majorant then
        return true;
    else
        return false;
    end if
end
\end{lstlisting}

\begin{figure*}[!htbp]
\vspace{-2em}
\centering
\begin{minipage}[c]{0.75\textwidth}
\centering
\includegraphics[width=\linewidth]{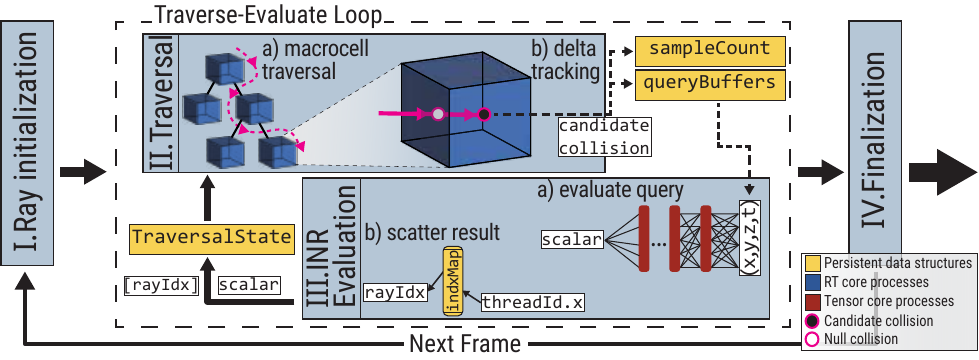}
\end{minipage}
\hspace{0.02\textwidth}
\begin{minipage}[c]{0.15\textwidth}
\centering
\includegraphics[width=1.15\linewidth]{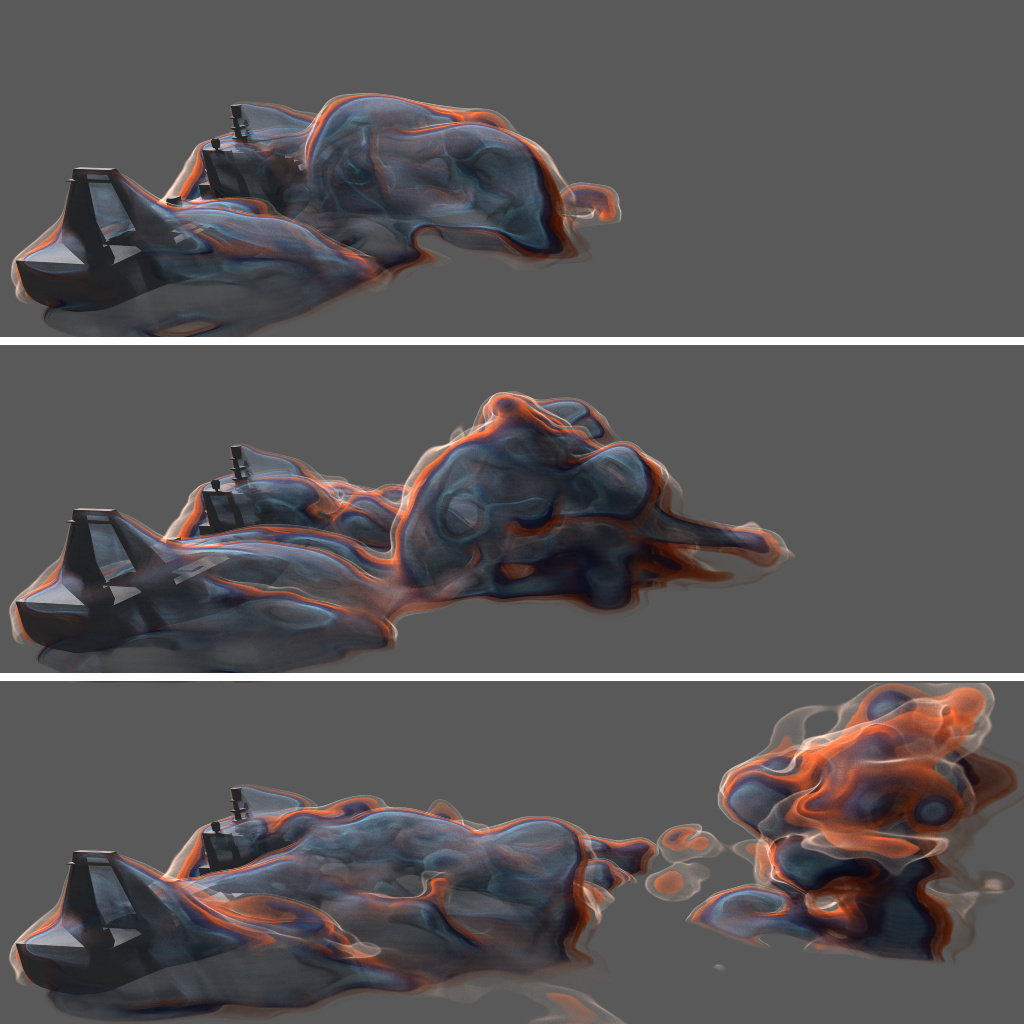}
\end{minipage}
\caption{Overview of our INR visualization framework: Each frame begins with \textbf{I. Ray initialization} (cf.~\autoref{sec:ray_init}), where rays and all other persistent data structures are initialized. Control then enters the \emph{traverse--evaluate} loop, which consists of two stages. In \textbf{II. Traversal} (\autoref{sec:traverse}), a) Ray tracing (RT) cores locate the current macrocell, and b) delta tracking within that macrocell generates a \emph{candidate collision}. In \textbf{III. INR evaluation} (\autoref{sec:tc_eval}), a) the recorded candidate queries are batched and evaluated on the GPU's tensor cores, and b) the resulting scalars are scattered back to the corresponding rays' \code{TraversalState}. Ultimately, \textbf{IV. Finalization} (cf.~\autoref{sec:finalize}) is invoked once all rays either find a collision or exit the volume, writing results to the frame buffer and collecting statistics.}
\label{fig:pipeline_overview}

\vspace{-1.2em}
\end{figure*}

If the candidate collision is accepted, emission is accumulated, and the ray terminates (emission-absorption model). If a \emph{null collision} occurs, another free-flight distance is drawn, and sampling continues from~\autoref{eq:sample_majorant}.

This formulation can be interpreted as rewriting the volume rendering equation in terms of interactions sampled from the majorant medium. The expected radiance along a ray starting at $x$ in direction $\omega$ can then be expressed as

\begin{equation}
\label{eq:vre_delta}
L(x,\omega)
=
\mathbb{E}\left[
\sum_{k} 
T(t_k)\, 
\frac{\sigma_t(x_{t_k})}{\bar{\sigma}}\,
L_e(x_{t_k},\omega)
\right].
\end{equation}
Here, $t_k$ is the $k$-th free-flight distance sampled under the majorant $\bar{\sigma}$, $x$ is the ray origin, and $\omega$ is the ray direction, so $x_{t_k} = x + t_k \omega$ denotes the sampled point along the ray. The term $T(t_k)$ is the accumulated transmittance up to $t_k$, and $\sigma_t(x_{t_k})/\bar{\sigma}$ is the probability that the interaction at $x_{t_k}$ is a real collision rather than a null event.

In more intuitive terms, delta tracking treats a heterogeneous medium as a homogeneous one with density $\bar{\sigma}$ and corrects the discrepancy through rejection sampling: each sampled interaction is accepted with probability $\sigma_t(x_t)/\bar{\sigma}$ and otherwise discarded as a null event. This mechanism enables traversal of spatially varying media without explicitly evaluating the heterogeneous transmittance. The core stepping logic is summarized in \autoref{lis:delta_tracking}.

\begin{lstlisting}[style=paperPseudo, label=lis:delta_tracking, caption={Core delta tracking loop for scientific volume visualization.}]
function DeltaTracking(volume, rayOrigin, rayDirection,
                    tMin, tMax, majorant, transferFunction)
    t = tMin;
    while t < tMax do
        t = t + FreeFlightStep(majorant);
        if t >= tMax then
            break;
        end if
        position = rayOrigin + t * rayDirection;
        scalar   = SampleVolume(volume, position);
        sampleColor  = transferFunction(scalar);

        if AcceptCollision(majorant, sampleColor.opacity) then
            return sampleColor;
        end if
    end while
    return backgroundColor;
end
\end{lstlisting}

A single global majorant can induce excessive null collisions, particularly when small regions exhibit high density. A common remedy is to subdivide the volume and assign each cell a local majorant $\bar{\sigma}(x)$. A traversal scheme such as a 3D digital differential analyzer (DDA) is then used to march between cells, applying the delta-tracking loop (\autoref{lis:delta_tracking}) within each region using its corresponding local majorant.

In scientific visualization (sciVis), the transfer function determines the extinction and therefore the local majorants. To compute these efficiently, we precompute the scalar minimum and maximum within each subvolume, commonly referred to as \emph{macrocells}. The majorants can be updated in parallel on the GPU by scanning each macrocell’s scalar minimum and maximum against the transfer function and recording the maximum resulting opacity. This enables dynamic, interactive majorant updates~\cite{Morrical2022QuickClusters}.

Unlike slicers or ray marching, where many partial samples are over-composited~\cite{OverOperator}, only a single sample per ray is accepted. This produces variance, yet accumulating more samples over multiple rays lets this noise converge. A common practice is to implement an \emph{accumulation buffer} next to a frame buffer. This buffer accumulates frames from a still scene and averages the results over time before copying the results back to the framebuffer.

Our motivation for adopting this method for INR rendering mirrors its original appeal in scientific volume rendering: fixed-step integration can quickly become non-interactive, whereas stochastic sampling remains efficient, extends naturally to secondary effects such as shadows, and converges in an unbiased manner. In the INR setting, the benefit is amplified, as each neural field query is substantially more expensive than a 3D texture fetch or finite-element interpolation, making query reduction essential for interactivity.

\section{Framework Design}
\label{sec:framework_design}
We design our framework around four constraints: (1) enabling interactive rendering of INRs, (2) allowing low-latency temporal exploration of time-varying INRs, (3) operating directly on INRs as-is, and (4) maintaining estimator correctness under practical approximations. 

Under these constraints, INR queries become the dominant bottleneck. Achieving interactivity, therefore, requires minimizing neural queries and efficiently executing the unavoidable ones. Dense sampling strategies, such as ray marching, entail excessive neural evaluations, and caching schemes~\cite{zavorotny2025CacheINR} break down for time-varying data, as temporal navigation would invalidate cached values. Allowing users to work directly with existing INRs, without added preprocessing or workflow overhead (as in established sciVis solutions), rules out approaches that require retraining or modifying the original network.


Overall, delta tracking has been shown to greatly reduce the number of samples, but na\"{\i}vely embedding neural inference into a ray tracing loop results in poor hardware utilization as neural inference and ray traversal exhibit fundamentally different parallelism patterns. Therefore, coordinated heterogeneous parallelism is required to tackle this problem efficiently. The following sections describe the pipeline (Section~\ref{sec:pipeline_overview}) and query-reduction strategies (Section~\ref{sec:query_reduction}), together enabling interactive visualization of time-varying INRs.

\subsection{Wavefront Pipeline Overview}
\label{sec:pipeline_overview}
A na\"{\i}ve implementation of delta tracking over a neural volume would evaluate the INR inside the innermost ray-marching loop, issuing one network inference per sample point.
On modern GPUs, this leads to severe under-utilization: delta tracking is inherently serial per ray, whereas neural inference benefits from large, uniform batches that can saturate tensor cores.
Our pipeline decouples the two workloads by splitting each frame into alternating \emph{traverse} and \emph{evaluate} stages that communicate through shared \code{INR query buffers}, following the wavefront ray tracing paradigm~\cite{laine2013MegaKernels}. Moreover, this design allows us to leverage dedicated hardware for each computation --- tensor cores for INR evaluations and ray tracing (RT) cores for traversal.

\autoref{fig:pipeline_overview} provides an overview of our rendering framework, which is organized into four stages: ray initialization, traversal, INR evaluation, and finalization. After ray initialization (\autoref{sec:ray_init}), a host-controlled traverse-evaluate loop alternates between RT-core macrocell traversal (\autoref{sec:traverse}), which advances rays and produces candidate collisions, and batched tensor-core INR evaluation (\autoref{sec:tc_eval}), which evaluates those candidates and updates ray state. The loop repeats until all rays terminate, typically after 18--22 iterations for a $64^3$ macrocell grid. Finalization (\autoref{sec:finalize}) writes the results to the framebuffer and updates per-pixel statistics; persistent cross-stage data structures are described in \autoref{sec:data_structures}.


\subsubsection{Data Structures}
\label{sec:data_structures}

\begin{figure}[!htbp]
    \centering
    \includegraphics[width=0.7\linewidth]{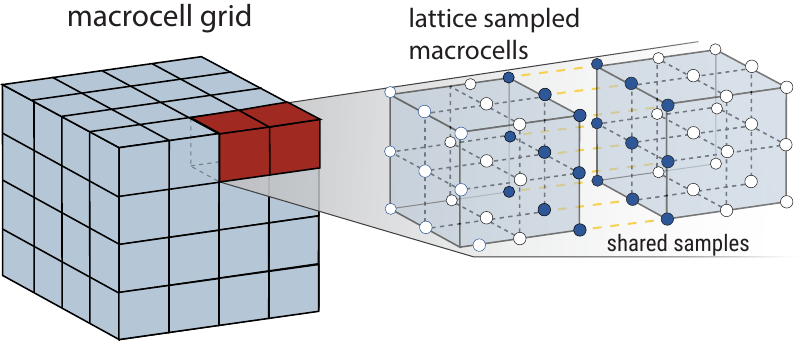}
    \caption{Illustration of the lattice-shaped sampling process for the macrocell grids. Blue colored (on the right) samples are shared between the two example macrocells, and all samples within a macrocell are used to find a minimum and a maximum of the scalars.}
    \label{fig:lattice_sampling}
    \vspace{-1.2em}
\end{figure}

The pipeline first constructs the spatial structures based on the macrocells outlined in \autoref{sec:delta_tracking_background}. Since INRs are continuous, exact per-cell extrema are not analytically tractable, so we approximate scalar bounds using GPU-based \emph{lattice} sampling. The macrocell grid is processed in $4^3$-cell tiles, where threads cooperatively evaluate the INR in shared memory over a $9^3$ lattice at cell boundaries and half-cell offsets. Each macrocell reduces its bounds from the local $3^3$ stencil covering the cell, storing per-cell minima and maxima for majorant construction. Overlapping lattices allow the reuse of many INR evaluations across adjacent macrocells. This process is illustrated in~\autoref{fig:lattice_sampling}).

For time-varying INRs, we precompute $T$ temporal macrocell grids at start-up, forming a $N_x \times N_y \times N_z \times T$ array of per-cell bounds. When the timestep changes, a \emph{working grid} is reconstructed from the two nearest temporal slices via linear interpolation. However, this interpolation may miss intermediate extrema. Increasing $T$ reduces these issues at the cost of memory and start-up time. One can add padding to the min and max values during interpolation to allow for error, but this would result in looser majorants and more null collisions.

\begin{figure}[!htbp]
    \centering
    \includegraphics[width=0.35\linewidth]{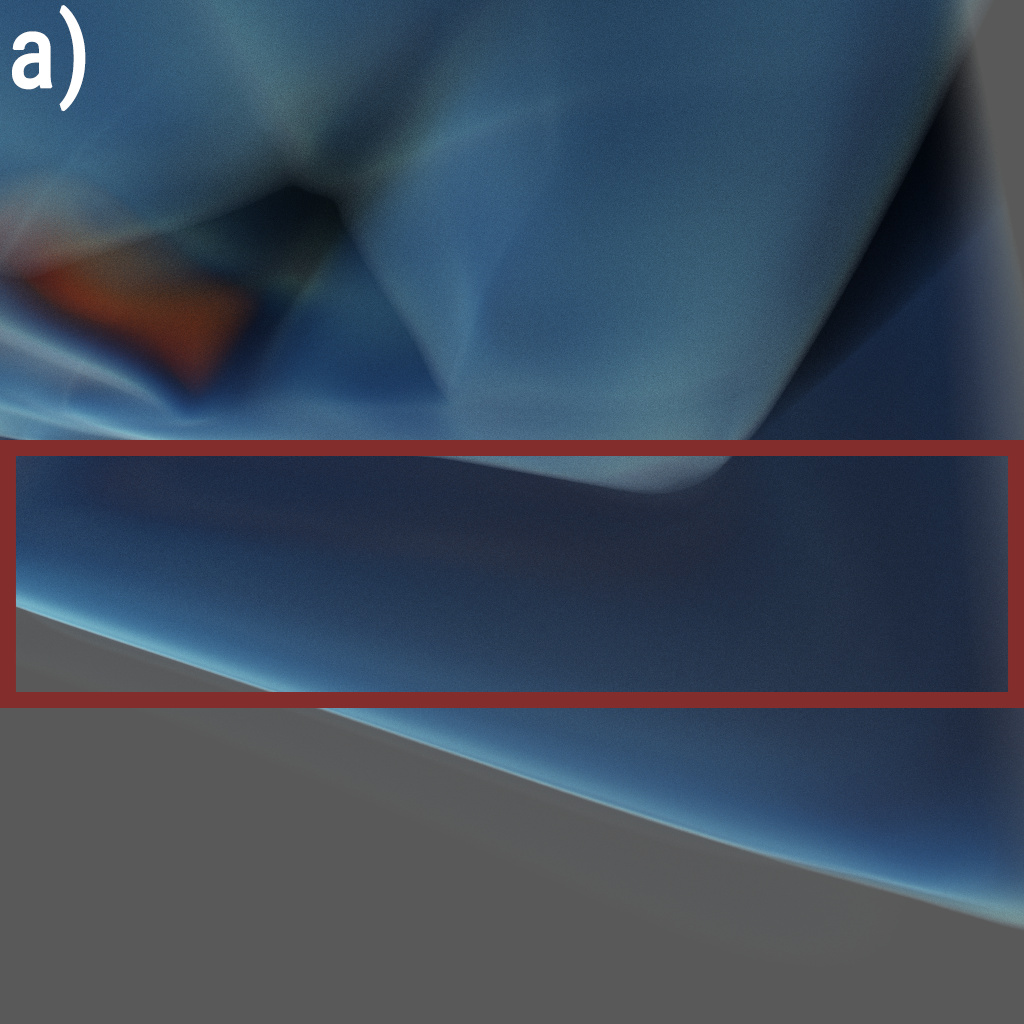}
    \includegraphics[width=0.35\linewidth]{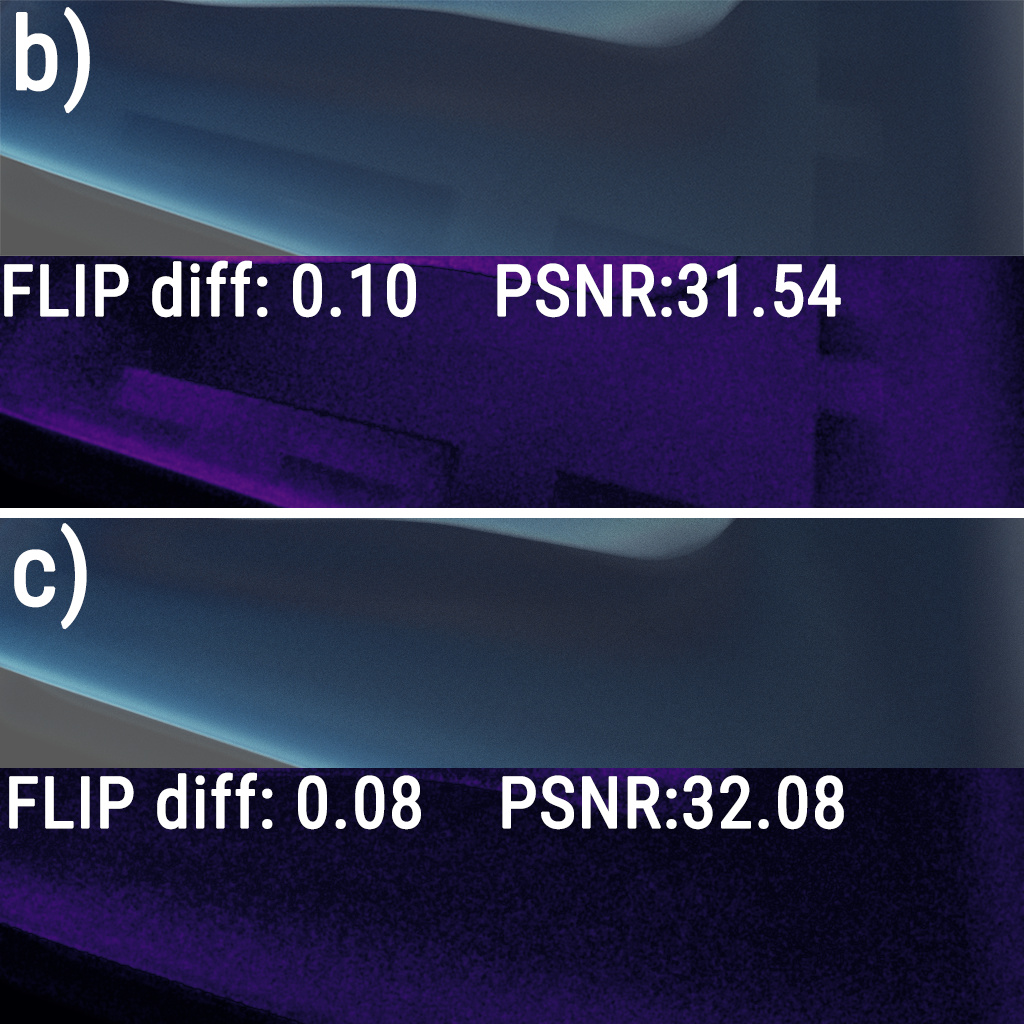}
    \includegraphics[height=0.35\linewidth]{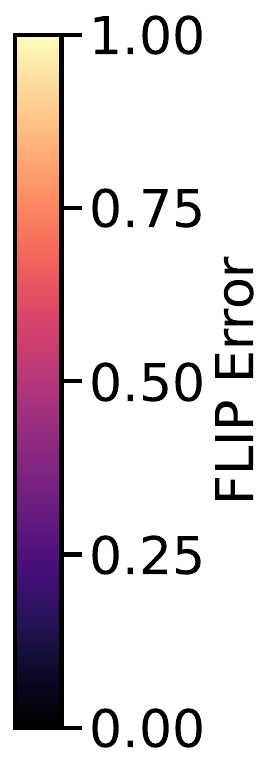}
    \caption{
    Comparison of ghost-pass macrocell refinement against ground truth on the S03\_001 dataset: a) ground truth rendered with a loosely allocated $1^3$ macrocell grid (2.80 FPS); b) crop of the red box from a) rendered using a $64^3$ macrocell grid without ghost-pass refinement (21.04 FPS); c) same as b) with ghost-pass refinement (18.65 FPS). Under b) and c), we report the FLIP~\cite{Andersson2021Flip} difference relative to the ground truth image, along with mean FLIP and Peak Signal-to-Noise Ratio (PSNR).}
    \label{fig:ghostpass}
\end{figure}

The main problem with these approximate macrocells is underestimation, which leads to visual artifacts resembling boxes. After the working grid is constructed, we can allow the renderer to progressively refine these underestimated bounds as the user explores the data over time. Yet this would only refine some of the macrocells if the user never changes the transfer function (i.e., majorants). As the free-flight distance (~\autoref{lis:free_flight}) is inversely proportional to the majorant, it causes some macrocells to be falsely mapped to a low majorant, causing rays to skip over that macrocell and never update. To address this, we introduce \emph{ghost passes}, in which the volume is rendered with a set of cosine-shaped transfer functions that sweep the scalar range. These passes are not displayed, but ensure that visible macrocells are probabilistically visited, enabling consistent refinement of macrocell bounds and mitigating persistent underestimation artifacts. ~\autoref{fig:ghostpass} depicts these artifacts and effects of the ghost pass refinement.


The majorants are converted into a sparse set of custom AABB primitives within the \optix ray tracing pipeline, emitting one primitive per macrocell whose majorant exceeds a small threshold, along with an \code{int3} array mapping primitive indices to cell coordinates. 
These primitives form a single Geometry Acceleration Structure (GAS) built with \code{OPTIX\_BUILD\_FLAG\_PREFER\_FAST\_TRACE} and subsequently compacted. 
Because only non-empty cells generate primitives, the resulting acceleration structure is sparse and cache-friendly. The GAS is quickly rebuilt within 2-3 ms whenever the working grid changes due to transfer-function edits or timestep updates.

In addition to these spatial structures, the pipeline allocates its persistent runtime buffers. The primary structure is a per-ray \code{TraversalState} record (as described in~\autoref{lis:traversal_state}), which stores all state required to suspend and resume a ray across kernel launches.

\begin{lstlisting}[language=c++, label=lis:traversal_state, style=rtKernel, caption=Per-ray persistent state that is preserved across traverse-evaluate iterations.] 
struct TraversalState { 
    Ray ray; 
    unsigned seed; // RNG state 
    uint8_t flags; // bools: active, hit, shadow_hit 
    float sample; // scalar from INR (NaN = pending) 
    float collision_product; // randFloat * majorant 
}; 
\end{lstlisting}


The pipeline also maintains the aforementioned INR query buffers: arrays for query positions (\code{float4}), and ray indices (\code{int}), each with a fixed capacity. An atomic counter keeps track of the number of samples generated per iteration. Results are written back to the \code{sample} in \code{TraversalState} without indirections.

\subsubsection{Ray Initialization}
\label{sec:ray_init}

Ray initialization is performed by a lightweight 2D CUDA kernel that launches one thread per pixel and prepares the persistent \code{TraversalState}. Each thread generates a jittered camera ray, intersects it with the volume bounding box, and stores the valid ray segment together with the per-ray random-number state. Rays that miss the volume are marked not active immediately; rays that intersect the volume are marked active and store the valid ray segment for subsequent traversal passes. After ray initialization, the host enters a loop that alternates between traversal and batched INR evaluation.


\subsubsection{RT-Core Macrocell Traversal and Delta Tracking}
\label{sec:traverse}

Unlike standard delta tracking implementations based on 3D DDA traversal, our method leverages ray tracing hardware. Macrocells with non-zero majorants are encoded as AABB primitives in a sparse \optix GAS. At the beginning of each traversal pass, a global sample counter is reset, and the \optix \emph{ray-generation program} is launched for all active rays. 

Each ray either resumes from its stored \code{TraversalState} or starts a new one. If a candidate collision is pending resolution from the previous iteration, the ray first performs the acceptance test (~\autoref{lis:real_col}) using the stored sample obtained from the INR query. If there is no pending candidate collision or the candidate yields a null collision, traversal continues within the macrocell. An accepted collision terminates the traversal.

A ray visits intersected macrocell bounding boxes in front-to-back order using the GAS. Within each macrocell, the custom \optix \emph{intersection program} executes delta tracking (see \autoref{lis:delta_tracking}), sampling candidate free-flight distances from the exponential distribution using the local majorant $\bar{\sigma}$. Traversal proceeds directly to the next macrocell primitive without host intervention if the tracking step occurs outside the macrocell.

When a candidate collision is reported, the \emph{closest-hit program} records the world-space sample position and stores the precomputed quantity \code{collision\_product} $= \xi \bar{\sigma}$ in the persistent \code{TraversalState}. Because the acceptance test depends only on this product, neither $\xi$ nor $\bar{\sigma}$ is stored separately, reducing per-ray state size. The \code{sample} field remains \code{NaN} until the INR evaluation completes. If INR evaluation is required for the sample, the program appends a query to the INR query buffers using \code{atomicAdd} on the global sample counter. Consequently, each ray generates at most one unresolved INR query per outer wavefront iteration.

After traversal, the host copies back only the sample counter. If it is zero, no unresolved samples remain and the primary traversal phase terminates. Otherwise, the collected sample positions are forwarded to the batched INR evaluation stage.

\subsubsection{Batched Tensor-Core INR Evaluation}
\label{sec:tc_eval}

The evaluation stage consumes the dense batch of sample coordinates produced by traversal and executes INR inference on tensor cores. Each warp processes a fixed-size tile of queries using WMMA (Warp Matrix Multiply-Accumulate) primitives, propagating them through the network’s fully connected layers in a single fused kernel. To maximize tensor-core utilization, weights are stored in 16-bit precision, and each warp maintains a private shared-memory workspace for staged inputs, weight tiles, and intermediate activations. The resulting scalars are written directly back into the corresponding \code{TraversalState} records via the ray-index array, eliminating a separate gather pass and completing the traverse–evaluate iteration in-place (see ~\autoref{lis:eval_scatter}).

\begin{lstlisting}[language=c++, label=lis:eval_scatter, style=rtKernel,
caption={Representative fused INR evaluation + scatter kernel. Each warp evaluates a tile of samples and writes scalar results directly into the per-ray \code{TraversalState} to avoid a separate gather pass.}]
__global__ void batchedEvalFused(
    const float4* coords, const int* rayIndices,
    TraversalState* states, int sampleCount)
{
    const int lane  = threadIdx.x & 31; // warp lane
    const int warp  = (blockIdx.x * blockDim.x + threadIdx.x) / 32;
    const int batch = warp * TILE; // TILE: queries per warp
    if (batch >= sampleCount) return;
    const int n = min(TILE, sampleCount - batch);
    // Evaluate tile (WMMA inference)
    float values[TILE];
    evaluateTile(coords + batch, values, n, model);
    //scatter each lane's result into the corresponding ray state
    if (lane < n) {
        const int rayIdx = rayIndices[batch + lane];
        states[rayIdx].sample = values[lane];
    }
}
\end{lstlisting}

When the evaluation kernel returns, every ray that requested a sample holds a resolved scalar in its \code{sample} field. The host resets the atomic counter and relaunches traversal (\autoref{sec:traverse}). Resumed rays apply the transfer function, perform a collision test against the stored \code{collision\_product}, and either terminate on a real event or clear the sample and continue stepping past a null collision.

\subsubsection{Finalization}
\label{sec:finalize}

Once the traverse--evaluate loop terminates (sample count reaches zero), a finalize kernel is called. The finalize kernel is responsible for maintaining the accumulation buffer and computing the per-pixel residual signal $r_i$ (\autoref{eq:divergence}) used by the adaptive ray budgeting schemes described in~\autoref{sec:query_reduction}.

Ray-traced shadows can be achieved by using an analogous traverse-evaluate loop that is executed after the primary rays have found a collision, and the finalize kernel incorporates the resulting visibility term.

\subsection{Query Reduction Strategies}
\label{sec:query_reduction}
Our key observation is that, under a limited ray budget, image quality improves more by selectively allocating rays than by uniformly tracing every pixel. In still or slowly changing regions, previously accumulated estimates are often sufficient, whereas unstable regions benefit most from new samples. We therefore treat ray allocation as a budgeting problem: decide which pixels to re-render, which to defer, and how to distribute those decisions so that skipped work produces acceptable noise and staleness patterns.

The strategies in~\autoref{sec:adaptive_budgeting} control \emph{where} the ray budget is spent at the pixel level. The techniques in~\autoref{sec:homogeneity_pruning} reduce the cost of \emph{each} traced ray by avoiding unnecessary INR queries inside near-uniform regions.
\subsubsection{Adaptive Ray Budgeting}
\label{sec:adaptive_budgeting}
Progressive refinement is already common in scientific visualization~\cite{hachisuka2008multidimensional, Rousselle2011AdaptiveSampling, ahrens2005paraview, wald2017ospray, Wu_VisItOSPRay_2018}. Since not all rays contribute equally, we use progressive refinement as a budgeting process: prioritizing unstable pixels, deferring stable ones, and avoiding staleness and artifacts in the resulting image.

We define a per-pixel \emph{convergence residual} signal $r_i$ for pixel $i$ at accumulation frame $n$ as the mean absolute channel difference between the newly rendered color $C_i^{(n)}$ and the running accumulation average $\bar{C}_i^{(n-1)}$ from the previous frame. To guide these decisions, we estimate the change in each pixel relative to the current accumulation.
\begin{equation}
    r_i = \frac{1}{3}\sum_{c \in \{R,G,B\}} \left| C_i^{(n)}[c] - \bar{C}_i^{(n-1)}[c] \right|, \quad r_i \in [0,1],
    \label{eq:divergence}
\end{equation}
where $C_i^{(n)}[c]$ is channel $c \in \{R,G,B\}$ of the color obtained by tracing pixel $i$ at frame $n$, and $\bar{C}_i^{(n-1)}[c]$ is the corresponding channel of the accumulated average from frame $n-1$. This residual is computed during finalization at negligible cost and drives most of the budgeting schemes. At ray initialization, each scheme decides whether pixel $i$ is \emph{budget-skipped}; budget-skipped pixels reuse their accumulated color and cast no rays or query the INR. Over these schemes, we explore different trade-offs between adaptivity and sampling pattern quality.

\paratitle{Residual-Histogram Thresholding.}
This scheme allocates the budget most aggressively to unstable pixels by ranking them by residual and skipping increasingly stable ones across multiple frames. Every rendered pixel contributes its residual to a 256-bin histogram, from which we derive a threshold retaining the top $\rho$ fraction of pixels for immediate rendering. After the frame, the host scans the histogram to build a cumulative distribution function (CDF) and finds the bin $b^*$ such that the cumulative fraction reaches $1 - \rho$, where $\rho$ is the user-specified re-render percentile (e.g.\ $\rho{=}0.6$ keeps the top 60\% most divergent pixels). The threshold $\tau = (b^* + 0.5)/256$ is uploaded for the next frame. During initialization, pixels whose residual $r_i$ meets or exceeds the threshold $\tau$ render immediately ($\text{skip}_i = 0$); pixels below receive a skip countdown proportional to their stability:
\begin{equation}
    \text{skip}_i = \left\lfloor \left(1 - \frac{r_i}{\tau}\right) \cdot S_{\max} + 0.5 \right\rfloor, \quad S_{\max} = \min\!\bigl(K,\; \lfloor n/4 \rfloor\bigr),
    \label{eq:skip_countdown}
\end{equation}
where $\text{skip}_i$ is the number of consecutive frames pixel $i$ will be skipped before re-entering the render set, $\tau$ is the residual threshold derived from the histogram CDF, $K$ is a user-specified maximum skip duration, and $n$ is the number of accumulated frames. The ratio $(1 - r_i/\tau)$ ranges from 0 (at the threshold) to 1 (fully converged), so the most stable pixels receive the longest skip.
The $\lfloor n/4 \rfloor$ ramp ensures that $S_{\max}$ grows gradually with $n$, preventing aggressive skipping during the first few frames when the accumulation buffer has not yet converged.
Each skipped frame decrements the counter by one; when it reaches zero, the pixel re-enters the render set. This guarantees that every pixel is rendered at least once every $S_{\max}+1$ frames, bounding worst-case staleness. 

Among the three schemes, this one is the most residual-driven and spatially concentrated, but also the most dependent on a reliable residuals.

\paratitle{STBN Thresholding.}
This scheme distributes a fixed render budget using spatiotemporal blue noise without using the residual signal. It therefore serves as the simplest baseline and tends to produce more favorable sparse-update patterns.

We preload $N$ spatiotemporal blue-noise (STBN) textures~\cite{Wolfe2022STBN} of resolution $128 \times 128$ and tile them across the screen. For a pixel at screen coordinate $x$ and frame index $n$, we fetch

\begin{equation}
\label{eq:stbn_lookup}
u(x,n) = \frac{\texttt{stbn}(x,\; n \bmod N)}{256},
\end{equation}
where $u(x,n) \in [0,1)$. A pixel renders if $u(x,n) < \rho$, where again $\rho$ denotes a user-defined render fraction. Over $N$ frames, each pixel cycles through the STBN sequence, producing approximately $\rho N$ render events per cycle with blue-noise spatial distribution.

Rendered pixels update a running average, while skipped pixels reuse the accumulated value. Because it is active from the first frame, it avoids the warm-up and stale-signal issues of residual-based schemes. The trade-off is that the budget is distributed independently of image instability, so updates are not concentrated on regions that would benefit most.

\paratitle{Residual-Weighted STBN.}
This scheme combines STBN-based allocation with residual-driven steering by modulating each pixel’s render probability according to its residual. Instead of a fixed fraction $\rho$, each pixel computes an effective render probability:
\begin{equation}
    \rho_i^{\text{eff}} = \text{clamp}\!\bigl(\rho + \alpha\,(r_i - \rho),\; \tfrac{2}{256},\; 1\bigr).
    \label{eq:var_stbn}
\end{equation}
Here $\alpha \in [0,1]$ controls the strength of residual steering. When $\alpha=0$, the scheme reduces to pure STBN; as $\alpha$ increases, more budget is directed toward unstable pixels, while the blue-noise mask still regularizes the spatial update pattern.
The floor at $2/256$ prevents permanent starvation.

A ray is cast when $u_i < \rho_i^{\text{eff}}$. Because decisions remain tied to the blue-noise texture, the pattern retains blue-noise characteristics for small $\alpha$, while larger $\alpha$ shifts it toward a more residual-driven allocation. The scheme thus offers an adjustable middle ground between stochastic allocation and fully residual-driven budgeting.

\paratitle{Residual reprojection.}
Camera motion resets the accumulation buffer, leaving the residual signal $r_i$ undefined. To preserve informed budgeting after interaction, we reproject residuals using stored world-space hit positions~\cite{Nehab2007ShadingReprojection, Yang2020TAA}, alongside pixel-value reprojection. Valid reprojected hits are used to set the histogram and skip state, while invalid or occluded pixels default to high residual and are re-rendered.

This mainly benefits the residual-driven schemes by restoring adaptive budgets immediately after a view change; pure STBN thresholding does not benefit because its decisions do not depend on residuals, but that also means the pure STBN mode does not lose performance when $r_i$ is undefined. The approximation degrades under large camera rotations, transfer-function edits, or timestep changes, so reprojected budgets should be viewed as heuristics rather than exact estimates under the new view.

\subsubsection{Homogeneity-Based Query Pruning}
\label{sec:homogeneity_pruning}

If a macrocell’s scalar range is sufficiently small, the INR queries could be avoided, as these macrocells will mostly behave as a uniform block. 
We exploit this observation to skip the INR query entirely for such cells, replacing it with an expected scalar $\hat{s}_c$.

\paratitle{Uniformity criterion.}
Each macrocell $c$ stores a precomputed scalar range $\Delta_c = s_c^{\max} - s_c^{\min}$ from its min--max bounds.
Given a user-controlled threshold~$\epsilon$, a candidate collision point in cell~$c$ is pruned (i.e., answered without an INR query) whenever $\Delta_c < \epsilon$.
To soften the transition between pruned and evaluated cells, a stochastic falloff band of width~$\delta$ linearly ramps the pruning probability from 1 to~0 over the interval $[\epsilon,\; \epsilon + \delta]$:
\begin{equation}
  p_{\text{skip}} =
  \begin{cases}
    1                                          & \Delta_c < \epsilon, \\
    1 - \frac{\Delta_c - \epsilon}{\delta}         & \epsilon \le \Delta_c < \epsilon + \delta, \\
    0                                          & \Delta_c \ge \epsilon + \delta.
  \end{cases}
  \label{eq:homog_skip}
\end{equation}

\begin{figure}[!htbp]
\vspace{-0.5em}
  \centering
  \begin{tabular}{@{}c@{\hspace{0.2em}}c@{\hspace{0.0em}}c@{\hspace{0.0em}}c@{}}
    \raisebox{-0.5\height}{\includegraphics[width=0.35\linewidth]{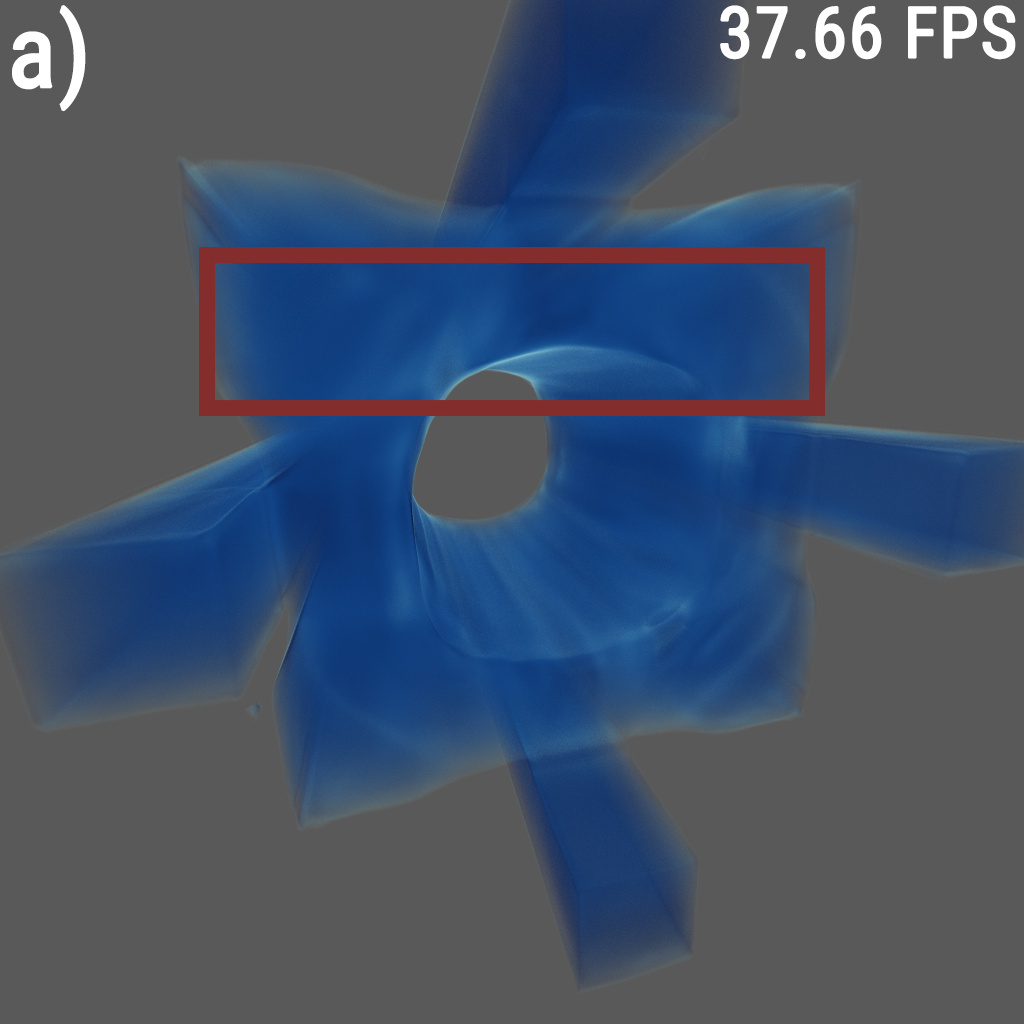}} &
    \raisebox{-0.5\height}{\includegraphics[height=0.35\linewidth]{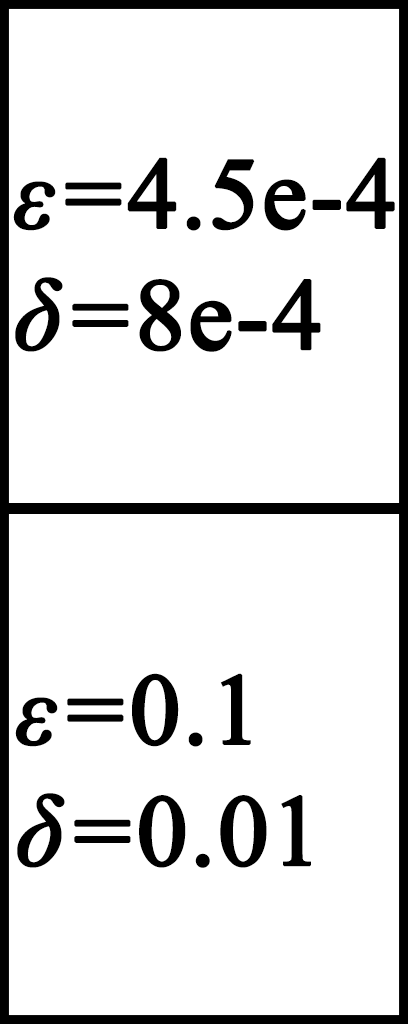}} &
    \raisebox{-0.5\height}{\includegraphics[width=0.35\linewidth]{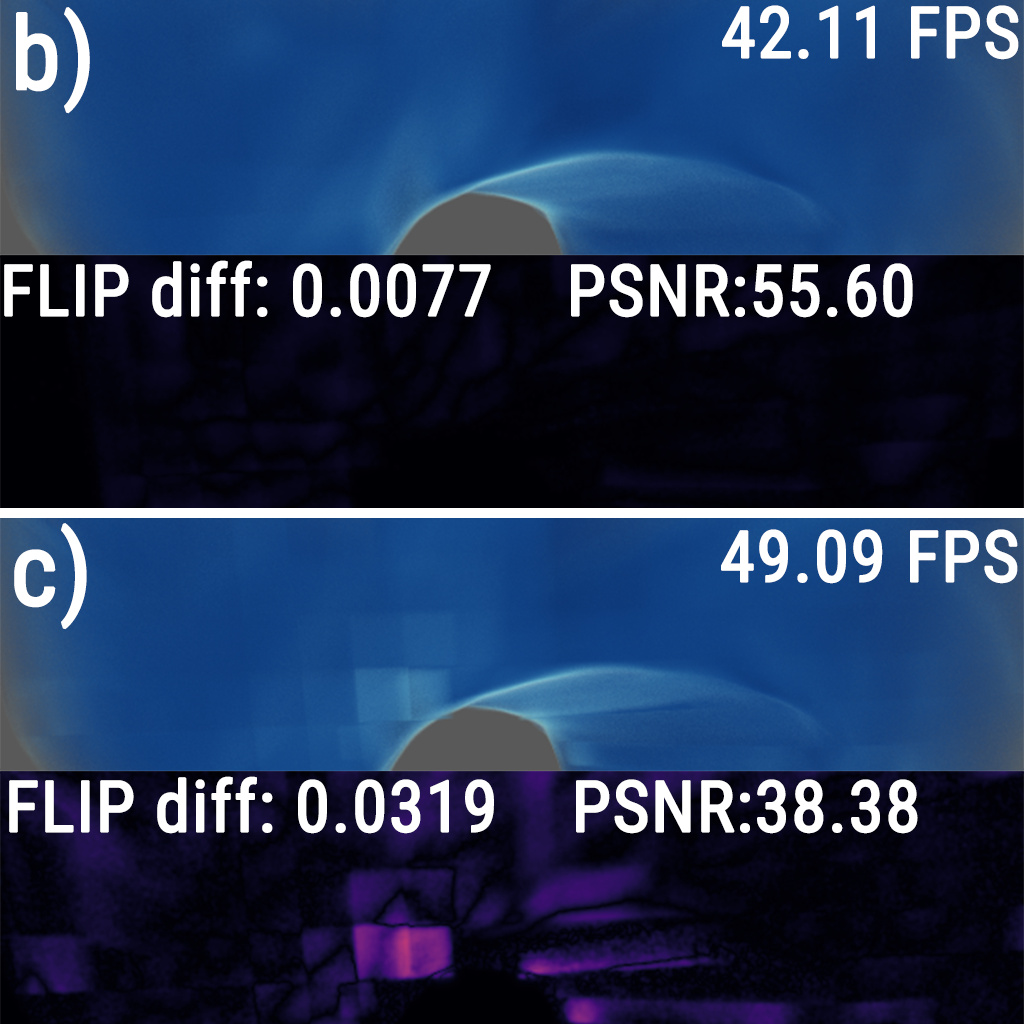}} &
    \raisebox{-0.5\height}{\includegraphics[height=0.35\linewidth]{figs/flip_colorbar.pdf}}
  \end{tabular}
  \caption{Homogeneity-based query pruning on the S05\_700 dataset. a) Ground truth without pruning. b) Crop (red box in a) using low threshold and falloff width. c) Same crop with a higher threshold and falloff width. FLIP and PSNR differences are reported relative to the full ground truth image.}
  \label{fig:homog_skip}
  \vspace{-0.8em}
\end{figure}

Because the substituted scalar is a per-cell constant rather than the true field value at the sample point, this approximation introduces bias: the collision test sees the cell’s expected value instead of the local densities, which can produce blocky artifacts at cell boundaries, particularly when the user-controlled threshold~$\epsilon$ is set aggressively. \autoref{fig:homog_skip} illustrates the effect of these user-controlled parameters on an example volume.

The stochastic falloff band mitigates hard transitions but does not eliminate the underlying bias.
In practice, this technique is best applied conservatively to cells with very small scalar ranges, where the substitution error is negligible relative to the transfer-function response. Notably, exploiting this fact makes sense only when sampling volumes is expensive, as is the case with the INRs.

\paratitle{Expected-value refinement.}
A na\"{\i}ve choice for $\hat{s}_c$ is the midpoint $(s_c^{\min} + s_c^{\max}) / 2$. However, by introducing another float per macrocell, we can refine this value. The idea is to maintain a running estimate $\hat{s}_c$ that is refined at runtime via an exponential moving average (EMA).
Whenever a real INR evaluation returns a scalar $s$ inside cell~$c$, the estimate is updated as
\begin{equation}
  \hat{s}_c \leftarrow \hat{s}_c + \alpha\,(s - \hat{s}_c),
  \label{eq:ev_ema}
\end{equation}
where $\alpha = 0.1\,\max(w_c,\, 0.01)$ and $w_c \in [0,1]$ is a center weight that attenuates updates from samples near cell boundaries, which may not be representative of the cell interior.
The initial estimate is set to the mean of the collected samples used during macrocell construction. The EMA refinement reduces the magnitude of the substitution error over successive frames, but does not eliminate it, as the pruned scalar remains spatially constant within each cell. While the stochastic falloff has a limited impact on image quality when the threshold $\epsilon$ is present, it is necessary to ensure sufficient sampling coverage for stable EMA updates. In our experiments, $\delta=0.006$ provides a robust balance across all tested datasets.

\section{Results and Discussion}
We evaluate the memory overhead and end-to-end performance of our system without
query reduction, and then assess the trade-offs of our query-reduction strategies.
As datasets, we use six INRs from three architectures, each mapping spatiotemporal
coordinates $(x,y,z,t)$ to a single scalar field value. Three INRs from Mohan et
al.~\cite{Mohan2025DINR, reed2021dynamicctreconstructionlimited} use a Fourier-feature network (FFN), which encodes
coordinates with Fourier features before a six-layer fully connected network with
Swish activations; they are reconstructed from dynamic X-ray CT scans of deforming
objects and represent energy-averaged linear attenuation coefficients. These
datasets follow the \emph{S0X\_XXX} naming convention. Two are sinusoidal
representation networks (SIRENs)~\cite{sitzmann2019siren} generated from CFD time
series~\cite{Popinet2004Tangaroa, Rojo1209SciVisa} produced with the Gerris Flow
Solver~\cite{gerrisflowsolver}; they use five width-$256$ sine layers followed by a
linear layer and represent simulated flow quantities over time. These are named
\emph{cylinder} and \emph{tangaroa}. The last is a coordinate-based network (CoordNet)~\cite{han2022coordnet}, a residual
sinusoidal MLP that grows the input to width $192$ over three residual blocks,
applies six width-$192$ residual blocks, and ends with a residual block projecting
to the scalar output; each block sums two $\omega_0{=}30$ sine layers through a skip
connection. It represents a vortical flow field and is named \emph{vorts}.

All experiments run at $1024^2$ resolution on an NVIDIA RTX 4090 using CUDA 13~\cite{cuda} and \optix 9~\cite{optix}. Unless stated otherwise, results use 400-frame converged renderings.
\subsection{Data Structure Sizes and Updates}
\label{sec:data_size_update}
We measure the memory footprint and update cost of the acceleration structures used for ray tracing and delta tracking.

\begin{table}[htbp!]
\centering
\caption{Data structure sizes for three example datasets.$^*$ indicates an average over 10 evenly spaced timesteps, as \optix GAS size depends on transfer function and timestep due to empty-cell culling.}
\label{tab:memory_usage}
\resizebox{\columnwidth}{!}{
\begin{tabular}{ll|llll|l}
\multicolumn{2}{l|}{\diagbox{data \& \\ macrocell res.}{Size (MB)}} & INR & \makecell{macrocell\\grid} & \makecell{working\\grid} & GAS & Total \\ \hline
\multicolumn{1}{l|}{\multirow{2}{*}{S03\_001 (FFN)}} & $64^3\times20$  & \multirow{2}{*}{1.26} & 40.0 & 3.0 & $0.49^*$ & 44.76\\ \cline{2-2} \cline{7-2}
\multicolumn{1}{l|}{} & $128^3\times20$ & & 320.0 & 24.0 & $3.59^*$ & 347.86\\ \hline
\multicolumn{1}{l|}{\multirow{2}{*}{Tangaroa (SIREN)}} & $64^3\times30$  & \multirow{2}{*}{1.01} & 60.0 & 3.0 & $0.21^*$  & 64.22\\ \cline{2-2} \cline{7-2}
\multicolumn{1}{l|}{} & $128^3\times30$ & & 480.0 & 24.0 & $1.24^*$ & 506.25\\ \hline
\multicolumn{1}{l|}{\multirow{2}{*}{Vorts (CoordNet)}} & $64^3\times20$  & \multirow{2}{*}{2.06} & 40.0 & 3.0 & $4.38^*$  & 49.44\\ \cline{2-2} \cline{7-2}
\multicolumn{1}{l|}{} & $128^3\times20$ & & 320.0 & 24.0 & $35.15^*$ & 381.21
\end{tabular}
}
\end{table}

One of the many advantages of an INR is its compact size, and the same neural architecture occupies the same amount of memory---unlike adaptive formats like unstructured meshes. This is the case for our datasets; each occupies 1--2 MB of GPU memory in isolation. In~\autoref{tab:memory_usage}, we chose to demonstrate two 4D macrocell grid sizes for three different INR architectures that also occupy constant memory. From those macrocell grids, we build a working grid: the interpolated and corrected 3D macrocell grid for the current timestep, along with the 3D majorant grid---again, much smaller, constant-size. Finally, a sparse \optix acceleration structure is built over those majorants, which is the only variable component of this memory usage. One thing we do not depict in this table is the INR query and \code{TraversalState}, yet their sizes are again trivial to calculate, since one of each is allocated per pixel (shadow and primary rays share the \code{TraversalState}). At $1024^2$ resolution, the \code{TraversalState} buffer occupies 42 MB, and query buffers collectively consume 20 MB.

\begin{figure}
    \vspace{-0.25em}
    \centering
    \includegraphics[width=0.99\linewidth]{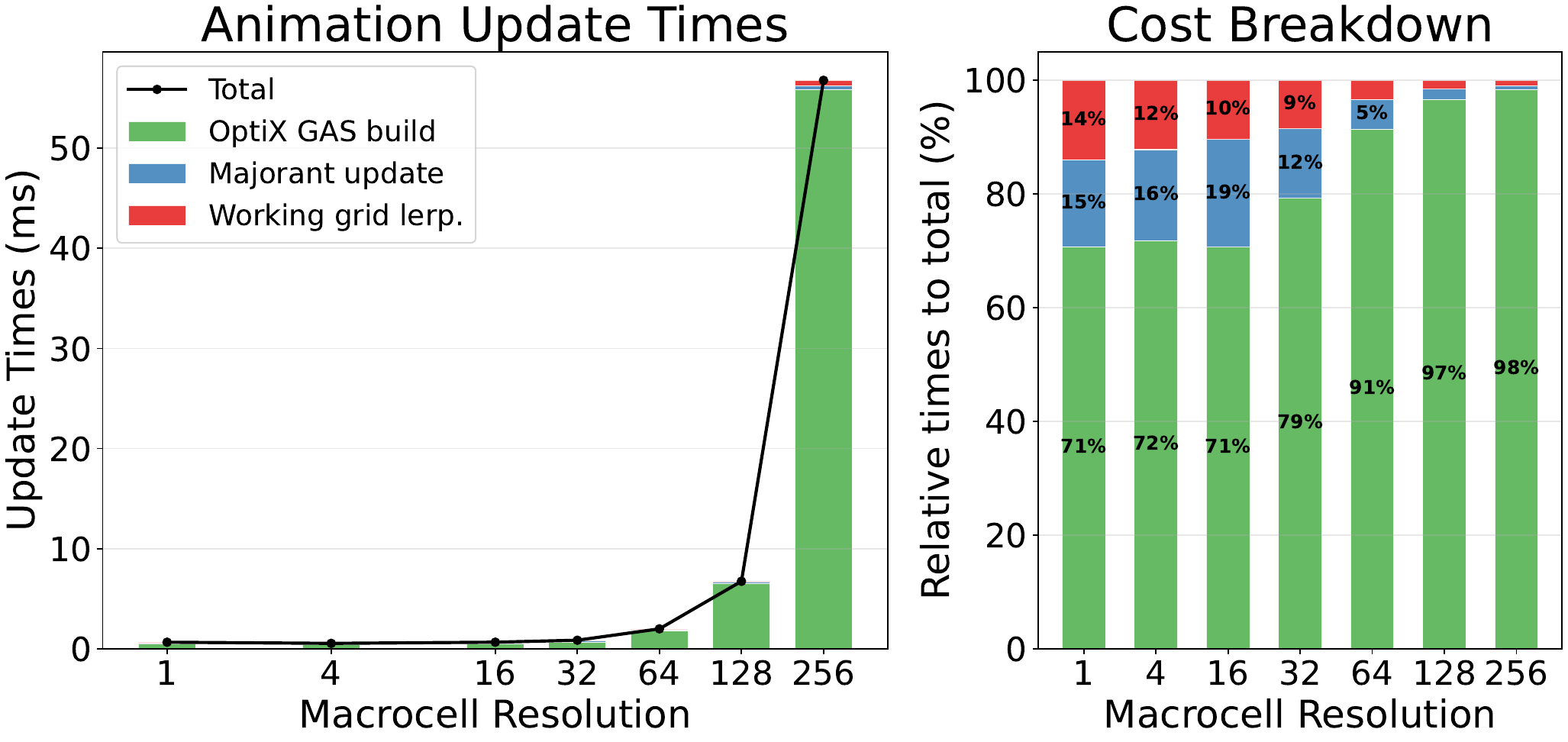}
    \caption{Data structure update time (lower is better) versus macrocell grid resolution. Times are averaged over 10 timestep interactions for S03\_001 and S04\_018. Left: absolute timings per update stage; right: timings normalized by total update time. The x-axis denotes per-dimension resolution, i.e., $\text{value}^3\times T$ with $T=20$.}
    \vspace{-0.4em}
    \label{fig:anim_update}
    \vspace{-1.4em}
\end{figure}

During animation, changing $t$ requires updating the macrocell-derived structures. \autoref{fig:anim_update} reports these costs as macrocell resolution increases; finer grids increase the number of cells and therefore the cost of interpolation, updates, and construction. Transfer-function edits follow the same update path, except that working-grid interpolation is skipped because cell extrema are unchanged.

As~\autoref{tab:memory_usage} shows, with a smaller macrocell resolution, rendering-related structures are much larger than the INR's base size. Even with the relatively larger, $128^3\times20$, grid, the acceleration structure memory remains within a 40-400 MB range. For the smaller macrocell grid resolution, measurements obtained using \code{nvidia-smi} indicate a total GPU memory usage of approximately 600 MB. The application explicitly accounts for $\sim$150 MB, including the trivial allocations for frame and accumulation buffers. We attribute the remaining memory to \optix and CUDA contexts, and their associated programs. This is substantially lower than typical scientific visualization systems~\cite{ahrens2005paraview, Childs2012visit,wald2017ospray}.

Despite the \optix acceleration structure build time dominating the animation updates, the total update time falls under 55 milliseconds (ms). With smaller macrocell grid resolutions, this time decreases to $\sim$2~ms. As a result, users can animate the continuous temporal deformations of the INR in real time.

\subsection{Framework Performance}
\label{sec:performance}
\begin{figure*}[htbp!]
    \vspace{-2em}
    \centering
    \setlength{\fboxsep}{1pt}
    \setlength{\tabcolsep}{0pt}
    \setlength{\fboxrule}{0.3pt}

    \newcommand{\pairbox}[2]{%
        \fbox{%
            \begin{minipage}[t]{0.175\textwidth}
                \centering
                \includegraphics[width=\linewidth,height=0.92\linewidth,keepaspectratio]{#1}\\[-0.5mm]
                \includegraphics[width=0.75\linewidth,height=0.75\linewidth,keepaspectratio]{#2}
            \end{minipage}%
        }%
    }
    
    \newlength{\figrowwidth}
    \setlength{\figrowwidth}{1.05\textwidth}
    
    \noindent
    \makebox[\textwidth][l]{%
        \hspace*{-0.7em}%
        \begin{minipage}{\figrowwidth}
            \centering
    
            \includegraphics[width=\figrowwidth]{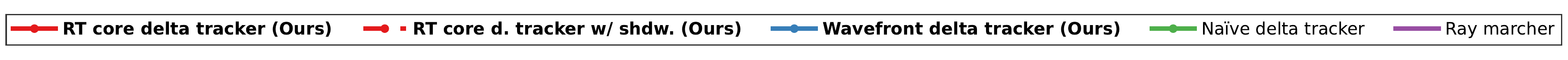}%
            \par\vspace{-0.6em}
    
            \resizebox{\figrowwidth}{!}{%
                \begin{tabular}{@{}cccccc@{}}
                    \pairbox{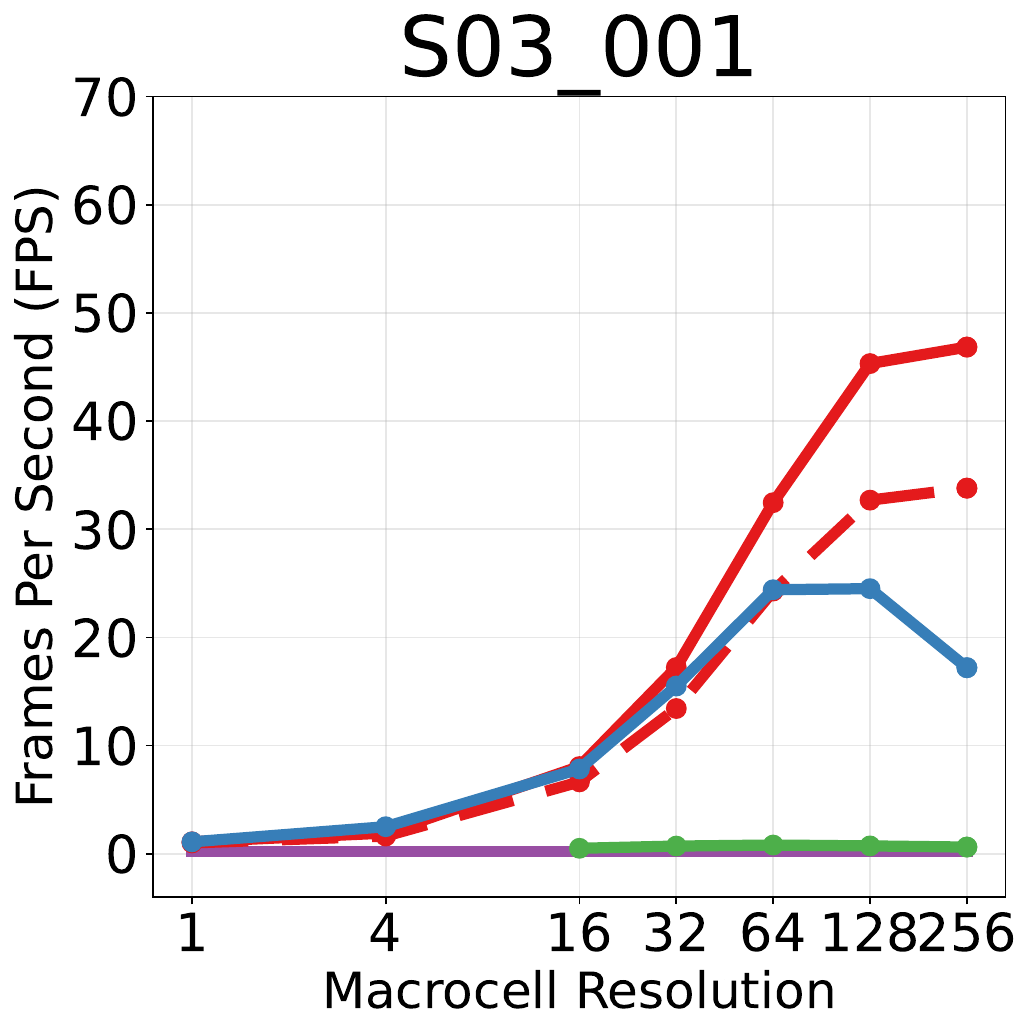}{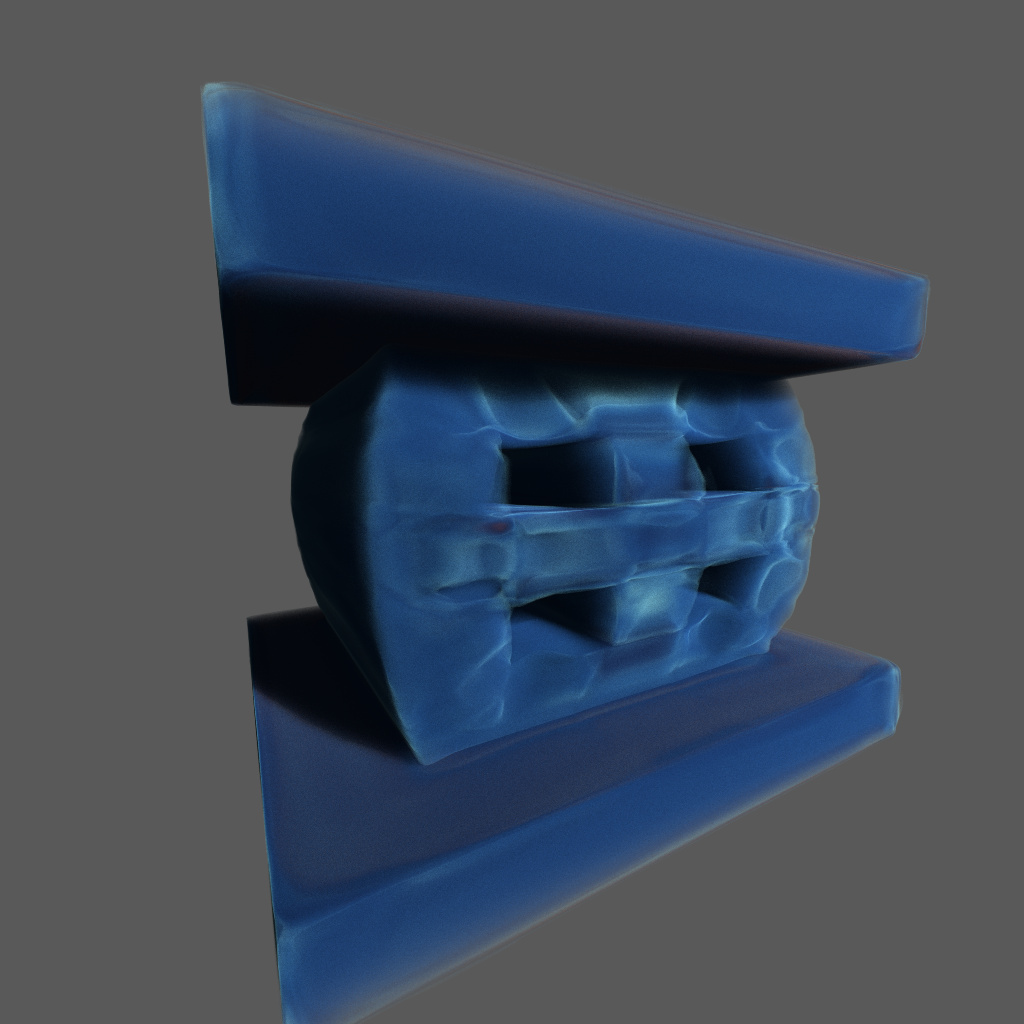} &
                    \pairbox{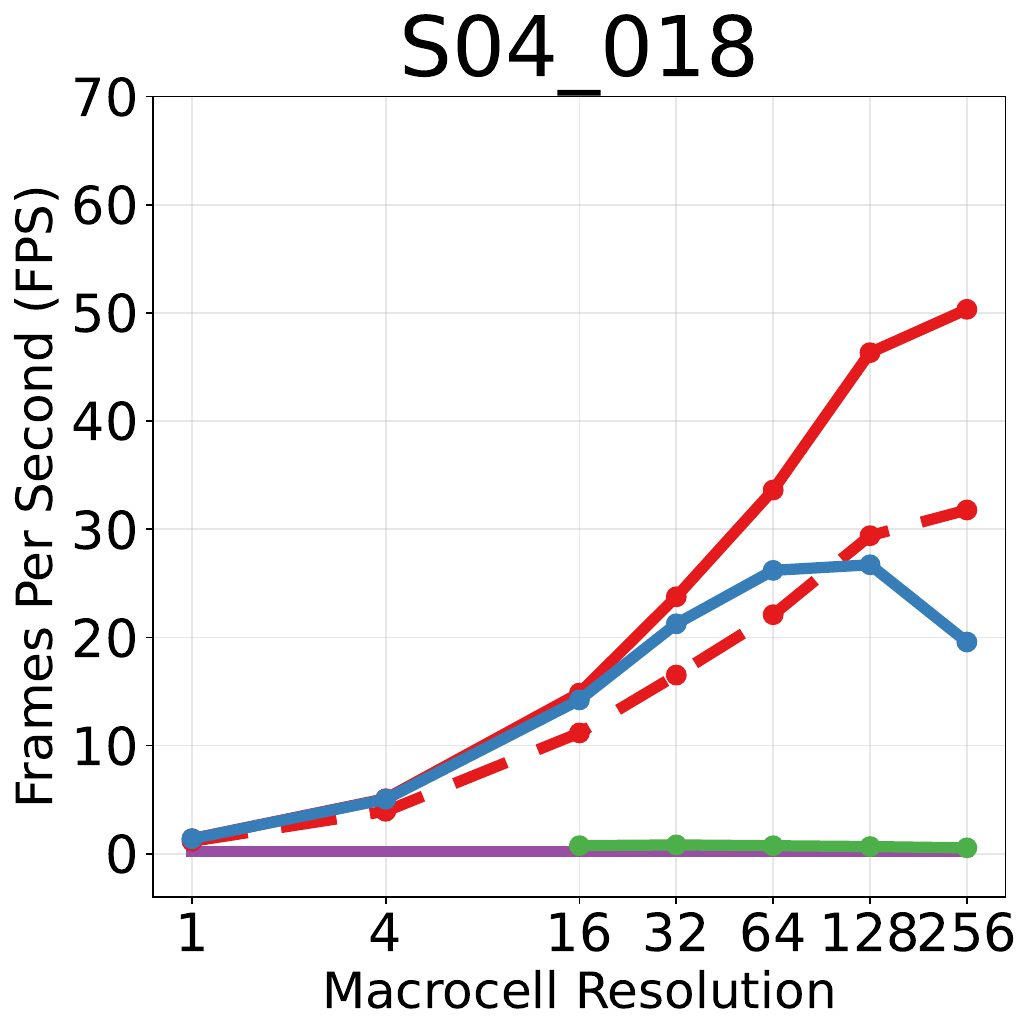}{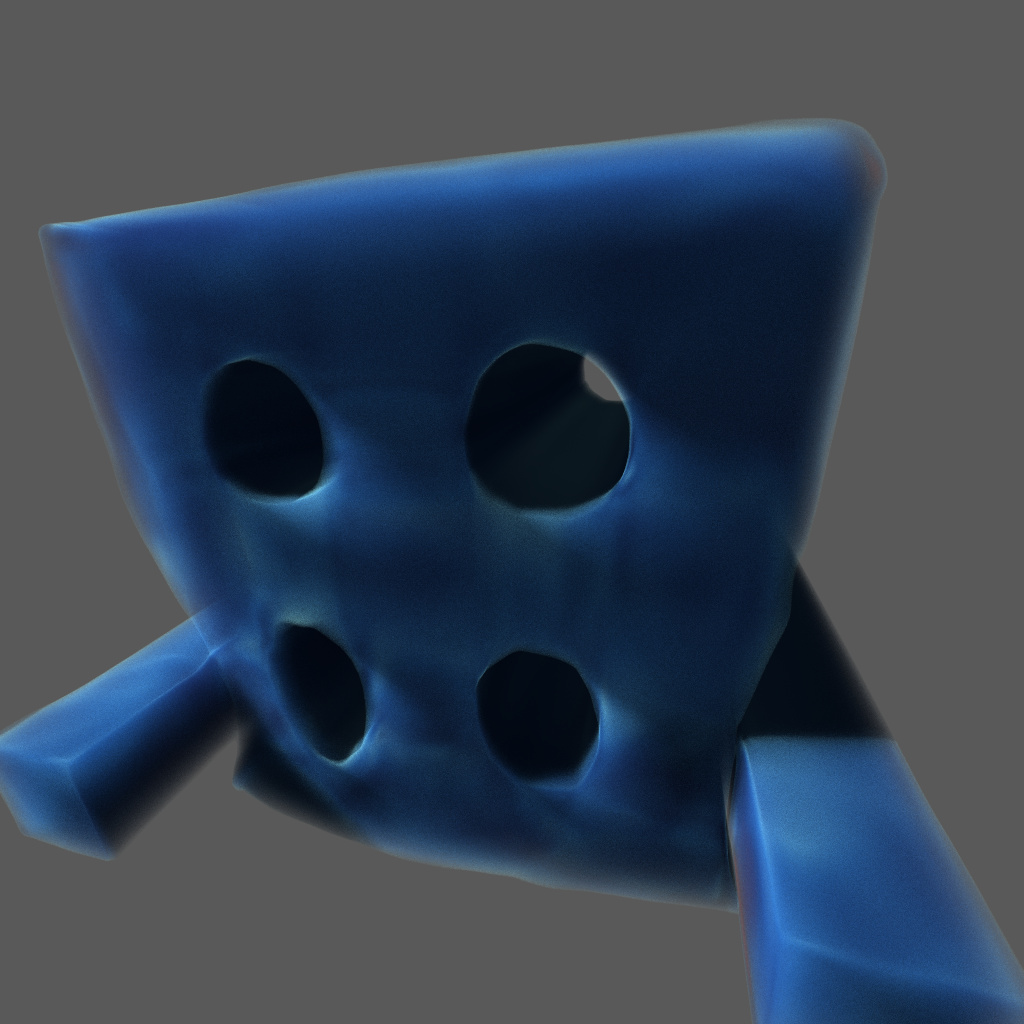} &
                    \pairbox{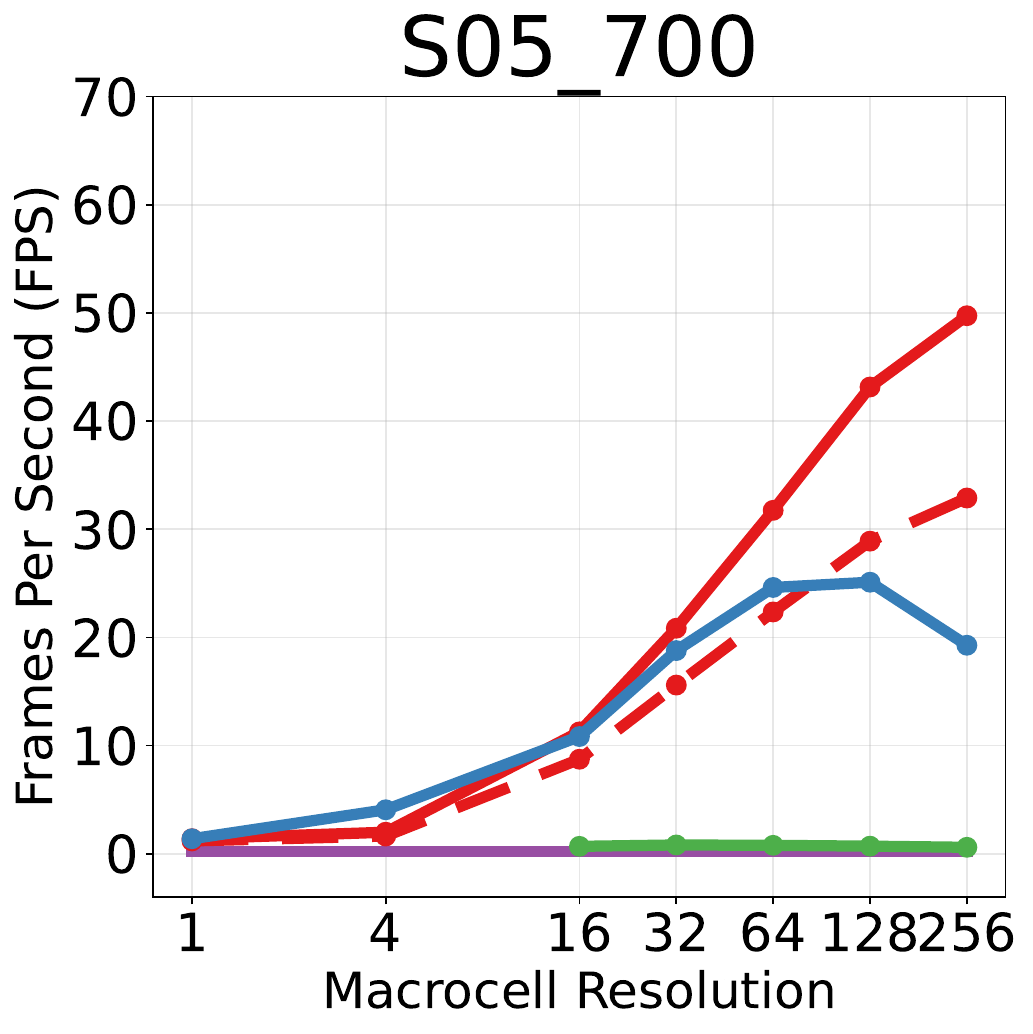}{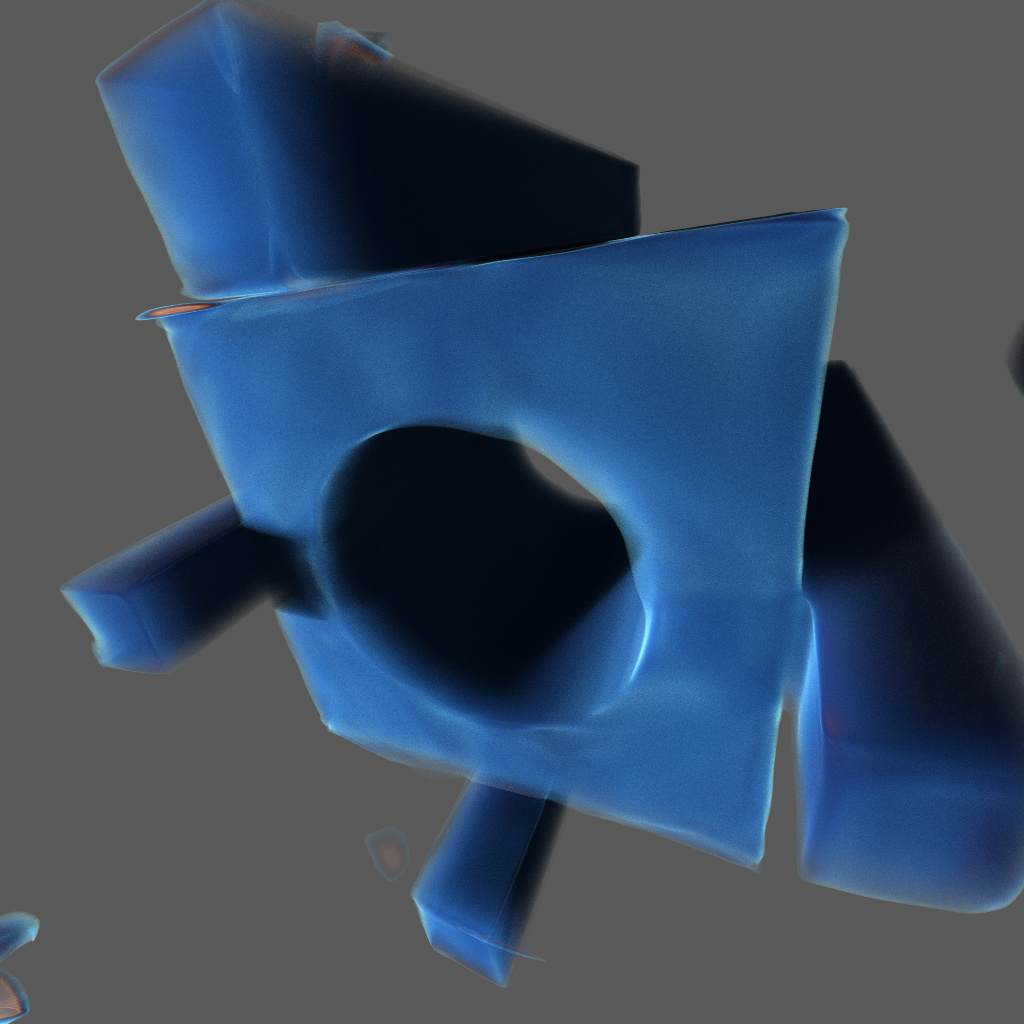} &
                    \pairbox{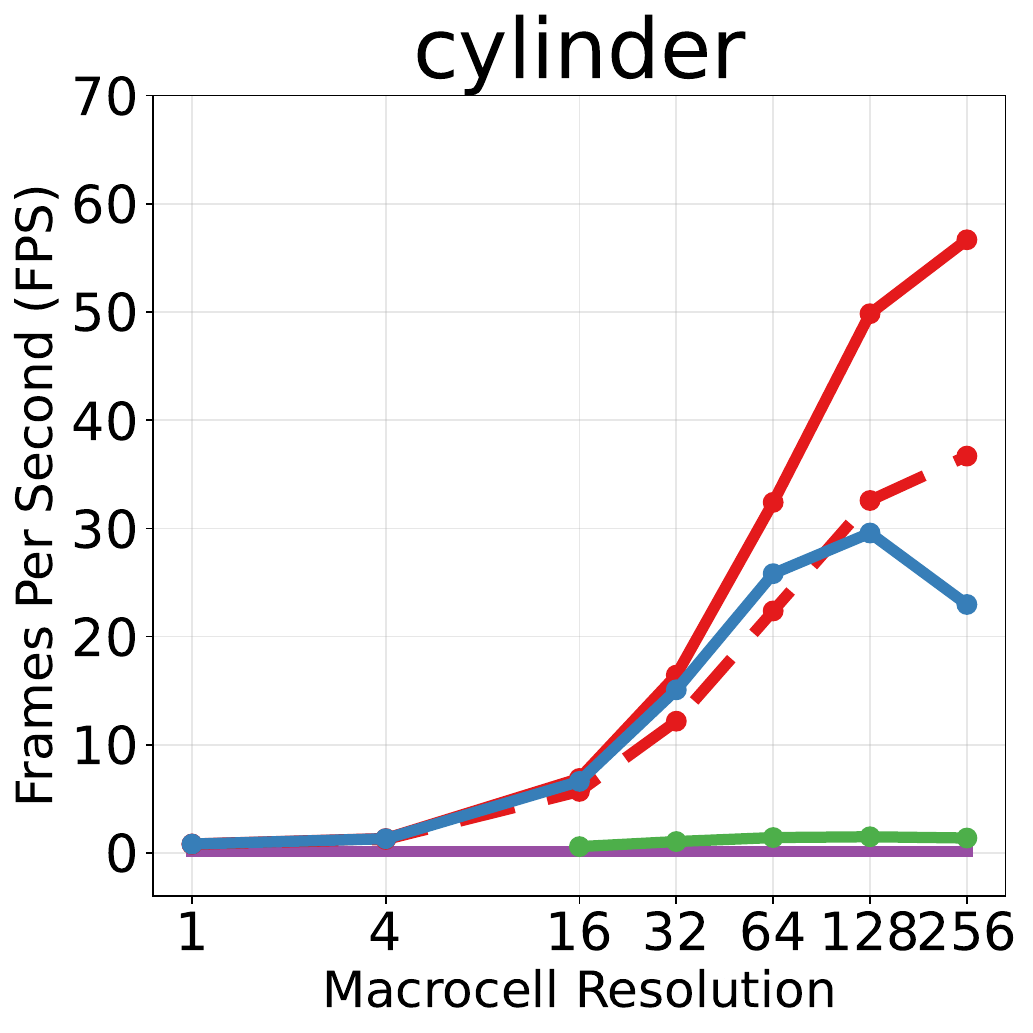}{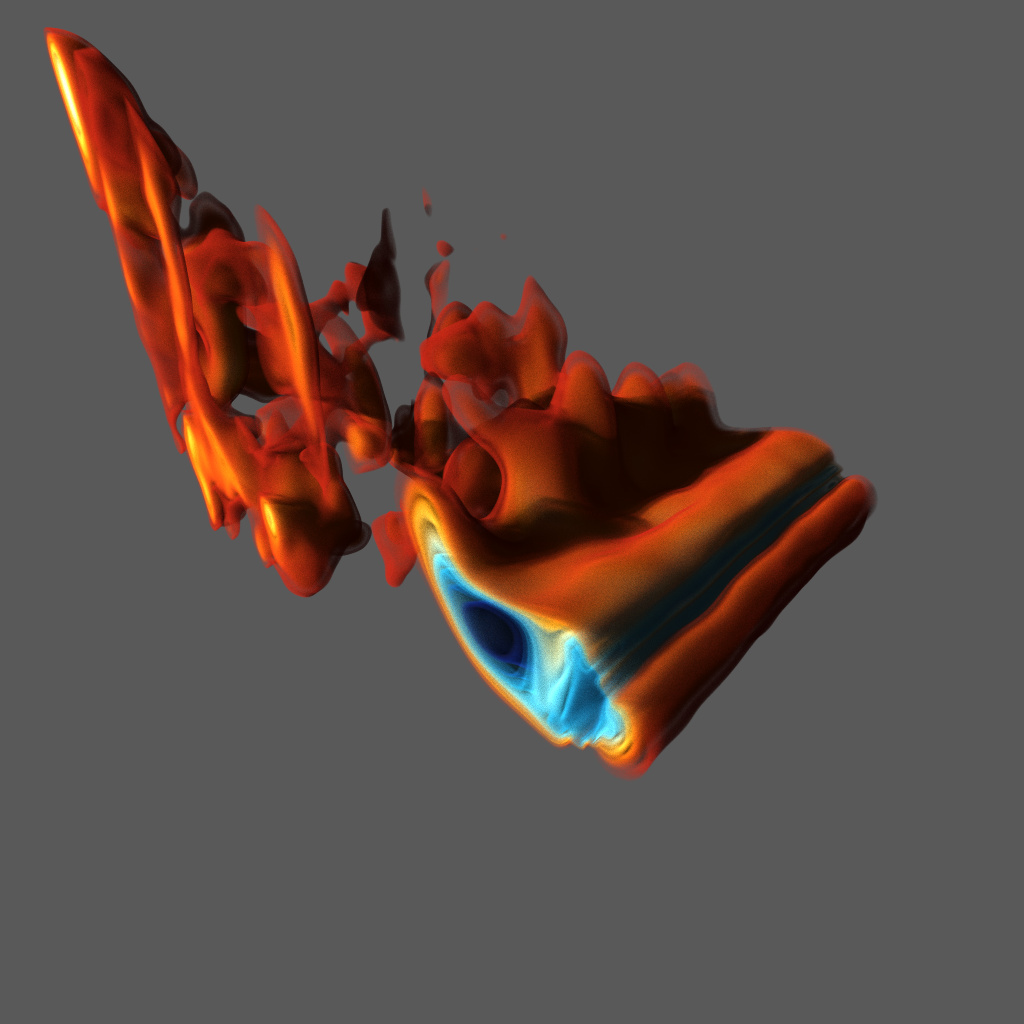} &
                    \pairbox{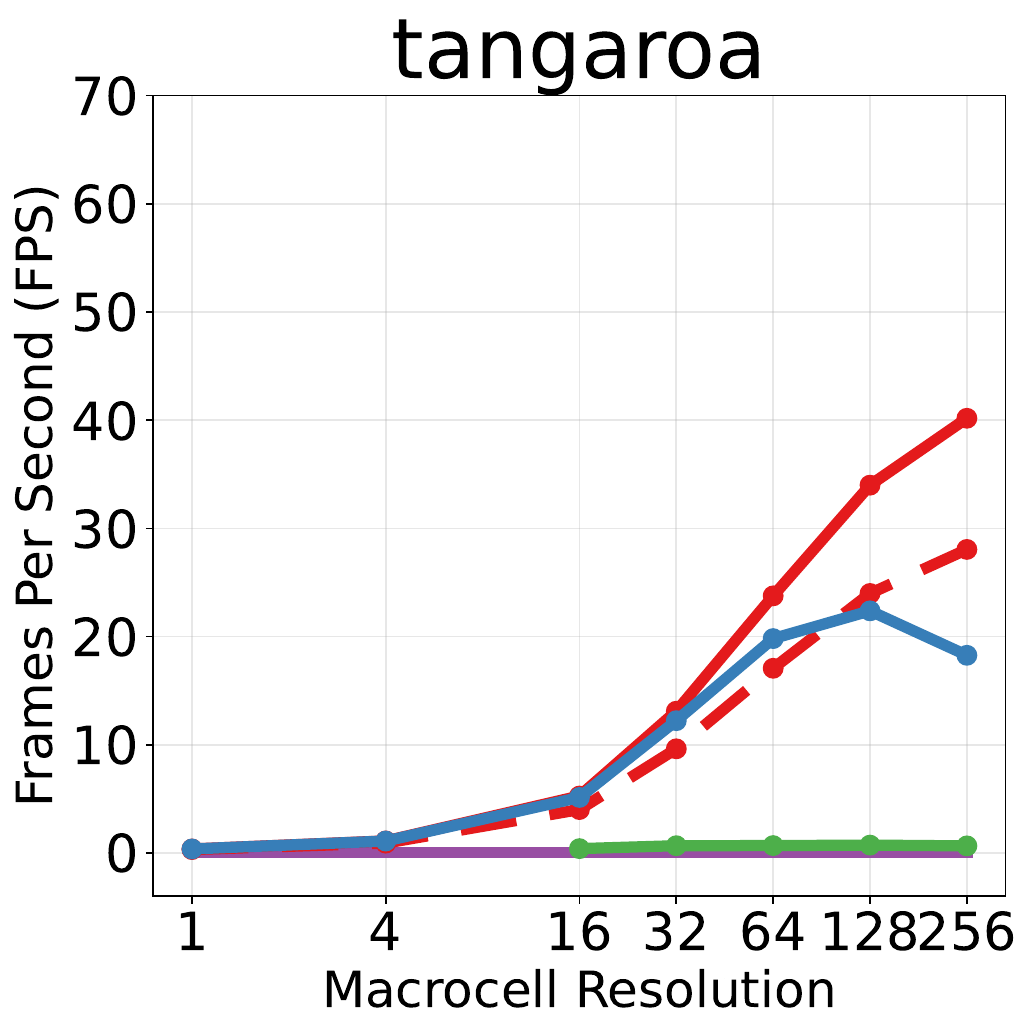}{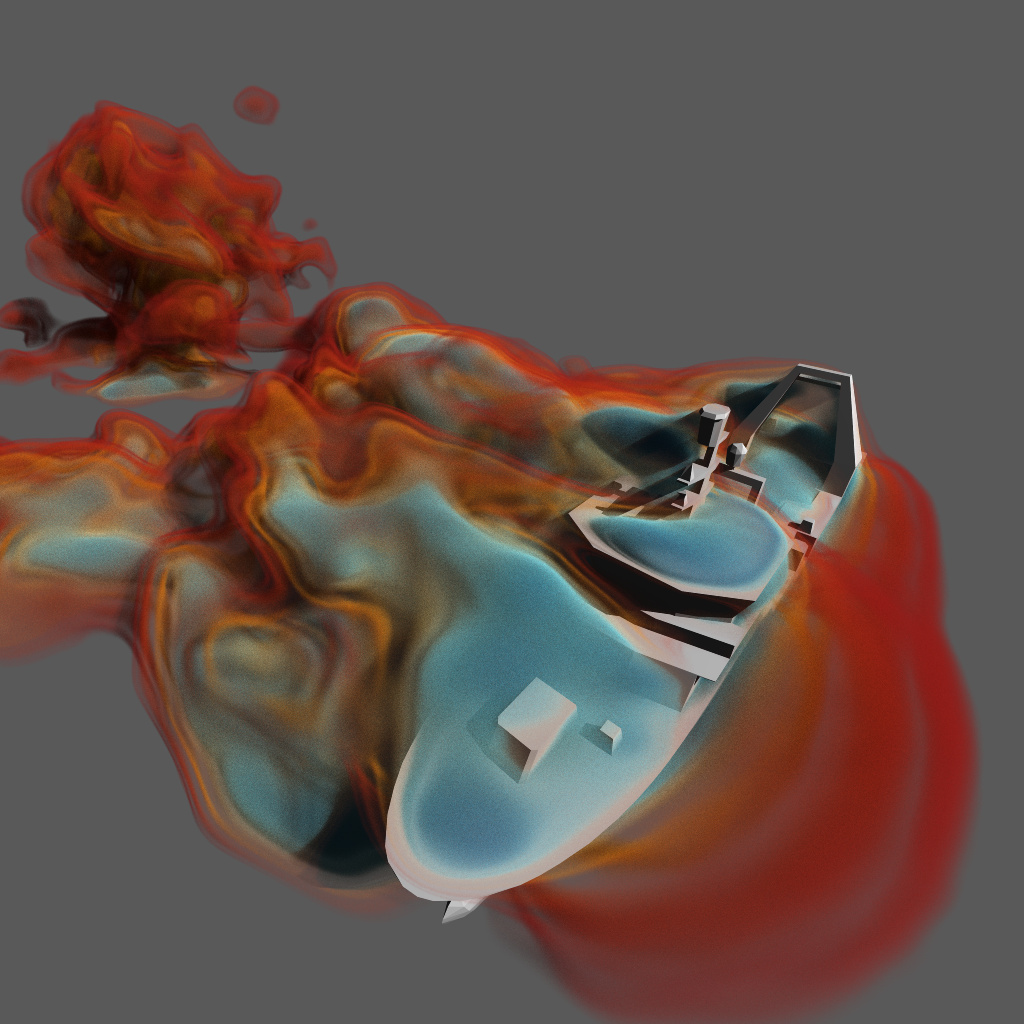} &
                    \pairbox{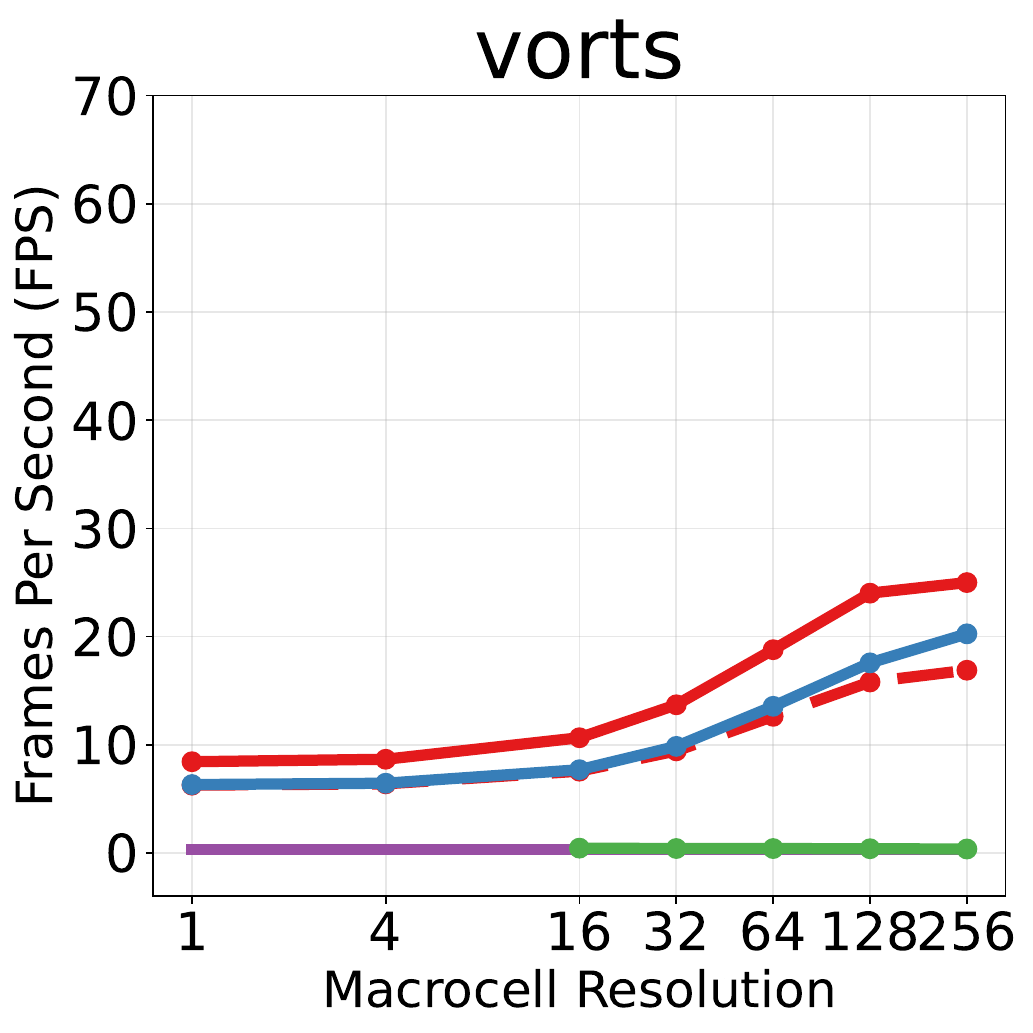}{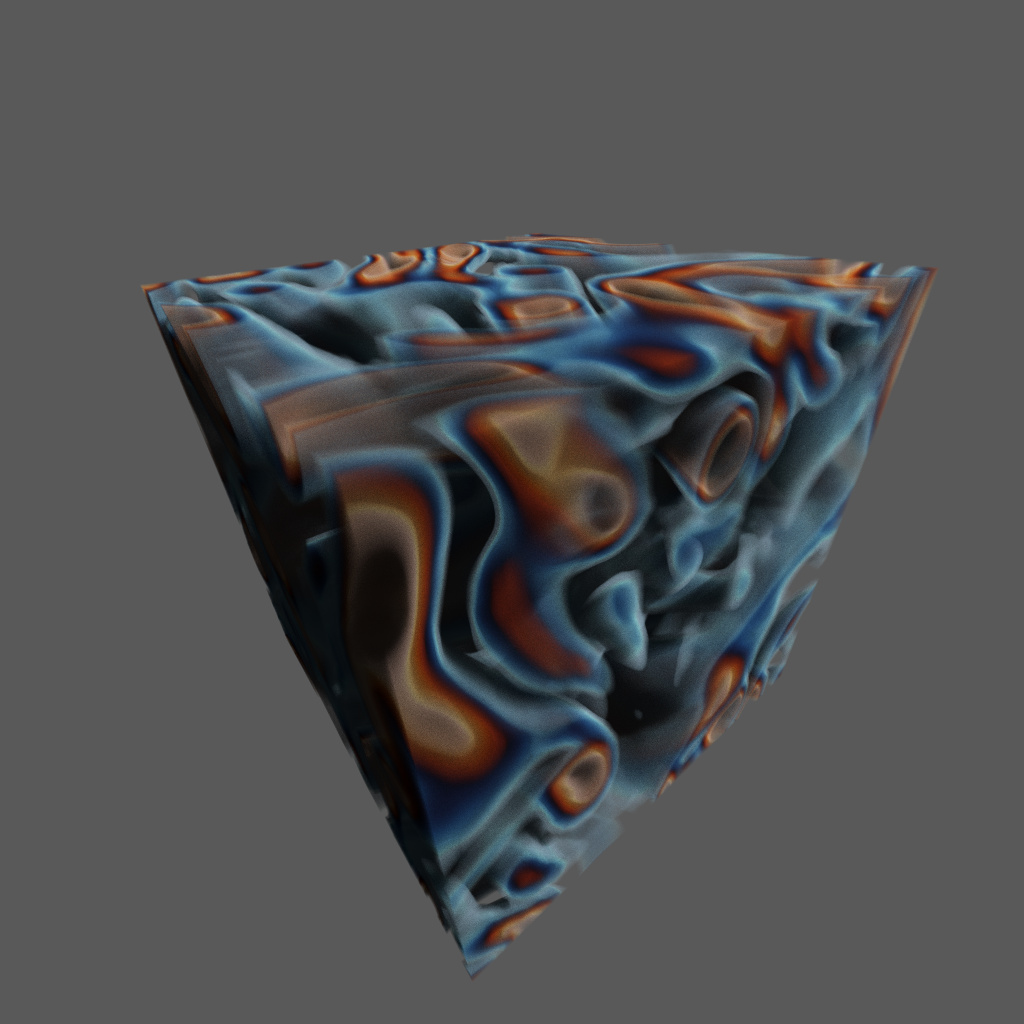}
                \end{tabular}%
            }%
        \end{minipage}%
    }

    \caption{Frames per second performance (higher is better) of various volume rendering methods against increasing macrocell grid resolution for six INR datasets. The methods are ray marching (which does not use macrocells and therefore remains nearly constant at $\sim0.2$ FPS), a na\"ive delta tracker (closely following ray marcher), wavefront delta tracker (ours), and ray tracing (RT) core accelerated wavefront delta tracker (ours). We present the performance of the RT core wavefront d. tracker with only emission+absorption and ray-traced shadows (as ``shdw.''). The 400-frame-converged images for each dataset are given below the plots.}
    \label{fig:performance}
\end{figure*}

We measure and compare the base efficacy of our framework without the query reduction strategies. In~\autoref{fig:performance}, we show various volume rendering implementations rendering the same scenes over six datasets. For each method, the frames per second (FPS) performance is measured against increasing macrocell grid resolution, as shown in~\autoref{fig:anim_update}.

To independently assess which components and hardware resources contribute to our framework’s performance, we evaluate four GPU-based renderer implementations: (1) a simple ray marcher; (2) a na\"ive delta tracker based on implementations designed for conventional explicit data formats (e.g., voxels and meshes); (3)  a wavefront delta tracker operating in the same four stages and using tensor cores, but without RT-core-based traversal (instead using 3D DDA); (4) our fully realized four stage framework, which utilizes RT cores for traversal (see~\autoref{fig:performance}). The results draw a consistent picture across different INR architectures, data shapes, and physical quantities represented, including CT- and CFD-derived scalars.

Ray marching is by far the slowest approach; even with sub-Shannon–Nyquist step sizes, it fails to exceed 1 FPS. As discussed earlier, such dense sampling strategies require a prohibitively large number of samples to achieve interactive performance on INR data. Since ray marching does not operate over macrocells, it is depicted as flat lines in~\autoref{fig:performance}.

A simple delta tracker without specialized hardware or query batching also performs below 1 FPS. Designed for conventional volume rendering with fast memory lookups, this approach underperforms when each sample instead requires a GPU thread to execute a six-layer neural inference independently per ray, failing to amortize the cost of neural evaluation across samples and leading to significant thread divergence and inefficient parallelism for its most expensive procedure.

In contrast, our four-stage framework applies more suitable parallelization for neural evaluation and leverages tensor cores, enabling delta tracking over INRs to reach approximately $\sim25$ FPS, making it $\sim125\times$ faster than the simple delta tracker.

Furthermore, replacing DDA traversal with RT-core-based traversal yields more than a $2\times$ performance improvement for most tests, achieving over 40 FPS for emission–absorption rendering and 30--40 FPS with shadows, which require approximately twice as many samples. 

Across these results, CoordNet is the most challenging representation for our framework, reaching only 19--24 FPS compared to roughly 30--40 FPS for the FFN and SIREN models. This stems from CoordNet’s deeper residual MLP, which introduces longer per-query dependency chains, reducing effective tensor-core and warp occupancy on the hardware.

\newcommand{\BigO}[1]{\ensuremath{O\bigl(#1\bigr)}}

Despite INR queries being the dominant bottleneck, using RT cores for traversal still provides a significant speed-up. We attribute this to two factors: (1) the sparse acceleration structure enables more effective empty-space skipping, with \BigO{n \log n} traversal rather than DDA’s \BigO{\text{cells-along-ray}}; and (2) DDA traversal must be suspended and resumed across kernel launches, requiring a larger \code{TraversalState} that persists per-ray DDA stepping state (current cell, axis distances, cell boundaries), increasing register pressure and memory traffic.

\autoref{tab:memory_usage},~\autoref{fig:anim_update}, and~\autoref{fig:performance} suggest that a macrocell grid resolution of $64^3$--$128^3$ provides a reasonable trade-off between end-to-end performance, memory consumption, and animation updates. For the following experiments, we use $64^3 \times 20$ for the FFN datasets and $64^3 \times 30$ for the SIREN datasets, using a slightly finer temporal discretization for the latter to better match their higher temporal variation while keeping the spatial resolution fixed.

\subsection{Query Reduction Tradeoffs}
We examine the trade-offs of the query-reduction schemes proposed in~\autoref{sec:adaptive_budgeting} and~\autoref{sec:homogeneity_pruning}. We explore the parameter space and the practicality of each scheme.

\begin{figure*}[htbp!]
    \centering
    \vspace{-0.9em}
    \setlength{\fboxsep}{2pt}
    \setlength{\fboxrule}{0.6pt}

    \newcommand{\truplebox}[3]{%
    \fbox{%
        \begin{minipage}[t]{0.22\textwidth}
            \vspace{0pt}
            \centering
            \includegraphics[width=\linewidth,height=\linewidth]{#1}\\[2mm]
            \includegraphics[width=0.65\linewidth,height=0.65\linewidth]{#2}\\[2mm]
            \includegraphics[width=0.65\linewidth,height=0.65\linewidth]{#3}
        \end{minipage}%
    }%
}

\scalebox{0.85}{%
\begin{tabular}{cccc}
    \begin{minipage}[t]{0.18\textwidth}
        \vspace{-0.5em}
        \centering
        {Ground Truth}\\[2mm]
        \includegraphics[width=\linewidth]{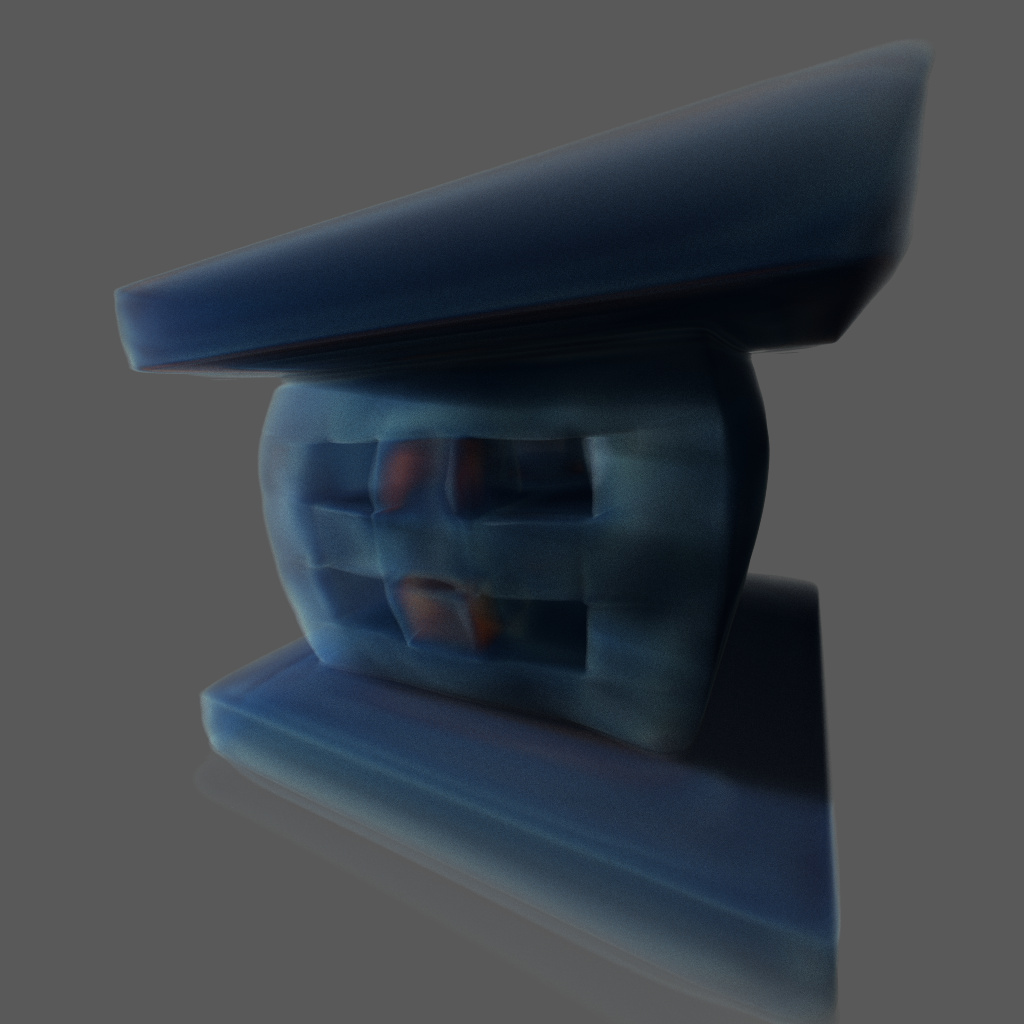}\\[4mm]
             \raggedleft
             \includegraphics[height=0.85\linewidth]{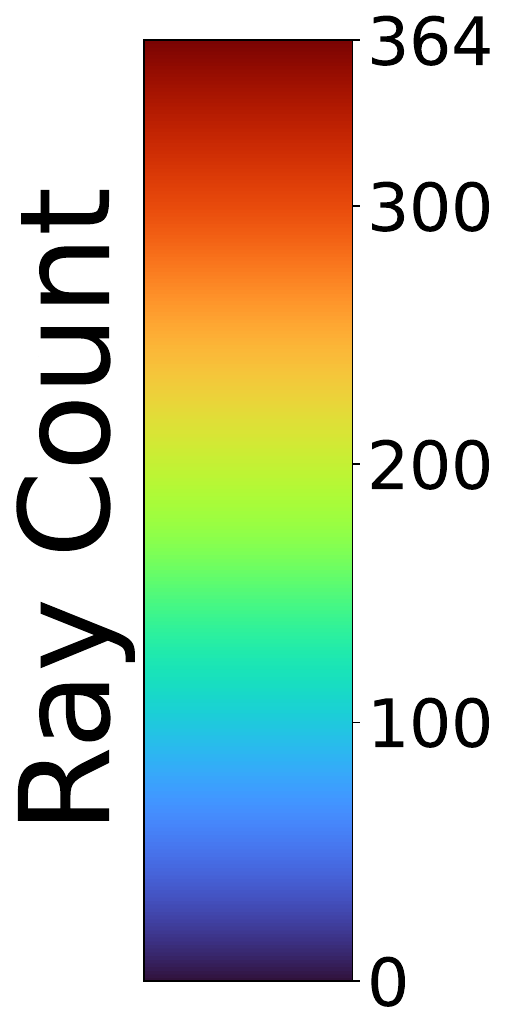}\\ [2.5mm]
             \begin{minipage}[c]{\linewidth} 
        \centering
        \begin{minipage}[c]{0.1\linewidth}
            \centering
            \rotatebox{90}{\textbf{At $\sim1.2$ seconds}}
        \end{minipage}
        \begin{minipage}[c]{0.85\linewidth}
            \centering
            \includegraphics[width=0.92\linewidth]{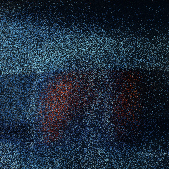}
        \end{minipage}
    \end{minipage}
        \end{minipage}&
        \truplebox{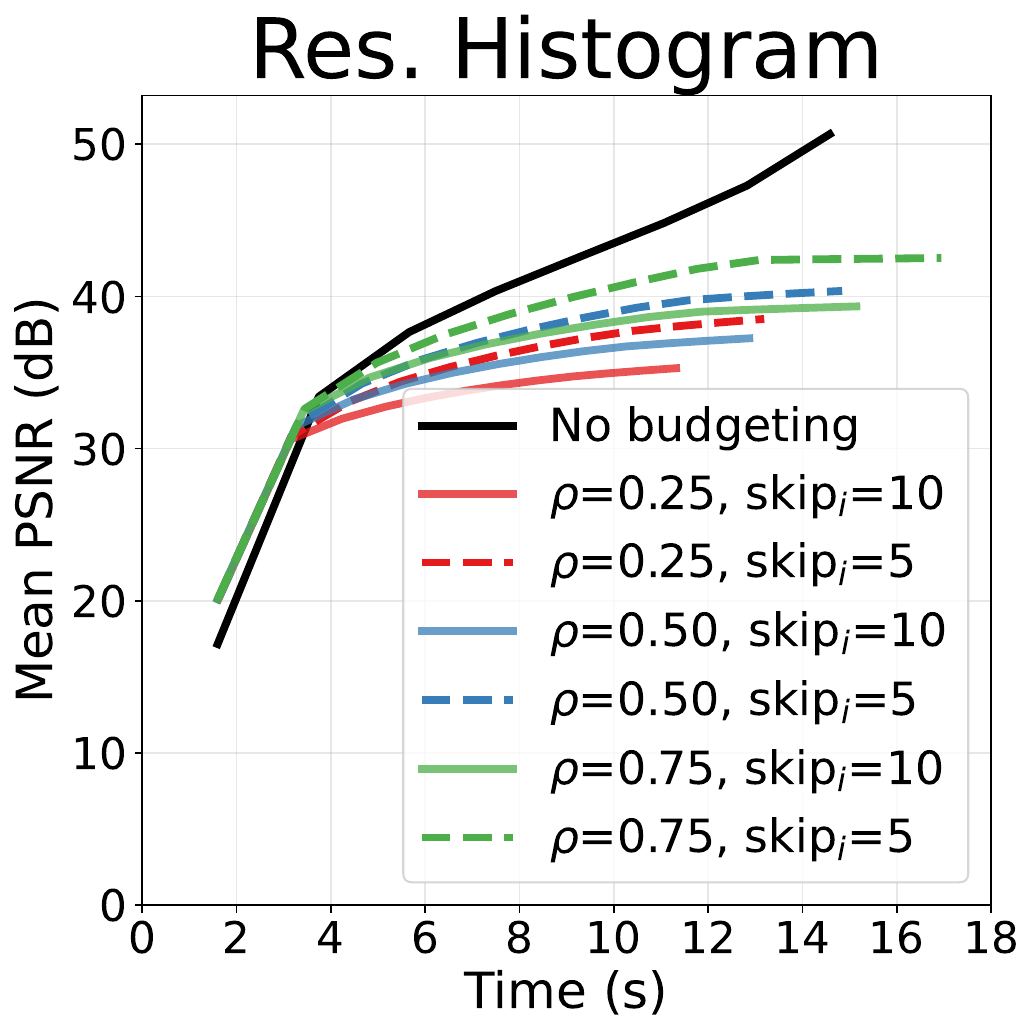}{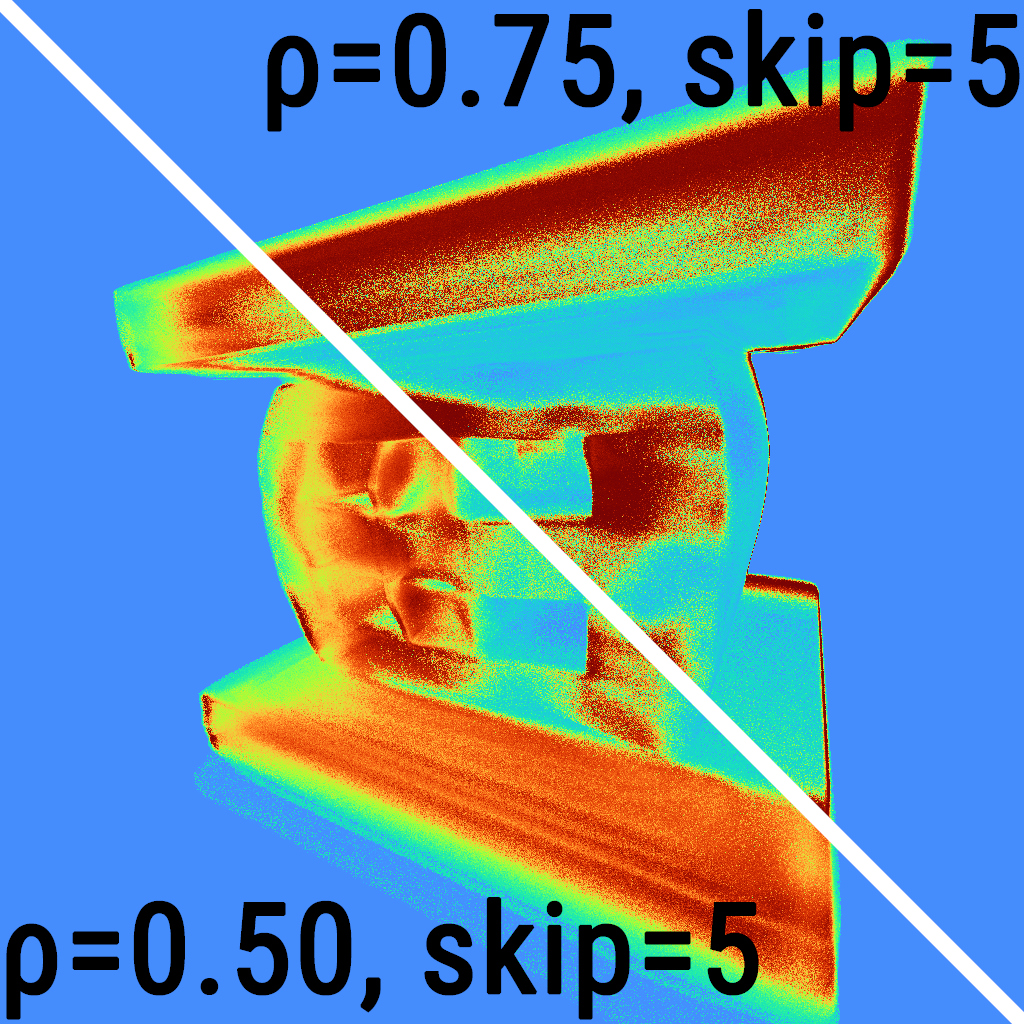}{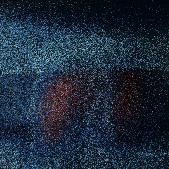} & \truplebox{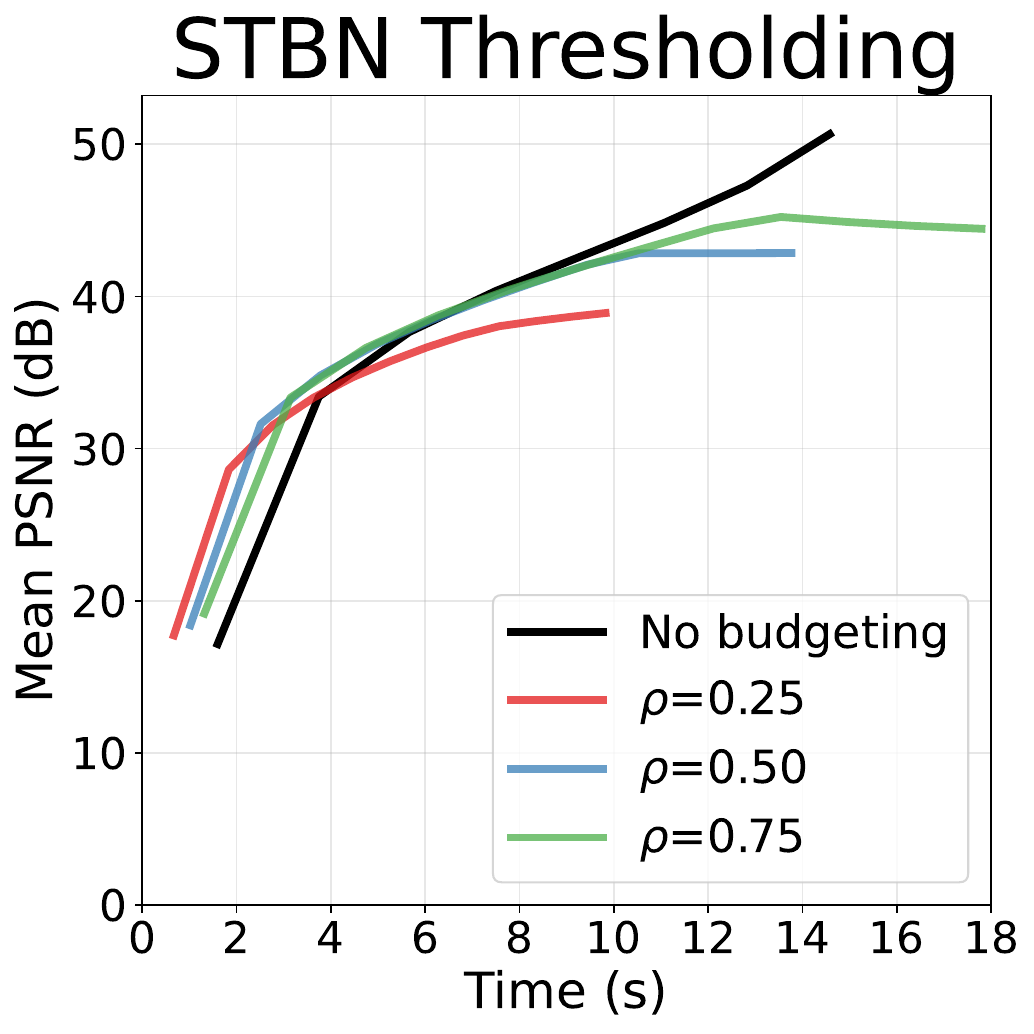}{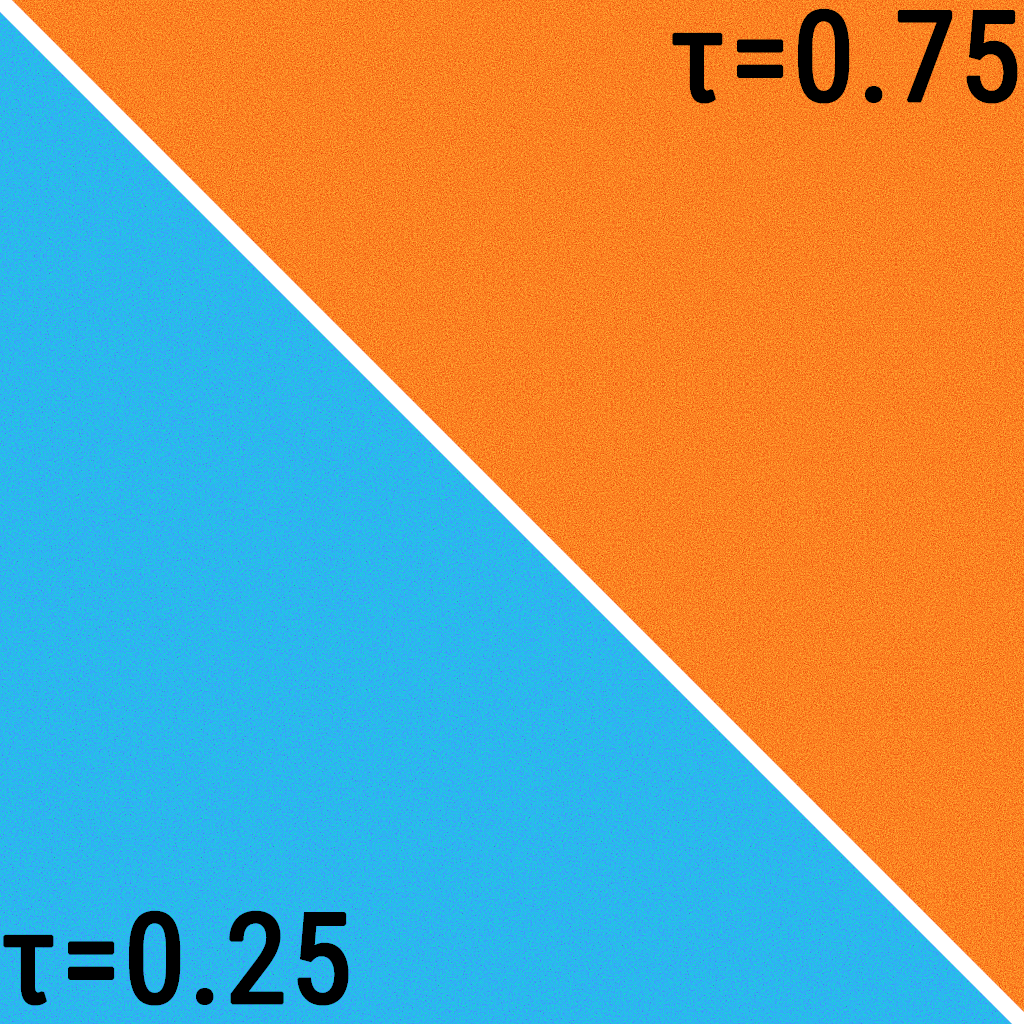}{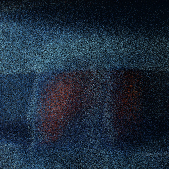} &
        \truplebox{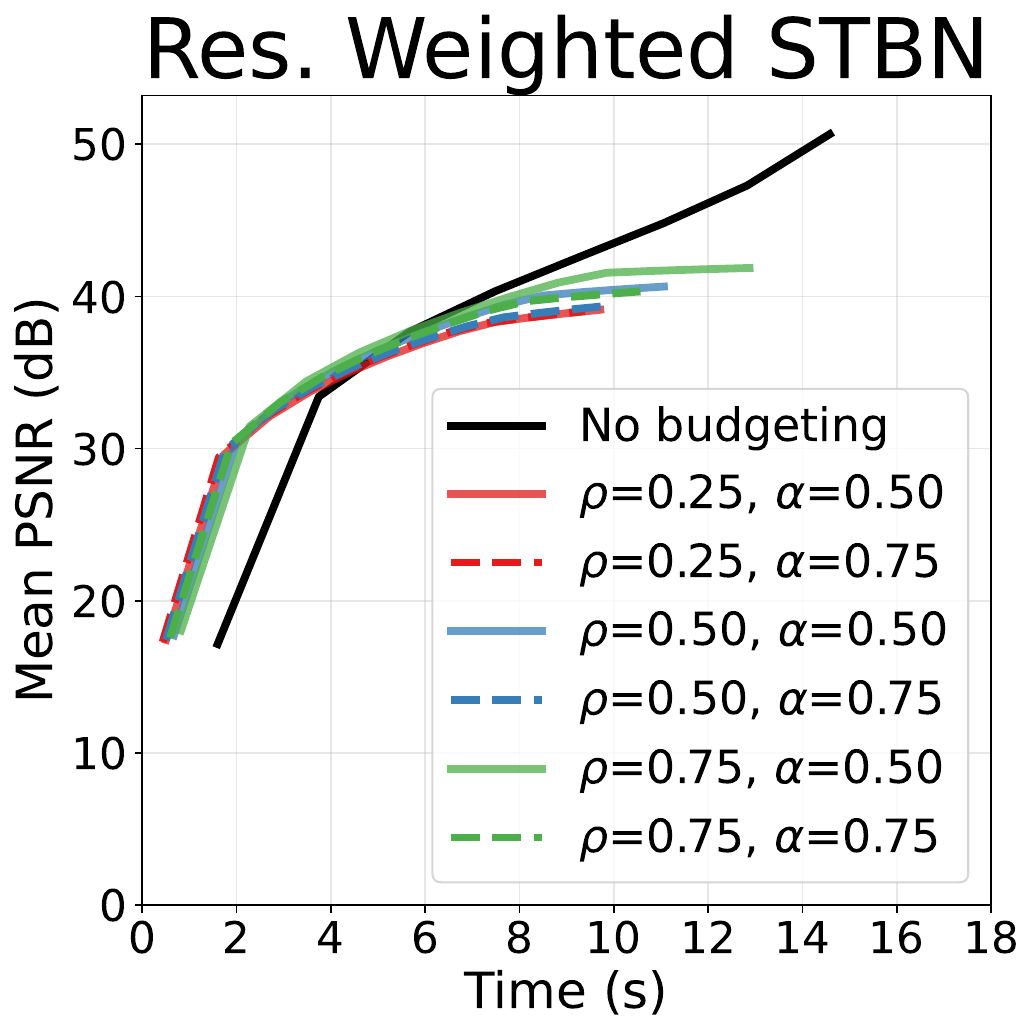}{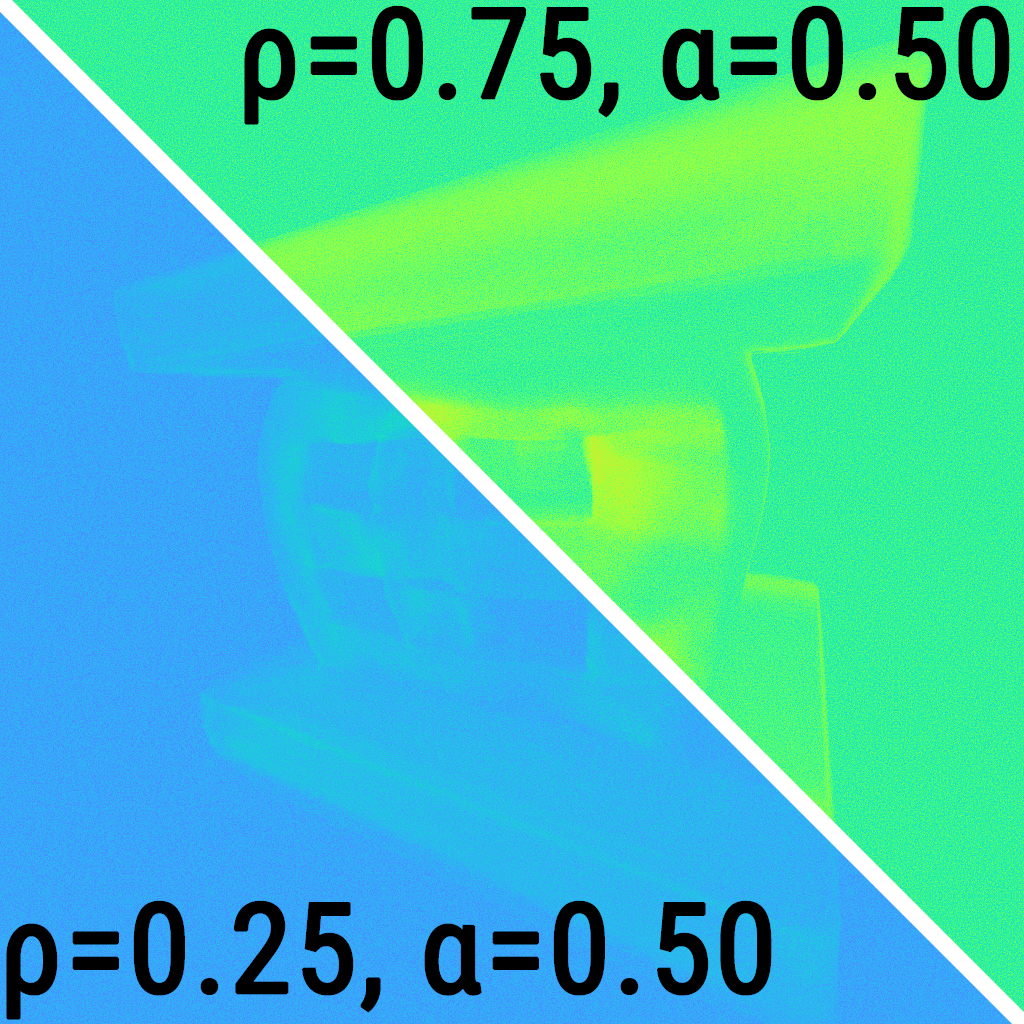}{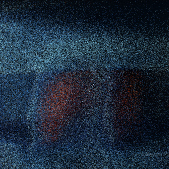} \\
    \end{tabular}
}

    \caption{Evaluation of ray budgeting strategies. The top row shows image quality (higher is better) as a function of elapsed time (seconds) for the three schemes described in~\autoref{sec:adaptive_budgeting} for various parameter configurations. Experiments are conducted on the S03\_001 dataset with a $64^3 \times 20$ macrocell grid resolution. Image quality is measured using the Peak Signal-to-Noise Ratio (PSNR) and the mean difference relative to a 400-frame ground-truth rendering without budgeting. The second row presents heatmaps of the average number of rays cast per pixel for two representative parameter settings per strategy. The final row shows close-ups of unconverged images at $\sim1.2$ seconds, highlighting noise patterns (using the same parameters as the top-right heatmaps in the second row).}
    \label{fig:converge_time}
    \vspace{-1.4em}
\end{figure*}

In~\autoref{fig:converge_time}, we evaluate three ray-budgeting strategies against a 400-frame ground truth image. Each plot shows Peak Signal-to-Noise Ratio (PSNR) relative to the ground truth over time for 3--6 representative parameter settings, while the heatmaps visualize the average number of rays cast per pixel, and the close-ups highlight early-frame noise patterns at $\sim1.2$ seconds.

Early plot points of the residual-histogram thresholding scheme exhibit closely clustered trends across parameter settings. Because it ranks pixels by residual and skips stable ones for multiple frames, it concentrates rays most aggressively on unstable regions, as reflected in the heatmaps. This yields lower error than the no-budget baseline in the first few seconds, but after approximately 3 seconds, the benefit diminishes as repeatedly skipped pixels reduce opportunities for further error correction. The initial frames retain the ``white-noise'' appearance of standard stochastic rendering.

\label{sec:query_reduction_tradeoff}

In contrast, STBN thresholding does not use residuals and distributes a fixed ray budget according to a high-frequency noise pattern. Its cost-effectiveness, therefore, depends mainly on the render fraction ($\rho$), trading reduced ray counts for increased error. Higher $\rho$ reduces early error, while lower $\rho$ saves more rays and increases effective frame rate by rendering only a fraction of pixels per frame. The curves reconverge after approximately 5 seconds as the image stabilizes. Among the tested settings, $\rho=0.75$ provides the best trade-off, maintaining effective amortization up to $\sim6$ seconds. Additionally, early frames perceptually benefit from the high-frequency structure of blue noise, as shown in the bottom row.

In the hybrid method (residual-weighted STBN), the base render fraction still controls the overall cost, while the residual weighting redistributes a portion of that budget toward unstable pixels. The resulting heatmaps show a concentration of rays on changing structures while retaining a well-distributed high-frequency pattern, making the scheme a more cost-effective compromise between pure STBN and residual-histogram thresholding. Its curves remain closely clustered, indicating the method is relatively insensitive to parameter choice. We observe it reaching less than 4\% error slightly faster than pure STBN in the first two seconds, while preserving similar early-frame noise characteristics.

The first set of frames is particularly relevant for interactive use, where rapid visual feedback influences user decisions. In this regime, STBN-based methods offer the clearest practical advantage, as blue-noise sub-sampling produces more favorable sparse-update patterns. Residual-weighted STBN is best viewed as a modest refinement: steering part of the budget toward unstable pixels makes it slightly more cost-effective in the first few seconds, but the gain over pure STBN is small and fades as the image stabilizes. The fully residual-driven histogram scheme is more aggressive, but on this dataset, its benefit is short-lived, and its early-frame noise patterns resemble ``white-noise,'' making it less attractive for perceptual quality. Nevertheless, it remains useful as a simple adaptive baseline and as an incremental step toward the hybrid methods, suggesting that residual guidance is better realized when combined with blue-noise allocation than when used alone. Residuals also remain valuable because they can be reprojected across frames during interaction, a benefit not captured by these static-view plots. All strategies produce images that converge to within 2\% of the ground truth.

\begin{figure}[!htbp]
\vspace{-0.8em}
\hspace{-3em}
    \centering
    \begin{minipage}[c]{0.44\linewidth}
        \centering
        \includegraphics[width=\linewidth]{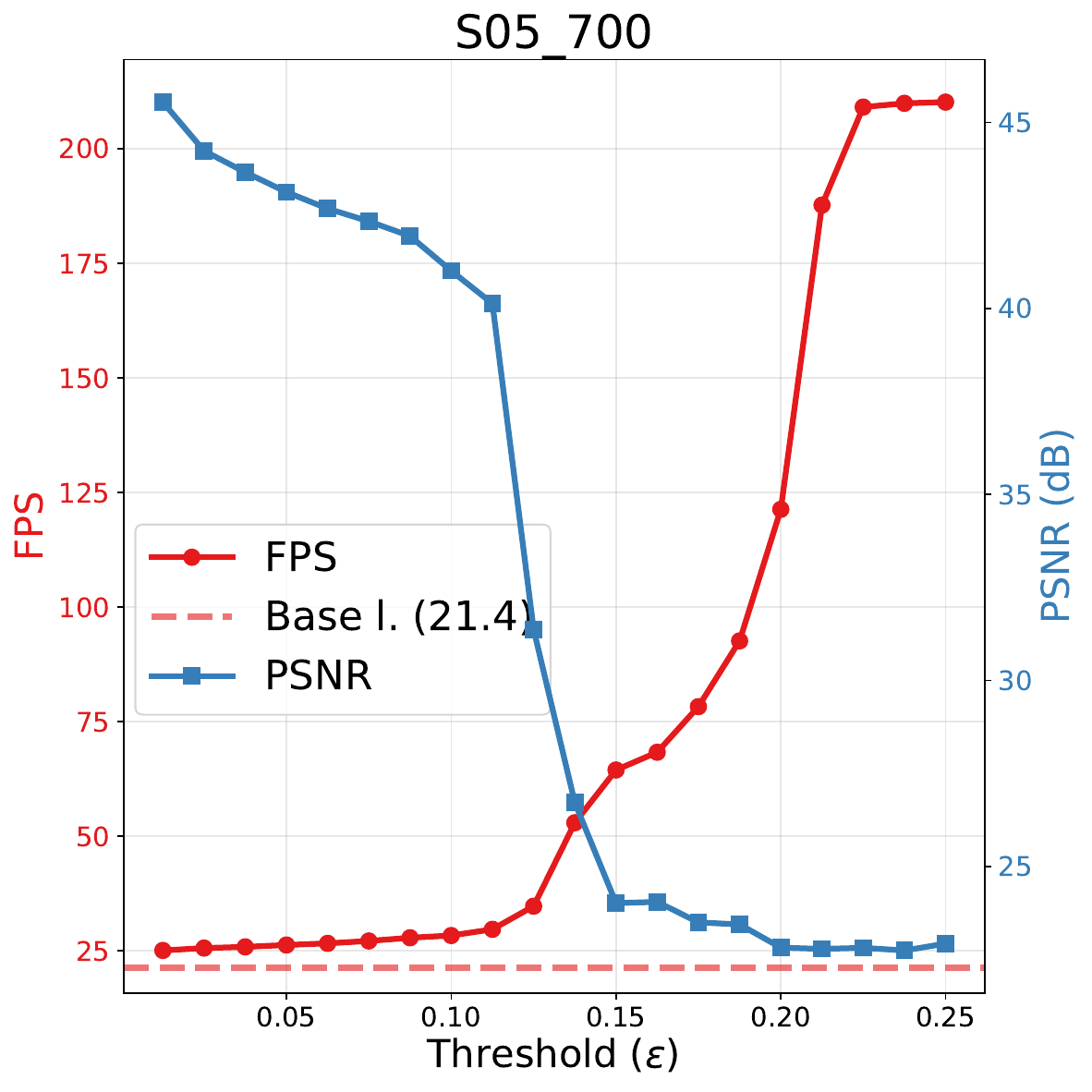}
    \end{minipage}
    \begin{minipage}[c]{0.18\linewidth}
        \centering
        \includegraphics[width=\linewidth]{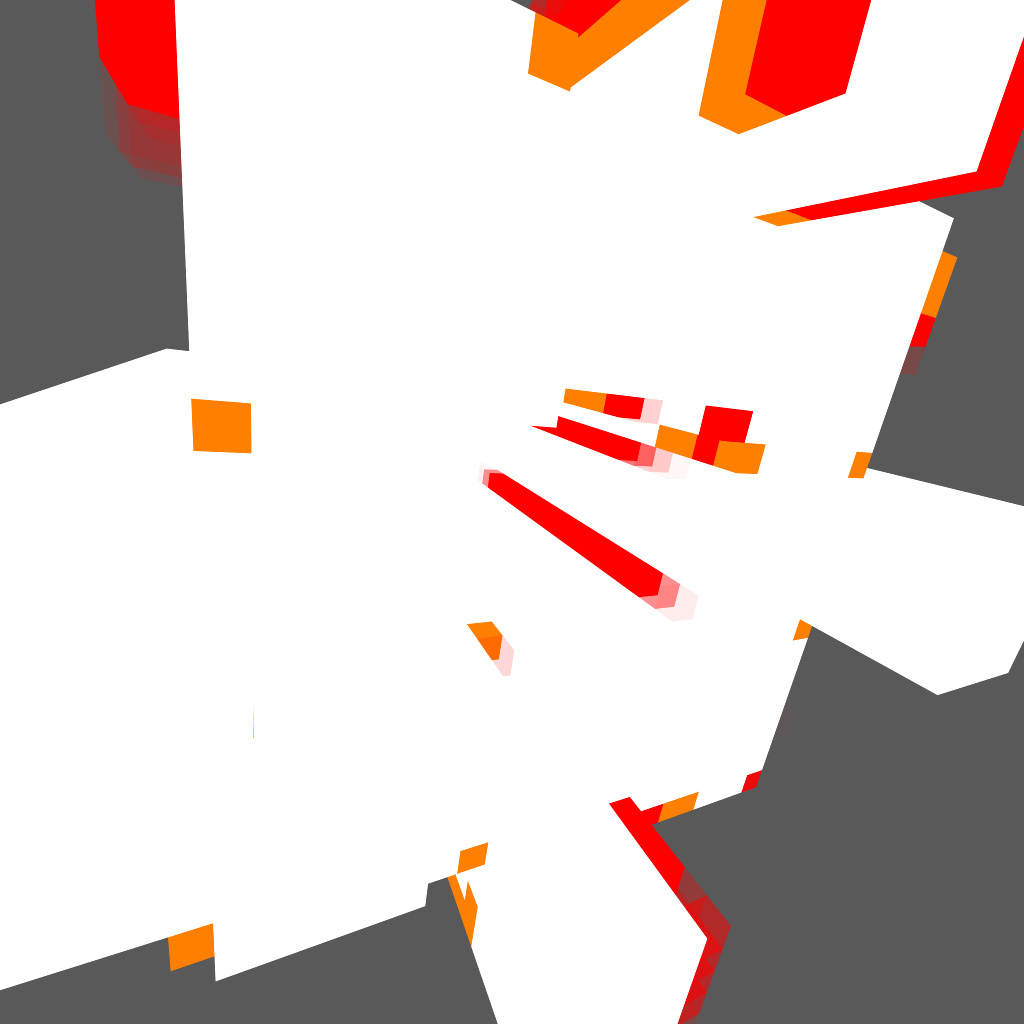}\\[2pt]
        \includegraphics[width=\linewidth]{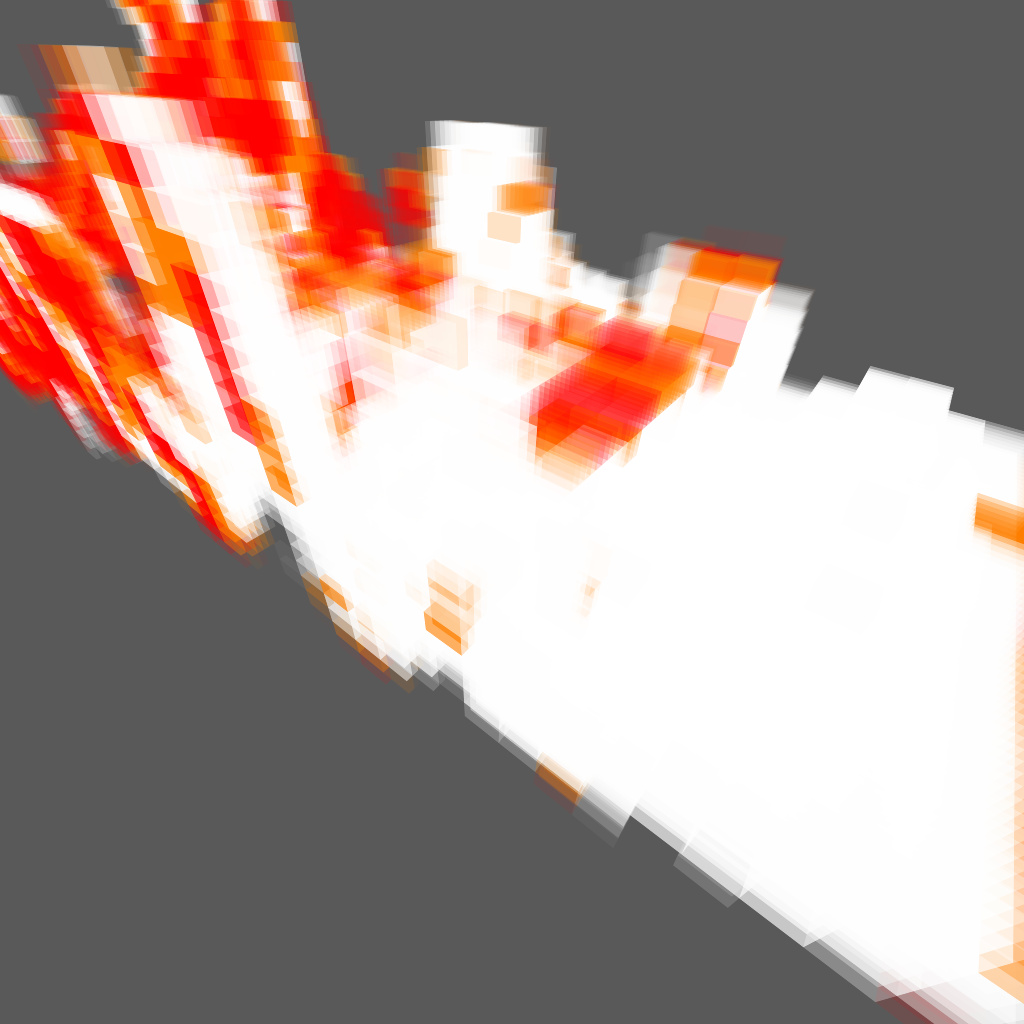}
    \end{minipage}
    \begin{minipage}[c]{0.44\linewidth}
        \centering
        \includegraphics[width=\linewidth]{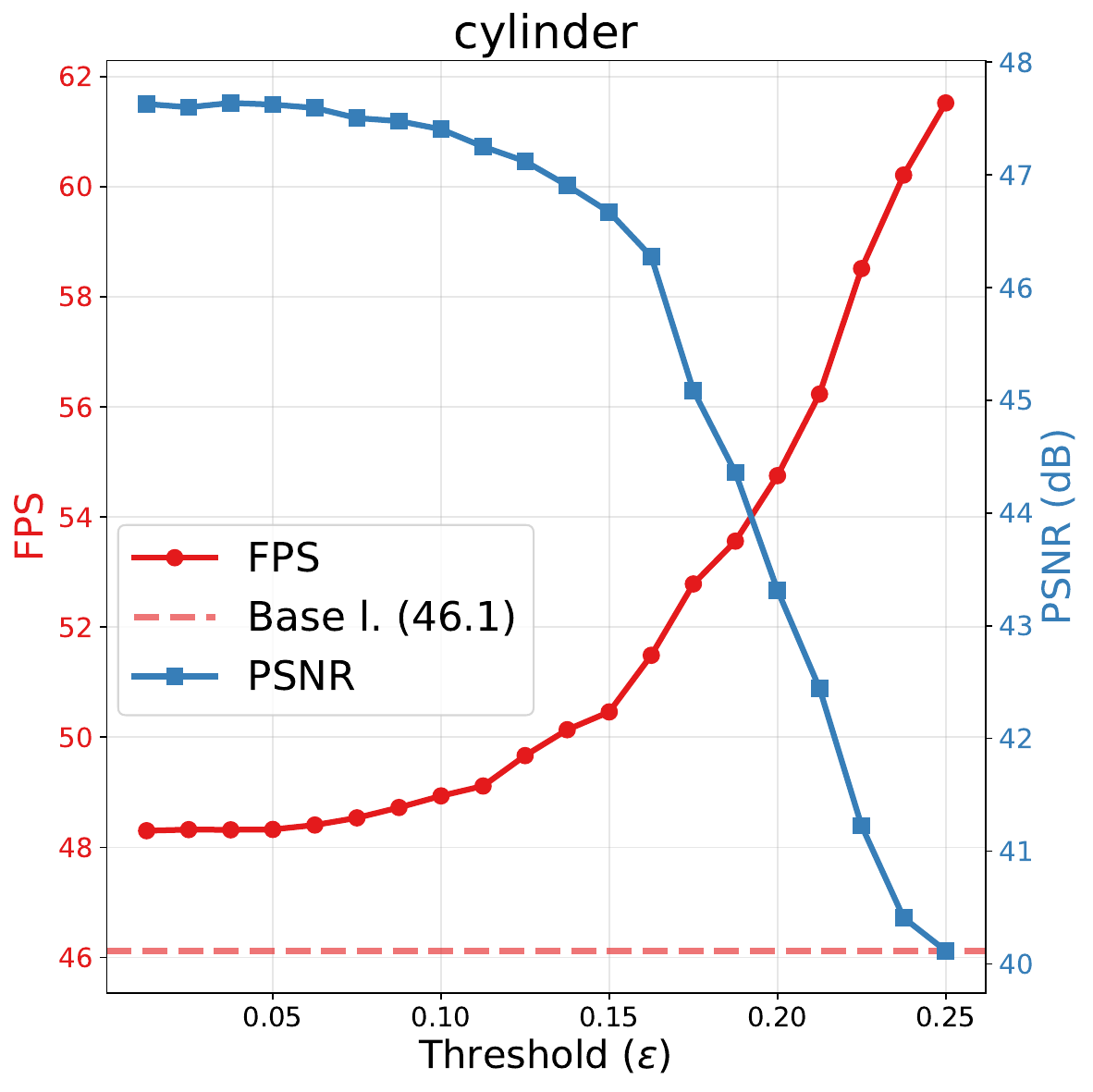}
    \end{minipage}
    \vspace{-1em}
    \caption{Homogeneity-based query pruning on S05\_700 and cylinder data. Dual-axis plots show performance (FPS, red) and image quality (Peak Signal-to-Noise Ratio, PSNR) versus increasing $\epsilon$ from~\autoref{sec:homogeneity_pruning}; higher is better on both y-axes. Middle visualizations mark pruned cells in red and stochastically pruned cells in orange for S05\_700 (top) and cylinder (bottom).}
    \label{fig:homog_cull}
    \vspace{-0.95em}
\end{figure}

Homogeneity-based query pruning trades image fidelity for FPS through $\epsilon$ as seen in ~\autoref{fig:homog_cull}. Larger values skip more homogeneous macrocells, but replace true samples with per-cell estimates. The main caveat we observe is a dataset-dependent critical threshold, $\epsilon_c$, after which quality drops sharply. We find $\epsilon_c=0.12$ for S05\_700 and $\epsilon_c=0.16$ for cylinder. At this point, the pruned fraction catches up with the empty-space fraction, already flattening ${\sim}58\%$ of S05\_700's and ${\sim}33\%$ of cylinder's visible cells. Further pruning then erodes the more heterogeneous macrocells, and rapidly degrades the image quality.

The degradation in quality is gradual at first, indicating that small $\epsilon$ values successfully prune near-uniform regions with minimal visual impact. This effect is particularly beneficial in scientific datasets, which often contain extended low-variance regions or zero-valued outer layers, allowing rays to traverse large portions of the volume without incurring additional INR queries. Beyond a critical threshold ($\epsilon \approx 0.15$), however, the error increases more noticeably as larger and more structurally significant regions are approximated. The visualizations on the right illustrate this behavior: deterministically pruned cells (red) and stochastically pruned regions (orange) concentrate in low-variance areas, while higher $\epsilon$ values begin to affect perceptually important structures. In practice, moderate thresholds strike a favorable balance, while larger thresholds may be more cost-effective when rendering more expensive neural models or when the rendering task is more demanding, such as path tracing.

\begin{figure}[!htbp]
    \centering
    \includegraphics[width=0.35\linewidth]{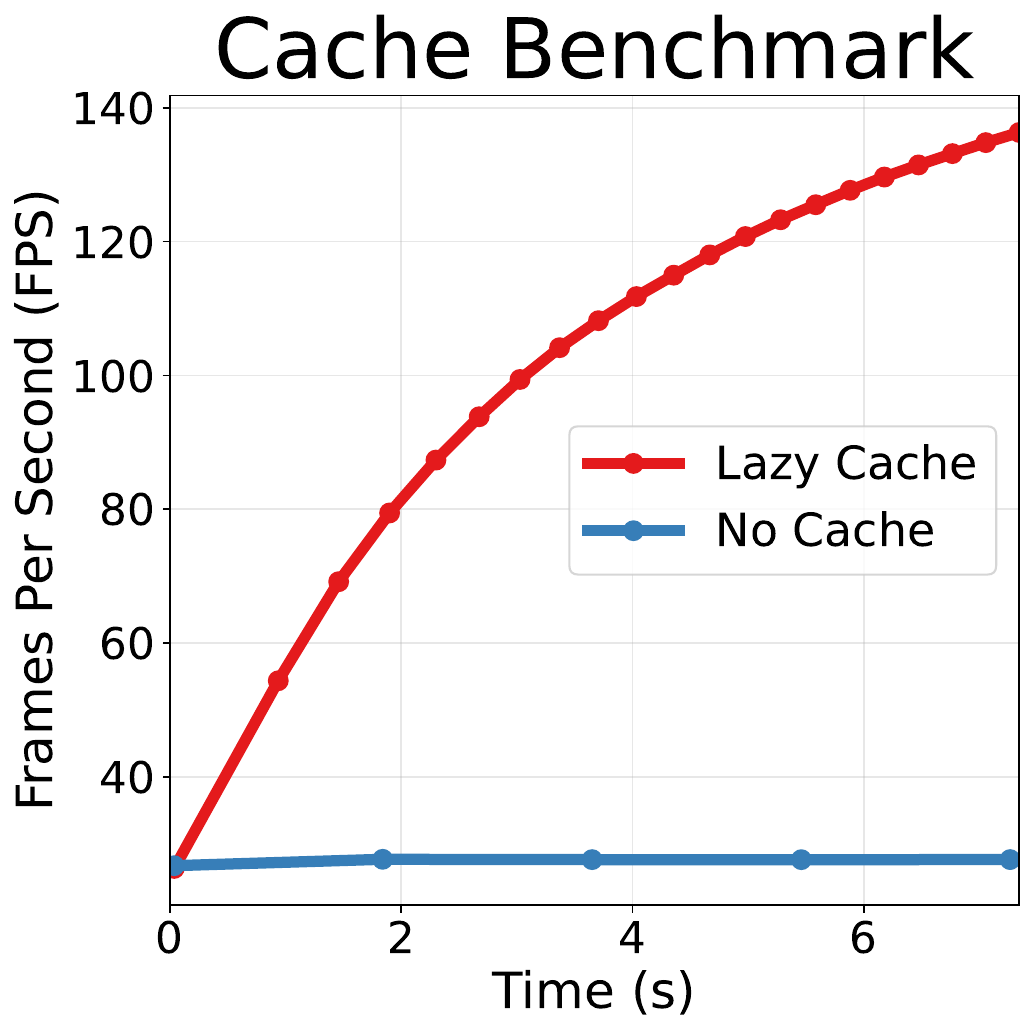}
    \includegraphics[width=0.35\linewidth]{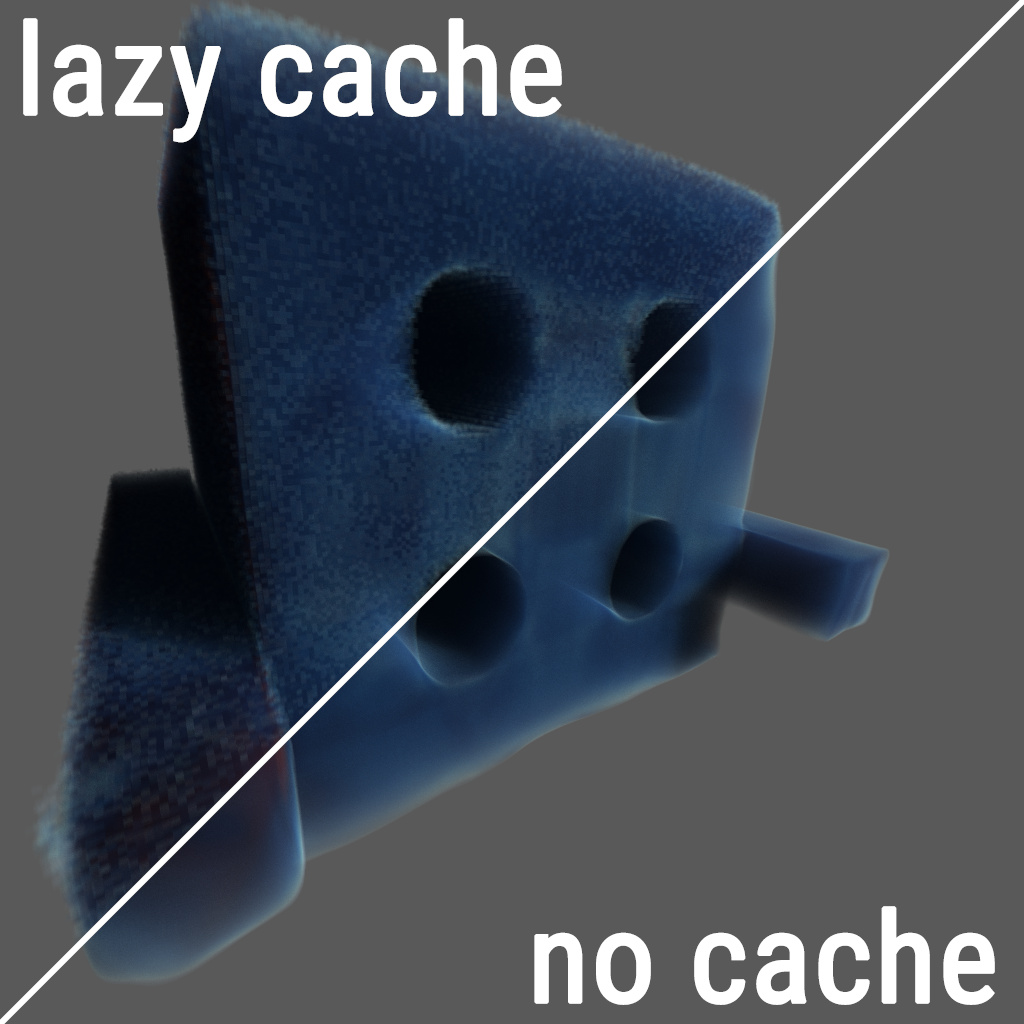}
    \caption{Frames per second performance over time with and without a lazy caching scheme (500 MB cache). The corresponding renderings after 400 frames are shown on the right.}
    \label{fig:cache_sampling}
    \vspace{-0.5em}
\end{figure}

A vital discussion concerns the use of cache-based acceleration (such as~\cite{zavorotny2025CacheINR}) as a complementary technique rather than an alternative to our framework. As a proof of concept, we implement a simple \emph{lazy caching} scheme on top of our pipeline; as shown in~\autoref{fig:cache_sampling}, it already provides performance gains that accumulate over time at a fixed timestep.

However, the temporal limitations noted in~\autoref{sec:framework_design} still apply: repeated navigation through time can invalidate cached samples and reduce their reuse for time-varying INRs. Moreover, caching introduces a spatial approximation trade-off, since finite cache resolution can bias the reconstruction. As visible in~\autoref{fig:cache_sampling}, artifacts remain even with a 500 MB cache (about $397\times$ the INR size). At that point, one could also consider other adaptive surrogate representations such as PruningAMR~\cite{Zvonek2025PruningAE}. Thus, caching can be beneficial under sufficient coherence, but it does not fully resolve the challenges of interactive INR rendering.

\section{Limitations and Future Work}
Our current implementation is tightly coupled to NVIDIA’s CUDA, \optix, and RT-core stack. While the pipeline is portable in principle, achieving comparable performance on other vendors would require careful mapping to their ray tracing and matrix-compute units; in such cases, the wavefront delta tracker in~\autoref{fig:performance} provides a viable alternative and starting point in the absence of RT cores. Our proof-of-concept uses a thin, templated CUDA evaluator with the INR architecture explicitly implemented; following this work with additional architectures would require adding the corresponding layers or relying on a library~\cite{Muller2021tinyCudaNN}.

\begin{wrapfigure}{r}{0.14\textwidth}
    \vspace{-0.5em}
    \centering
    \includegraphics[width=0.12\textwidth]{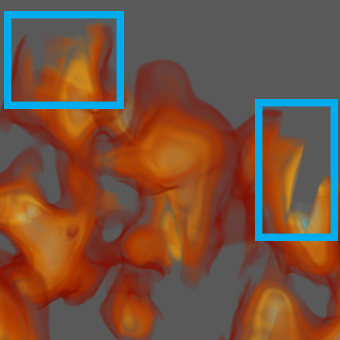}
    \caption{Artifacts in the cylinder data caused by a reduced temporal macrocell count
    ($T=5$) and no ghost passes.}
    \label{fig:macrocell-limitations}
    \vspace{-0.8em}
\end{wrapfigure}

A fundamental limitation lies in the construction of macrocells. Extrema are approximated via lattice sampling and temporal interpolation, which can misestimate extrema and lead to loose majorants, increasing null collisions, or producing structured artifacts (e.g., block-like regions as seen in~\autoref{fig:macrocell-limitations}). These temporal artifacts occur when temporal macrocell bounds are too coarse or insufficiently refined, which is especially relevant for INRs with high temporal variability, such as those for CFD simulations. Ghost passes mitigate underestimation but do not guarantee uniform refinement. In addition, both ray budgeting and homogeneity-based pruning are heuristics that rely on stable residuals and well-behaved scalar ranges; during interactions or when aggressive thresholds are used, homogeneity-based pruning may introduce bias. Finally, INRs tend to favor low-frequency structure, which can produce smooth or ``network-shaped'' artifacts independent of the rendering method and reflects a broader limitation of the representation.

Future work could address both representation and rendering efficiency. On the representation side, tighter macrocell bounds, improved temporal interpolation, or learned bound estimation could reduce underestimation and null collisions. In particular, the INR training process could expose additional structure, such as conservative sub-region bounds via interval networks or auxiliary predictors for local extrema, enabling more accurate majorants. On the rendering side, replacing heuristic query-reduction schemes with learned or temporally coherent strategies could improve robustness under interaction, provided they remain lightweight. Gradient-based shading could also be incorporated by estimating local gradients using the central-difference method~\cite{carson2021rtgem}. Additionally, neurally predicting candidate-hit regions to guide sampling could further reduce unnecessary INR evaluations. Finally, testing our framework on physics-informed neural networks (PINNs) is a promising direction for future work.
\vspace{-0.2em}
\section{Conclusion}
We presented a query-efficient stochastic volume rendering framework for time-varying implicit neural representations (INRs), enabling interactive visualization and low-latency exploration of the continuous time domain without requiring resampling, retraining, or caching as a prerequisite. By reinterpreting delta tracking as a query-reduction mechanism and combining it with a four-stage ray tracing pipeline, our method decouples traversal from neural evaluation and exploits heterogeneous GPU hardware, including ray tracing and tensor cores. This design minimizes and amortizes neural inference costs while maintaining estimator correctness under practical approximations, enabling direct rendering and interactive animation of continuous time-varying INRs. Our results show that the framework achieves interactive performance ($\sim30$--$40$ FPS at $1024^2$ in most benchmarks) with ray-traced shadows while supporting animation updates within a few milliseconds.

We further demonstrated query reduction can be introduced as a cost-effective extension to our framework in two complementary ways. Adaptive ray budgeting reduces image-level costs by selectively casting rays to pixels that benefit most from new samples, exposing trade-offs among convergence behavior, spatial adaptivity, and perceptual noise characteristics. Homogeneity-based query pruning reduces cost at the sample level by avoiding unnecessary INR evaluations in near-uniform regions, providing a trade-off between rendering performance and approximation error. Used conservatively, both techniques improve efficiency with limited impact on image quality, while STBN-based and hybrid budgeting schemes further improve the perceptual quality of early frames.

Ultimately, these results show that time-varying INRs can be rendered directly and explored continuously without intermediate resampling or surrogate structures. We believe this framework broadens the practical use of neural representations in scientific visualization and enables richer exploration and higher-quality rendering of dynamic volumetric data.

\acknowledgments{%
This research was supported in part by NSF OAC award 2138811, 2609465 (58503144, 58503145, and 58503708), NSF OISE award 2330582 (58503534), and NASA JPL Subcontract No. 56000334. This research was also supported by the U.S. Department of Energy (DOE) under grants DE-SC-0023320 and DE-SC-0023319. This work was further supported in part by the U.S. DOE Lawrence Livermore National Laboratory Laboratory Directed Research and Development program under grant LLNL-LDRD 25-FS-002, and was performed under the auspices of the U.S. Department of Energy by Lawrence Livermore National Laboratory under Contract DE-AC52-07NA27344 (LLNL-CONF-2017782).

This work used Distributed Implicit Neural Representation (DINR), v1.0, licensed from Lawrence Livermore National Security, LLC.

Generative AI tools were used in this work in a limited capacity, including debugging, script generation for benchmarking and plotting, and grammar editing and trimming of human-written text.

The code for this work is available at \href{https://github.com/alpers-git/VIZ-INR}{\nolinkurl{github.com/alpers-git/VIZ-INR}}.
  
}

\bibliographystyle{abbrv-doi-hyperref}

\bibliography{template}

@inproceedings{Wolfe2022STBN,
booktitle = {Eurographics Symposium on Rendering},
editor = {Ghosh, Abhijeet and Wei, Li-Yi},
title = {{Spatiotemporal Blue Noise Masks}},
author = {Wolfe, Alan and Morrical, Nathan and Akenine-Möller, Tomas and Ramamoorthi, Ravi},
year = {2022},
publisher = {The Eurographics Association},
DOI = {10.2312/sr.20221161}
}

@incollection{ahrens2005paraview,
  title     = {{ParaView: An} End-User Tool for Large-Data Visualization},
  editor    = {Charles D. Hansen and Chris R. Johnson},
  booktitle = {Visualization~Handbook},
  year      = {2005},
  author    = {Ahrens, James and Geveci, Berk and Law, Charles}
}

@incollection{Andersson2021Flip,
  author    = {Pontus Andersson and Jim Nilsson and Tomas Akenine{-}M{\"{o}}ller},
  editor    = {Adam Marrs and Peter Shirley and Ingo Wald},
  title     = {{Visualizing and Communicating Errors in Rendered Images}},
  booktitle = {Ray Tracing Gems II},
  year      = {2021},
  chapter   = {19},
  pages     = {301--320}
}

@incollection{carson2021rtgem,
  author    = {Carson Brownlee and David Demarle},
  editor    = {Adam Marrs and Peter Shirley and Ingo Wald},
  title     = {Fast Volumetric Gradient Shading Approximations for Scientific Ray Tracing},
  booktitle = {Ray Tracing Gems II},
  year      = {2021},
  doi            = {https://doi.org/10.1007/978-1-4842-7185-8},
  publisher      = {Apress Berkeley, CA}
}

@misc{Childs2012visit,
  author    = {Hank Childs and Eric Brugger and Brad Whitlock and Jeremy Meredith and Sean Ahern and David Pugmire and Kathleen Biagas and Mark Miller and Cyrus Harrison
               and Gunther H. Weber and Hari Krishnan and Thomas Fogal and Allen Sanderson
               and Christoph Garth and E. Wes Bethel and David Camp and Oliver R\"{u}bel
               and Marc Durant and Jean M. Favre and Paul Navr\'{a}til},
  title     = {{VisIt: An End-User Tool For Visualizing and Analyzing Very Large Data}},
  year      = {2012},
  booktitle = {High Performance Visualization--Enabling Extreme-Scale Scientific Insight}
}

@misc{cuda,
 author={NVIDIA and Vingelmann, Péter and Fitzek, Frank H.P.},
  title  = {{CUDA release: 13}},
  author = {{NVIDIA Corporation}},
  url   = {{https://developer.nvidia.com/cuda-toolkit}, Accessed: 27 March 2026}
}

@inproceedings{fong:2017,
  author    = {Fong, Julian and Wrenninge, Magnus and Kulla, Christopher and Habel, Ralf},
  title     = {{Production} {Volume} {Rendering}: {SIGGRAPH} 2017 {Course}},
  year      = {2017},
  doi       = {10.1145/3084873.3084907},
  booktitle = {ACM SIGGRAPH 2017 Courses},
  series    = {SIGGRAPH '17}
}

@article{Gunther:2016,
  title   = {{MCFTLE: Monte Carlo Rendering of Finite-Time {Lyapunov} Exponent Fields}},
  volume  = {35},
  doi     = {10.1111/cgf.12914},
  author  = {G{\"u}nther, Tobias and Kuhn, Alexander and Theisel, Holger},
  journal = {Computer Graphics Forum},
  year    = {2016}
}

@article{Hadwinger2005RealtimeIsosurface,
  author  = {Hadwiger, Markus and Sigg, Christian and Scharsach, Henning and Bühler, Khatja and Gross, Markus},
  title   = {{Real-Time Ray-Casting and Advanced Shading of Discrete Isosurfaces}},
  volume  = {24},
  journal = {Computer Graphics Forum},
  doi     = {10.1111/j.1467-8659.2005.00855.x},
  year    = {2005}
}

@article{Hofmann:2020,
  title   = {{Neural Denoising for Path Tracing of Medical Volumetric Data}},
  author  = {Hofmann, Nikolai and Martschinke, Jana and Engel, Klaus and Stamminger, Marc},
  volume  = {3},
  number  = {2},
  journal = {ACM Transactions on Graphics (Proceedings of SIGGRAPH~'20)},
  doi     = {10.1145/3406181},
  year    = {2020}
}

@article{Kalos2010EfficientFreePath,
  journal = {Computer Graphics Forum},
  title   = {{Free Path Sampling in High Resolution Inhomogeneous Participating Media}},
  author  = {Szirmay-Kalos, László and Tóth, Balázs and Magdics, Milán},
  year    = {2011},
  doi     = {10.1111/j.1467-8659.2010.01831.x},
  volume  = {30},
  number  = {1}
}

@inproceedings{kruger2003accelerationgpu,
  author    = {Kr{\"u}ger, J. and Westermann, R.},
  booktitle = {Proceedings of IEEE Visualization},
  title     = {{Acceleration Techniques for GPU-based Volume Rendering}},
  year      = {2003},
  series    = {IEEE Visualization Conference},
  doi       = {10.1109/VISUAL.2003.1250384}
}

@article{Kutz2017SDT,
  title   = {Spectral and Decomposition Tracking for Rendering Heterogeneous Volumes},
  author  = {Kutz, Peter and Habel, Ralf and Li, Yining Karl and Nov{\'a}k, Jan},
  journal = {ACM Transactions on Graphics (Proceedings of SIGGRAPH 2017)},
  year    = {2017},
  volume  = {36},
  number  = {4},
  issn    = {0730-0301},
  doi     = {10.1145/3072959.3073665}
}

@inproceedings{laine2013MegaKernels,
author = {Laine, Samuli and Karras, Tero and Aila, Timo},
title = {Megakernels considered harmful: wavefront path tracing on GPUs},
year = {2013},
publisher = {Association for Computing Machinery},
doi = {10.1145/2492045.2492060},
booktitle = {Proceedings of the 5th High-Performance Graphics Conference},
pages = {137–143},
numpages = {7},
series = {HPG '13}
}

@inproceedings{ljung2006multiresolution,
  author    = {Ljung, Patric and Lundström, Claes and
               Ynnerman, Anders},
  booktitle = {EUROVIS - Eurographics/IEEE VGTC Symposium on Visualization},
  title     = {{Multiresolution Interblock Interpolation in Direct
               Volume Rendering}},
  year      = {2006},
  doi       = {10.2312/VisSym/EuroVis06/259-266}
}

@article{Martschinke:2019,
  title   = {{Adaptive Temporal Sampling for Volumetric Path Tracing of Medical Data}},
  author  = {Martschinke, Jana and Hartnagel, Stefan and Keinert, Benjamin and Engel, Klaus and Stamminger, Marc},
  journal = {Computer Graphics Forum},
  year    = {2019},
  doi = {https://doi.org/10.1111/cgf.13771},
}

@inproceedings{morrical2019spaceskip,
  author    = {Morrical, Nate and Usher, Will and Wald, Ingo and Pascucci, Valerio},
  booktitle = {2019 IEEE Visualization Conference},
  title     = {{Efficient Space Skipping and Adaptive Sampling of Unstructured Volumes Using Hardware Accelerated Ray Tracing}},
  year      = {2019},
  doi       = {10.1109/VISUAL.2019.8933539}
}

@article{morrical2020rtxpointlocext,
  author  = {Morrical, Nathan and Wald, Ingo and Usher, Will and Pascucci, Valerio},
  journal = {IEEE Transactions on Visualization and Computer Graphics},
  title   = {Accelerating Unstructured Mesh Point Location with {RT} Cores},
  year    = {In Press},
  volume  = {},
  number  = {},
  pages   = {1-14},
  doi={10.1109/TVCG.2020.3042930}
}

@article{Morrical2022QuickClusters,
  author  = {Morrical, Nate and Sahistan, Alper and Güdükbay, Uğur and Wald, Ingo and Pascucci, Valerio},
  journal = {IEEE Transactions on Visualization and Computer Graphics},
  title   = {{Quick} {Clusters}: {A} {GPU}-{Parallel} {Partitioning} for {Efficient} {Path} {Tracing} of {Unstructured} {Volumetric} {Grids}},
  year    = {2023},
  volume  = {29},
  number  = {01},
  doi     = {10.1109/TVCG.2022.3209418}
}

@article{Morrical2023AARBF,
  author  = {Morrical, Nate and Zellmann, Stefan and Sahistan, Alper and Shriwise, Patrick and Pascucci, Valerio},
  journal = {IEEE Transactions on Visualization and Computer Graphics},
  title   = {{Attribute-Aware RBFs: Interactive Visualization of Time Series Particle Volumes Using RT Core Range Queries}},
  year    = {2023},
  volume  = {30},
  number  = {1},
  doi     = {10.1109/TVCG.2023.3327366}
}

@article{Novak2014Residual,
  author    = {Nov\'{a}k, Jan and Selle, Andrew and Jarosz, Wojciech},
  title     = {Residual Ratio Tracking for Estimating Attenuation in Participating Media},
  year      = {2014},
  publisher = {Association for Computing Machinery},
  journal   = {ACM Trans. Graph.},
  volume    = {33},
  number    = {6},
  doi       = {10.1145/2661229.2661292}
}

@misc{optix,
  title  = {{NVIDIA OptiX Ray Tracing Engine}},
  author = {{NVIDIA Corporation}},
  note   = {Available at \url{https://developer.nvidia.com/optix}, Accessed: 27 March 2026}
}

@misc{Zvonek2025PruningAE,
  title={Pruning {AMR}: Efficient Visualization of Implicit Neural Representations via Weight Matrix Analysis},
  author={Jennifer Zvonek and Andrew Gillette},
  journal={ArXiv},
  year={2025},
  volume={abs/2512.02967},
   doi           = {10.48550/arXiv.2512.02967},
  url           = {https://arxiv.org/abs/2512.02967}
}

@incollection{OverOperator,
  author    = {Milan Ikits and Joe Kniss and 
               Aaron Lefohn and Charles Hansen},
  title     = {{Volume Rendering Techniques}},
  booktitle = {{GPU} Gems},
  note      = {Available at \url{https://developer.nvidia.com/sites/all/modules/custom/gpugems/books/GPUGems/gpugems_ch39.html}, Accessed: 24 June 2022},
  year      = {2004}
}

@article{Perlin1989Hypertexture,
  author  = {Perlin, K. and Hoffert, E. M.},
  title   = {Hypertexture},
  year    = {1989},
  journal = {SIGGRAPH Comput. Graph.},
  volume  = {23},
  number  = {3},
  doi     = {10.1145/74334.74359}
}

@ARTICLE{sahistan2026MDWT,
  author={Sahistan, Alper and Zellmann, Stefan and Miao, Haichao and Morrical, Nate and Wald, Ingo and Pascucci, Valerio},
  journal={IEEE Transactions on Visualization and Computer Graphics}, 
  title={Materializing Inter-Channel Relationships With Multi-Density Woodcock Tracking}, 
  year={2026},
  volume={32},
  number={3},
  pages={2726-2740},
  doi={10.1109/TVCG.2026.3653310}}

@inproceedings{Georgiev2016BlueNoiseDithering,
author = {Georgiev, Iliyan and Fajardo, Marcos},
title = {Blue-noise dithered sampling},
year = {2016},
publisher = {Association for Computing Machinery},
doi = {10.1145/2897839.2927430},
booktitle = {ACM SIGGRAPH 2016 Talks},
articleno = {35},
numpages = {1},
series = {SIGGRAPH '16}
}

@ARTICLE{Bauer2023FoVoNet,
author={Bauer, David and Wu, Qi and Ma, Kwan-Liu},
journal={ IEEE Transactions on Visualization \& Computer Graphics },
title={{ FoVolNet: Fast Volume Rendering using Foveated Deep Neural Networks }},
year={2023},
volume={29},
number={01},
pages={515-525},
doi={10.1109/TVCG.2022.3209498},
publisher={IEEE Computer Society}}

@inproceedings{Koskela2017FoveatedInstant,
author = {Koskela, Matias and Immonen, Kalle and Viitanen, Timo and J\"{a}\"{a}skel\"{a}inen, Pekka and Multanen, Joonas and Takala, Jarmo},
title = {Foveated instant preview for progressive rendering},
year = {2017},
publisher = {Association for Computing Machinery},
doi = {10.1145/3145749.3149423},
booktitle = {SIGGRAPH Asia 2017 Technical Briefs},
articleno = {10},
}

@article{Guenter2012Foveated3D,
author = {Guenter, Brian and Finch, Mark and Drucker, Steven and Tan, Desney and Snyder, John},
title = {Foveated 3D graphics},
year = {2012},
publisher = {Association for Computing Machinery},
volume = {31},
number = {6},
doi = {10.1145/2366145.2366183},
journal = {ACM Trans. Graph.},
articleno = {164},
}

@incollection{Pharr2023Volume,
  edition   = {4},
  title     = {{11 Volume Scattering}},
  booktitle = {{Physically based rendering: From theory to implementation}},
  publisher = {The MIT Press},
  author    = {Pharr, Matt and Jakob, Wenzel and Humphreys, Greg},
  year      = {2023},
  pages     = {697-736}
}

@article{Patney2016FoveatedVR,
author = {Patney, Anjul and Salvi, Marco and Kim, Joohwan and Kaplanyan, Anton and Wyman, Chris and Benty, Nir and Luebke, David and Lefohn, Aaron},
title = {Towards foveated rendering for gaze-tracked virtual reality},
year = {2016},
publisher = {Association for Computing Machinery},
volume = {35},
number = {6},
doi = {10.1145/2980179.2980246},
journal = {ACM Trans. Graph.},
articleno = {179}
}

@INPROCEEDINGS {Ye2022RectFoveated,
author = { Ye, Jiannan and Xie, Anqi and Jabbireddy, Susmija and Li, Yunchuan and Yang, Xubo and Meng, Xiaoxu },
booktitle = { 2022 IEEE on Conference Virtual Reality and 3D User Interfaces (VR) },
title = {{ Rectangular Mapping-based Foveated Rendering }},
year = {2022},
pages = {756-764},
doi = {10.1109/VR51125.2022.00097},
url = {https://doi.ieeecomputersociety.org/10.1109/VR51125.2022.00097},
publisher = {IEEE Computer Society}}

@inproceedings{Rottger2003SmartHAVR,
  author    = {Röttger, Stefan and Guthe, Stefan and Weiskopf, Daniel and Ertl, Thomas and Straßer, Wolfgang},
  year      = {2003},
  title     = {{Smart Hardware-Accelerated Volume Rendering}},
  booktitle = {Proceedings of the Symposium on Data Visualisation 2003},
  series    = {VISSYM '03},
  volume    = {3}
}

@inproceedings{sahistan2025MDWT,
booktitle = {Eurographics Symposium on Parallel Graphics and Visualization},
title = {{Multi-Density Woodcock Tracking: Efficient \& High-Quality Rendering for Multi-Channel Volumes}},
author = {Sahistan, Alper and Zellmann, Stefan and Morrical, Nate and Pascucci, Valerio and Wald, Ingo},
year = {2025},
DOI = {10.2312/pgv.20251150}
}

@article{Sahistan2024VisNonTrivialPart,
  author={Sahistan, Alper and Demirci, Serkan and Wald, Ingo and Zellmann, Stefan and Barbosa, João and Morrical, Nate and Güdükbay, Uğur},
  journal={IEEE Transactions on Visualization and Computer Graphics}, 
  title={Visualization of Large Non-Trivially Partitioned Unstructured Data with Native Distribution on High-Performance Computing Systems}, 
  year={2024},
  volume={},
  number={},
  pages={1-14},
  doi={10.1109/TVCG.2024.3427335}}

@article{wald2017ospray,
  author  = {Wald, I and Johnson, GP and Amstutz, J and Brownlee, C and Knoll, A and Jeffers, J and G\"{u}nther, J and Navratil, P},
  journal = {IEEE Transactions on Visualization and Computer Graphics},
  title   = {{OSPRay - A CPU} Ray Tracing Framework for Scientific Visualization},
  year    = {2017},
  volume  = {23},
  number  = {1},
  pages   = {931-940},
doi={10.1109/TVCG.2016.2599041}
}

@inproceedings{wald2019rtxpointloc,
  booktitle = {Proceedings of High-Performance Graphics - Short Papers},
  series    = {HPG~'19},
  title     = {{RTX} Beyond Ray Tracing: Exploring the Use of Hardware Ray Tracing Cores for Tet-Mesh Point Location},
  author    = {Wald, Ingo and Usher, Will and Morrical, Nathan and Lediaev, Laura and Pascucci, Valerio},
  year      = {2019},
  doi = {10.2312/hpg.20191189}
}

@article{wald2021exabrick,
  title   = {{Ray Tracing Structured {AMR} Data Using {ExaBricks}}},
  journal = {IEEE Transactions on Visualization and Computer Graphics},
  author  = {Wald, Ingo and Zellmann, Stefan and Usher, Will and Morrical, Nate and Lang, Ulrich and Pascucci, Valerio},
  year    = {2021},
  volume  = {27},
  number  = {2},
  doi     = {10.1109/TVCG.2020.3030470}
}

@techreport{woodcock:1965,
  title       = {{Techniques used in the {GEM} Code for {Monte} {Carlo} Neutronics Calculation in Reactors and Other Systems of Complex Geometry}},
  series      = {Applications of Computing Methods to Reactor Problems},
  institution = {Argonne National Laboratory},
  author      = {Woodcock, E. and Murphy, T. and Hemmings P. and T.C., L},
  year        = {1965}
}

@inproceedings{Wu_VisItOSPRay_2018,
  booktitle = {Eurographics Symposium on Parallel Graphics and Visualization},
  editor    = {Hank Childs and Fernando Cucchietti},
  title     = {{VisIt}-{OSPRay}: {Toward} an {Exascale} {Volume} {Visualization} {System}},
  author    = {Wu, Qi and Usher, Will and Petruzza, Steve and Kumar, Sidharth and Wang, Feng and Wald, Ingo and Pascucci, Valerio and Hansen, Charles D.},
  year      = {2018},
  publisher = {The Eurographics Association},
  issn      = {1727-348X},
  isbn      = {978-3-03868-054-3},
  DOI = {10.2312/pgv.20181091}
}

@article{Yue:2010,
  title   = {{Unbiased, Adaptive Stochastic Sampling for Rendering Inhomogeneous Participating Media}},
  author  = {Yue, Yonghao and Iwasaki, Kei and Chen, Bing-Yu and Dobashi, Yoshinori and Nishita, Tomoyuki},
  journal = {ACM Transactions on Graphics},
  year    = {2010},
  doi     = {10.1145/1882261.1866199},
  volume  = {29},
  number  = {6}
}

@article{zellmann:2022,
  author  = {Zellmann, Stefan and Seifried, Daniel and Morrical, Nathan and Wald, Ingo and Usher, Will and Law-Smith, Jamie and Walch-Gassner, Stefanie and Hinkenjann, Andr{\'e}},
  journal = {Computing in Science   Engineering},
  title   = {Point Containment Queries on Ray Tracing Cores for {AMR} Flow Visualization},
  year    = {2022},
  volume  = {},
  number  = {},
  DOI={10.1109/mcse.2022.3153677}
}

@article{Zellmann2024Beyond,
  journal = {Computer Graphics Forum},
  title   = {{Beyond ExaBricks: GPU Volume Path Tracing of AMR Data}},
  author  = {Zellmann, Stefan and Wu, Qi and Sahistan, Alper and Ma, Kwan-Liu and Wald, Ingo},
  year    = {2024},
  volume  = {43},
  number  = {3},
  doi     = {10.1111/cgf.15095}
}

@ARTICLE{Mohan2025DINR,
  author={Mohan, K. Aditya and Ferrucci, Massimiliano and Divin, Chuck and Stevenson, Garrett A. and Kim, Hyojin},
  journal={IEEE Transactions on Computational Imaging}, 
  title={Distributed Stochastic Optimization of a Neural Representation Network for Time-Space Tomography Reconstruction}, 
  year={2025},
  volume={11},
  number={},
  pages={362-376},
  doi={10.1109/TCI.2025.3547265}}

@inproceedings{zavorotny2025CacheINR,
booktitle = {Eurographics Symposium on Parallel Graphics and Visualization},
editor = {Reina, Guido and Rizzi, Silvio and Gueunet, Charles},
title = {{From Cluster to Desktop: A Cache-Accelerated INR framework for Interactive Visualization of Tera-Scale Data}},
author = {Zavorotny, Daniel and Wu, Qi and Bauer, David and Ma, Kwan-Liu},
year = {2025},
publisher = {The Eurographics Association},
ISSN = {1727-348X},
ISBN = {978-3-03868-274-5},
DOI = {10.2312/pgv.20251153}
}

@article{hachisuka2008multidimensional,
    author  = {Hachisuka, Toshiya and Jarosz, Wojciech and Weistroffer, Richard Peter and Dale, Kevin and Humphreys,
               Greg and Zwicker, Matthias and Jensen, Henrik Wann},
    title   = {Multidimensional Adaptive Sampling and Reconstruction for Ray Tracing},
    journal = {ACM Transactions on Graphics (Proceedings of SIGGRAPH)},
    volume  = {27},
    number  = {3},
    year    = {2008},
    pages   = {33:1--33:10},
    doi     = {10/fm6c2w}
}

@inproceedings{Rousselle2011AdaptiveSampling,
author = {Rousselle, Fabrice and Knaus, Claude and Zwicker, Matthias},
title = {Adaptive sampling and reconstruction using greedy error minimization},
year = {2011},
publisher = {Association for Computing Machinery},
doi = {10.1145/2024156.2024193},
booktitle = {Proceedings of the 2011 SIGGRAPH Asia Conference},
articleno = {159},
numpages = {12},
location = {Hong Kong, China},
series = {SA '11}
}

@INPROCEEDINGS{Viola2004ImportanceVolume,
  author={Viola, I. and Kanitsar, A. and Groller, M.E.},
  booktitle={IEEE Visualization 2004}, 
  title={Importance-driven volume rendering}, 
  year={2004},
  volume={},
  number={},
  pages={139-145},
  doi={10.1109/VISUAL.2004.48}}

@article{han2025dcinr,
  title={DCINR: a Divide-and-Conquer Implicit Neural Representation for Compressing Time-Varying Volumetric Data in Hours},
  author={Han, Jun and Yang, Fan},
  journal={IEEE Transactions on Visualization and Computer Graphics},
  year={2025},
  publisher={IEEE},
  doi={10.1109/TVCG.2025.3564255},
}

@article{tang2023ecnr,
  title={ECNR: Efficient compressive neural representation of time-varying volumetric datasets},
  author={Tang, Kaiyuan and Wang, Chaoli},
  journal={arXiv preprint arXiv:2311.12831},
  year={2023}
}

@ARTICLE{11264349,
  author={Han, Jun and Tang, Kaiyuan and Wang, Chaoli},
  journal={IEEE Transactions on Visualization and Computer Graphics}, 
  title={MoE-INR: Implicit Neural Representation with Mixture-of-Experts for Time-Varying Volumetric Data Compression}, 
  year={2025},
  volume={},
  number={},
  pages={1-11},
  doi={10.1109/TVCG.2025.3633893}}

@article{han2022coordnet,
  title={Coordnet: Data generation and visualization generation for time-varying volumes via a coordinate-based neural network},
  author={Han, Jun and Wang, Chaoli},
  journal={IEEE Transactions on Visualization and Computer Graphics},
  volume={29},
  number={12},
  pages={4951--4963},
  year={2022},
  publisher={IEEE},
  doi={10.1109/TVCG.2022.3197203}
}

@inproceedings{di2016fast,
  title={Fast error-bounded lossy HPC data compression with SZ},
  author={Di, Sheng and Cappello, Franck},
  booktitle={2016 ieee international parallel and distributed processing symposium (ipdps)},
  pages={730--739},
  year={2016},
  organization={IEEE}
}

@article{lindstrom2014fixed,
  title={Fixed-rate compressed floating-point arrays},
  author={Lindstrom, Peter},
  journal={IEEE transactions on visualization and computer graphics},
  volume={20},
  number={12},
  pages={2674--2683},
  year={2014},
  publisher={IEEE},
   doi={10.1109/TVCG.2014.2346458}
}

@article{lu2021compressive,
  author  = {Yuzhe Lu and Kairong Jiang and Joshua A. Levine and Matthew Berger},
  journal = {Computer Graphics Forum},
  month   = {6},
  number  = {3},
  pages   = {135--146},
  title   = {Compressive Neural Representations of Volumetric Scalar Fields},
  volume  = {40},
  year    = {2021},
  doi = {10.1111/cgf.14295}
}

@inproceedings{saragadam2022miner,
  title={Miner: Multiscale implicit neural representation},
  author={Saragadam, Vishwanath and Tan, Jasper and Balakrishnan, Guha and Baraniuk, Richard G and Veeraraghavan, Ashok},
  booktitle={European Conference on Computer Vision},
  pages={318--333},
  year={2022},
  organization={Springer},
  doi = {10.1007/978-3-031-20050-2_19},
}

@inproceedings{reiser2021kilonerf,
  title={Kilonerf: Speeding up neural radiance fields with thousands of tiny {MLPs}},
  author={Reiser, Christian and Peng, Songyou and Liao, Yiyi and Geiger, Andreas},
  booktitle={Proceedings of the IEEE/CVF international conference on computer vision},
  pages={14335--14345},
  year={2021},
  doi={10.1109/ICCV48922.2021.01407}
}

@article{reiser2023merf,
  title={Merf: Memory-efficient radiance fields for real-time view synthesis in unbounded scenes},
  author={Reiser, Christian and Szeliski, Rick and Verbin, Dor and Srinivasan, Pratul and Mildenhall, Ben and Geiger, Andreas and Barron, Jon and Hedman, Peter},
  journal={ACM Transactions on Graphics (ToG)},
  volume={42},
  number={4},
  pages={1--12},
  year={2023},
  publisher={ACM New York, NY, USA},
  doi = {10.1145/3592426}
}

@inproceedings{Nehab2007ShadingReprojection,
author = {Nehab, Diego and Sander, Pedro V. and Lawrence, Jason and Tatarchuk, Natalya and Isidoro, John R.},
title = {Accelerating real-time shading with reverse reprojection caching},
year = {2007},
publisher = {Eurographics Association},
booktitle = {Proceedings of the 22nd ACM SIGGRAPH/EUROGRAPHICS Symposium on Graphics Hardware},
pages = {25–35},
numpages = {11},
location = {San Diego, California},
series = {GH '07},
isbn = {9781595936257}
}

@article{Yang2020TAA,
author = {Yang, Lei and Liu, Shiqiu and Salvi, Marco},
title = {A Survey of Temporal Antialiasing Techniques},
journal = {Computer Graphics Forum},
volume = {39},
number = {2},
pages = {607-621},
doi = {https://doi.org/10.1111/cgf.14018},
year = {2020}
}

@article{Mildenhall2021NeRF,
author = {Mildenhall, Ben and Srinivasan, Pratul P. and Tancik, Matthew and Barron, Jonathan T. and Ramamoorthi, Ravi and Ng, Ren},
title = {NeRF: representing scenes as neural radiance fields for view synthesis},
year = {2021},
publisher = {Association for Computing Machinery},
volume = {65},
number = {1},
doi = {10.1145/3503250},
journal = {Commun. ACM},
pages = {99–106},
numpages = {8}
}

@article{kerbl20233d,
  title={3d gaussian splatting for real-time radiance field rendering.},
  author={Kerbl, Bernhard and Kopanas, Georgios and Leimk{\"u}hler, Thomas and Drettakis, George and others},
  year={2023}
}

@article{bauer2025gscache,
  title={GSCache: Real-Time Radiance Caching for Volume Path Tracing using 3D Gaussian Splatting},
  author={Bauer, David and Wu, Qi and Gadirov, Hamid and Ma, Kwan-Liu},
  journal={arXiv preprint arXiv:2507.19718},
  year={2025}
}

@inproceedings{han2025toward,
  title={Toward Distributed 3D Gaussian Splatting for High-Resolution Isosurface Visualization},
  author={Han, Mengjiao and Sewell, Andres and Insley, Joseph and Knowles, Janet and Mateevitsi, Victor A and Papka, Michael E and Petruzza, Steve and Rizzi, Silvio},
  booktitle={2025 IEEE International Conference on eScience (eScience)},
  pages={331--332},
  year={2025},
  organization={IEEE}
}

@article{dyken2025volume,
  title={Volume Encoding Gaussians: Transfer Function-Agnostic 3D Gaussians for Volume Rendering},
  author={Dyken, Landon and Sewell, Andres and Usher, Will and Debardeleben, Nathan and Petruzza, Steve and Kumar, Sidharth},
  journal={arXiv preprint arXiv:2504.13339},
  year={2025}
}

@ARTICLE{Rojo1209SciVisa,
  author = { Baeza Rojo, Irene and G{\"u}nther, Tobias },
  title = { Vector Field Topology of Time-Dependent Flows in a Steady Reference Frame },
  journal = { IEEE Transactions on Visualization and Computer Graphics (Proc. IEEE Scientific Visualization) },
  location = { Vancouver, Canada },
  year = { 2019 },
}

@article{Popinet2004Tangaroa,
  author = {Popinet, St{\'e}phane and Smith, Murray and Stevens, Craig},
  title = {Experimental and Numerical Study of the Turbulence Characteristics of Airflow around a Research Vessel},
  journal = {Journal of Atmospheric and Oceanic Technology},
  volume = {21},
  number = {10},
  pages = {1575-1589},
  year = {2004},
  doi = {10.1175/1520-0426(2004)021<1575:EANSOT>2.0.CO;2},
}

@ARTICLE{gerrisflowsolver,
  author = {S. Popinet},
  title = {Free Computational Fluid Dynamics},
  journal = {ClusterWorld},
  year = {2004},
  volume = {2},
  number = {6},
  url = {http://gfs.sf.net/}
}

@inproceedings{sitzmann2019siren,
    author = {Sitzmann, Vincent
              and Martel, Julien N.P.
              and Bergman, Alexander W.
              and Lindell, David B.
              and Wetzstein, Gordon},
    title = {Implicit Neural Representations
              with Periodic Activation Functions},
    booktitle = {Proc. NeurIPS},
    year={2020}
}

@misc{Muller2021tinyCudaNN,
  author       = {Thomas M{\"u}ller},
  title        = {{tiny-cuda-nn}},
  year         = {2021},
  howpublished = {\url{https://github.com/NVlabs/tiny-cuda-nn}},
  note         = {Version 2.0, released April 21, 2021}
}

@misc{reed2021dynamicctreconstructionlimited,
      title={Dynamic CT Reconstruction from Limited Views with Implicit Neural Representations and Parametric Motion Fields}, 
      author={Albert W. Reed and Hyojin Kim and Rushil Anirudh and K. Aditya Mohan and Kyle Champley and Jingu Kang and Suren Jayasuriya},
      year={2021},
      eprint={2104.11745},
      archivePrefix={arXiv},
      primaryClass={eess.IV},
      url={https://arxiv.org/abs/2104.11745},
       doi={10.1109/ICCV48922.2021.00226}
}

\end{document}